%% file: DS3Ver1.tex
\documentclass[10pt,a4paper,article]{amsart}
\usepackage{amssymb,stmaryrd}
\usepackage{amsthm,amstext}
\usepackage{amssymb,amsmath}
\usepackage{amsfonts}
\usepackage{MnSymbol,slashed}
\usepackage{fancyhdr}
\usepackage{graphicx}
\usepackage{pdfpages}
\usepackage{float}
\usepackage{subfig}
\usepackage{simplewick}
\usepackage{graphics,psfrag}
\usepackage{pstricks}
\usepackage{latexsym}
\usepackage{changepage}
\usepackage{caption}
\usepackage{paralist}
\usepackage{tikzit}
\input{basic.tikzstyles}
\input{quantum.tikzstyles}
\allowdisplaybreaks[1]

\theoremstyle{plain}
\newtheorem{theorem}{Theorem}[section] 

\newtheorem{lemma}[theorem]{Lemma}
\newtheorem{corollary}[theorem]{Corollary}
\newtheorem{proposition}[theorem]{Proposition}
\newtheorem{definition}[theorem]{Definition}
\newtheorem{example}[theorem]{Example}
\newtheorem{remark}[theorem]{Remark}
\newcommand{\CC}{{\hbox{{$\mathcal C$}}}}

\newcommand{\CH}{\hbox{{$\mathcal H$}}}

\newcommand{\CM}{\hbox{{$\mathcal M$}}}

\newcommand{\CL}{\hbox{{$\mathcal L$}}}

\newcommand{\CR}{\hbox{{$\mathcal R$}}}
\newcommand{\CQ}{\hbox{{$\mathcal Q$}}}

\newcommand{\C}{\mathbb{C}}
\newcommand{\R}{\mathbb{R}}

\newcommand{\Z}{\mathbb{Z}}

\newcommand{\sign}{\mathrm{sign}}

\newcommand{\del}{\partial}

\newcommand{\isom}{{\cong}}
\newcommand{\eps}{{\epsilon}}
\newcommand{\tens}{\mathop{{\otimes}}}
\newcommand{\la}{{\triangleright}}
\newcommand{\ra}{{\triangleleft}}

\newcommand{\id}{\mathrm{id}}

\newcommand{\<}{\langle}
\renewcommand{\>}{\rangle}
\newcommand{\End}{\mathrm{ End}}
\newcommand{\Tr}{\mathrm{ Tr}}

\def\rcross{{\triangleright\!\!\!<}}
\def\lcross{{>\!\!\!\triangleleft}}

\def\bicross{{\blacktriangleright\!\!\triangleleft}}
\def\dcross{{\bowtie}}

\newcommand{\vac}{{|{\rm vac}\>}}
\newcommand{\vacket}{\<\mathrm{vac}|}
\renewcommand{\o}{{}_{(1)}}
\renewcommand{\t}{{}_{(2)}}

\begin{document}

\author{Alexander Cowtan and Shahn Majid${}^1$}
\address{Department of Computer Science,  University of Oxford\\
\& School of Mathematical Sciences, Queen Mary University of London \\ 
\& Cambridge Quantum Computing}
\email{akcowtan@gmail.com\\
s.majid@qmul.ac.uk}
\thanks{${}^1$ On sabbatical at Cambridge Quantum Computing}

\title{Quantum double aspects of surface code models}
	\begin{abstract} We revisit the Kitaev model for fault tolerant quantum computing on a square lattice with underlying quantum double $D(G)$ symmetry, where $G$ is a finite group. We provide projection operators for its quasiparticles content as irreducible representations of $D(G)$ and combine this with $D(G)$-bimodule properties of open ribbon excitation spaces $\CL(s_0,s_1)$ to show how open ribbons can be used to teleport information between their  endpoints $s_0,s_1$. We give a self-contained account that builds on earlier work but emphasises applications to quantum computing as surface code theory, including gates on $D(S_3)$. We show how the theory reduces to a simpler theory for toric codes in the case of $D(\Z_n)\isom\C\Z_n^2$, including toric ribbon operators and their braiding. In the other direction, we show how our constructions generalise to $D(H)$ models based on a finite-dimensional Hopf algebra $H$, including site actions of $D(H)$ and partial results on ribbon equivariance even when the Hopf algebra is not semisimple.  \end{abstract}
\maketitle

\section{Introduction}

The idea of fault-tolerant quantum computing using topological methods has been around for some years now, notably the Kitaev model in the original work\cite{Kit} and important sequels such as \cite{Bom,BSW,BMCA,Meu}. Here we add to this growing body of literature with a renewed focus on the quantum double $D(G)$ Hopf algebra symmetry  implicit in the original Kitaev model, where $G$ is a finite group. The model here is built on a suitable oriented graph but for our purposes we focus on a fixed oriented square lattice. The Hilbert space $\CH$ of the system is then the tensor product over all arrows of a vector space with basis $G$ at every arrow. Every site, by which we mean a choice of a face and vertex on it, carries a representation of the quantum group $D(G)$. In general `quasiparticles' in the model are defined as irreducible representations of this quantum group and we explain how these can be detected using certain projection operators $P_{\CC,\pi}$,  where $\CC$ is a conjugacy class in $G$ and $\pi$ is an irreducible representation of the isotropy group. We then study quasiparticles at the end-points $s_0,s_1$ of an open ribbon $\xi$, again taking a $D(G)$ approach to the ribbon operator commutation relations. Most of these results are in Section~\ref{secG} but a more sophisticated view of ribbon operators as left and right module maps $F_\xi:D(G)^*\to \End(\CH)$ is deferred to Section~\ref{secH} as a warm up for the generalisation there. 

Of particular interest in the paper is the space $\CL(s_0,s_1)$ of states created from a local vacuum by all possible ribbon operations $F_\xi$ for a fixed $\xi$. This was a key ingredient in \cite{Kit} and its independence as a subspace of $\CH$ on deformations of the ribbon expresses the topological nature of the model. Our results here build on ideas in \cite{BSW} whereby this space carries the left action of $D(G)$ at $s_0$ and another action, which we view as a right action, at $s_1$. The space is then isomorphic to $D(G)$ itself as a bimodule under left and right multiplication and hence subject to its Peter-Weyl decomposition as a direct sum of $\End(V_{\CC,\pi})$ over all quasiparticle irrep spaces $V_{\CC,\pi}$. We use this to create a state $|{\rm Bell};\xi\>\in\CL(s_0,s_1)$ and show that this can be used to teleport information between $s_0,s_1$. We illustrate the theory further as well as give more details and examples of quantum computations for $D(S_3)$ in Section~\ref{secS3}, where $S_3$ is the group of permutations on 3 elements. Likewise, the theory simplifies but carries some of the same structures in the toric case $D(\Z_n)$ for which the ribbon theory is in Section~\ref{secredZn}. 

The paper begins with a preliminary warm up Section~\ref{secZn} which sets up the basic ideas as this easier level of $D(\Z_n)$ but from the point of view of this as $\Z_n\times\Z_n$ with a certain factorisable quastiriangular structure in the sense of Drinfeld\cite{Dri}. The body of the paper concludes in Section~\ref{secH} with some partial results going the other way to $D(H)$, where $H$ is a finite-dimensional Hopf algebra. The Kitaev theory at this level but with $H$ semisimple so that (over $\C$) we have $S^2=\id$ was introduced in \cite{BMCA} where it was was shown that one has a $D(H)$ action at every site, but without explicitly considering ribbons. The latter, however, are special cases of `holonomy maps' in the follow-up work \cite{Meu}, again in the semisimple case. This work focusses more on the topological and `gauge theory' aspects rather than ribbon operators specifically, thus at the very least we aim in the semisimple case for a much more explicit treatment of what is already known in some form. Thus, our main result of the section on ribbon operators as left and right module maps $D(H)^*\to\End(\CH)$ is similar to \cite[Thm~8.1]{Meu} except that that applies to a special class of holonomy operators that explicitly do not include ribbon ones, and our proofs are much more explicit. For example, the equivariance of the smallest open ribbons (which are base for our induction) is proven in Figures~\ref{figTLpf},~\ref{figLTpf} by Hopf algebra calculations.

The bottom line, however, is that the theory is known to generalise well to $H$ semisimple and the most novel aspect of Section~\ref{secH} is that we do as much as we can without assuming this. Computationally speaking, the $H$ non-semisimple case loses the interpretation of the integral actions as check operators which are measured to detect unwanted excitations. In addition, ribbon operators on the vacuum are no longer in general equivalent up to isotopy. For this reason, the logical space does not enjoy the same `topological protection' as the semisimple case. On the other hand, we find that there is no problem with a $D(H)$ action at every site, but for dual-triangle operators and ribbon operators involving them, we need two versions ${}^{(\pm)}L$ depending one whether we use $S^{\pm 1}$ at the relevant incoming arrow. That means that the same ribbon operator is not a module map from both the left and the right at the same time. This obstruction can also be put on the faces and is  not a deal breaker, but requires more study for a fully worked out theory. For example, in the quasitriangular Hopf algebra case the two are equivalent by conjugation, $S=u S^{-1}(\ )u^{-1}$ for Drinfeld's element $u\in H$ in \cite{Dri}. Thus, this aspect of Section~\ref{secH} should be seen as first steps in a fully general Kitaev theory. 

In fact such a more general theory is needed in order to apply to quantum groups such as $u_q(sl_2)$ at roots of unity, which in turn would be needed to connect up to ideas for quantum computing based on modular tensor categories associated to such non-semi-simple quantum groups. For example, the Fibonacci anyons surveyed in \cite{fib} are based on $u_q(sl_2)$ at $q^5=1$. The double $D(u_q(sl_2))$ here  also underlies the Turaev-Viro invariant of 3-manifolds and hence this should certainly be a source of topological stability if the Kitaev model can be extended to such cases. If so, it would then be related closely to $2+1$ quantum gravity with point sources, which is a viable theory and another reason to expect that this is ultimately possible. There are many other obstacles also, however, to such a programme, some of which are discussed in the final Section~\ref{secrem}. We also discuss there other issues for topological quantum computing and possible links with ZX-calculus.

\subsection*{Acknowledgements}  While completing the writing of other parts of the paper, there appeared the preprint  \cite{Chen} which covers some of the same ground as Section 4 with regard to the ribbon operators in the semisimple case where $S^2=\id$. Our approach is different and is, moreover, directed to exposing the issues for the general non-semisimple case. We thank the team at CQC for helpful discussions. 

\section{Preliminaries: Toric code $D(\Z_n)$ model}\label{secZn}

 Let $\C Z_n$ denote the group Hopf algebra with generator $h$ where $h^n=1$ and $\Delta h=h\tens h$, $\eps h=1$, $Sh=h^{-1}$ for the coproduct, counit and antipode. Let $\C(\Z_n)$ be the Hopf algebra of functions on $\Z_n$ with a basis of $\delta$-functions on $\Z_n={0,1,\cdots,n-1}$ and $\Delta\delta_i=\sum_j \delta_j\tens\delta_{i-j}$, $\eps(\delta_i)=\delta_{i,0}$, $S\delta_i=\delta_{-i}$ for the Hopf algebra structure. The normalised integrals in these Hopf algebras are
 \[ \Lambda={1\over n}\sum_i h^i \in \C\Z_n,\quad \Lambda^*=\delta_0\in \C(\Z_n).\]
 
 The quantum double $D(\Z_n)=\C(\Z_n)\tens\C\Z_n\isom\C \Z_n\times\C\Z_n$ as Hopf algebras, since the groups are Abelian, and since (over $\C$) $\C\Z_n\isom \C(\Z_n)$ by the
Fourier isomorphism 
\begin{equation}\label{Zisom} g\mapsto \sum_i q^{i}\delta_i,\quad \delta_i\mapsto {1\over n} \sum_k q^{-ik}g^k,\end{equation}
where $q$ is a primitive $n$th root of unity. Now the double is $\C \Z_n\times\C\Z_n$ with generators $g,h$ respectively for the two copies, commuting and obeying $h^n=g^n=1$. Under this isomorphism, the general $D(G)$ theory in Section~\ref{secG} looks much simpler  and we therefore treat this toric case first as a model for the later sections.

Denoting the generators of the two copies of $\C\Z_n$ in this form of the double  by $g,h$ respectively, the $D(\Z_n)$ quasitriangular structure and quantum Killing form are
\[ \CR={1\over n}\sum_{i,j} q^{-ij}g^i\tens h^j,\quad \CQ=\sum_{i,j,k,l} q^{-ij-kl}g^ih^l\tens g^kh^j\]
to see that the latter is nondegenerate (the derivation of this from the usual form $D(\Z_n)$ will be given in Section~\ref{secredZn} later.) 

Now let $\Sigma = \Sigma(V, E, P)$ be a square lattice viewed as a directed graph with its usual (cartesian) orientation. The Hilbert space will be a tensor product of vector spaces with one copy of $\C\Z_n$ at each arrow $e \in E$. We denote the basis of each copy by $|i\>$. Next, for each vertex $v \in V$ and each face $p \in P$ we define an action of $\Z_n$ which acts on the vector spaces around the vertex or around the face, and trivially elsewhere, according to
\[ \includegraphics[scale=0.7]{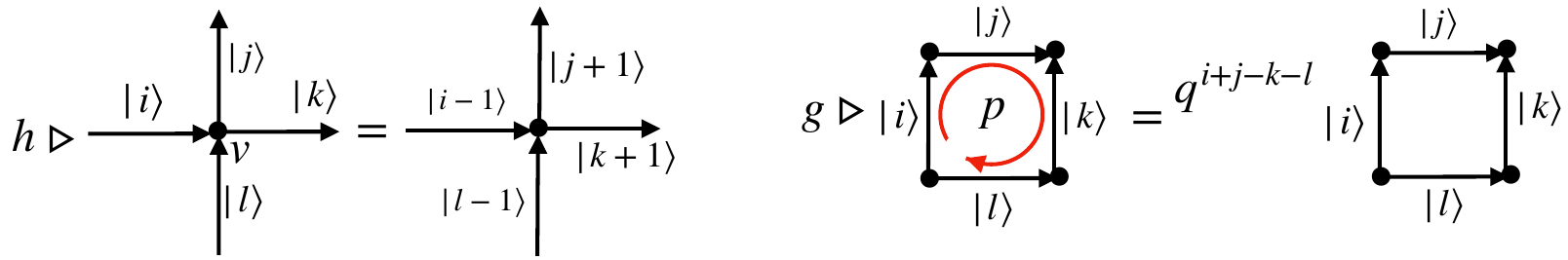}\]
These are built from four-fold copies  of the operator $X$ and its adjoint and of $Z$ and its adjoint, where $X|i\>=|i+1\>$ and $Z|i\>=q^i|i\>$ obey $ZX=qXZ$. Here $h\la$ subtracts in the case of arrows pointing towards the vertex and $g\la$ has $k,l$ entering negatively in the exponent because these are contra to a {\em clockwise} flow around the face in our conventions.  These combine to an action of $\Z_n\times\Z_n$ at every `site' $(v,p)$ defined as a vertex $v$ and an adjacent face $p$ (the exact placement of $v$ in relation to $p$ is not relevant in an Abelian group model such as this).   

\begin{lemma} For every site $(v,p)$, the operators $g\la$ and $h\la$ commute and give a representation of $\Z_n\times\Z_n$
on the Hilbert space $\CH$. 
\end{lemma}
\proof This is a direct calculation acting on the 6 relevant vector spaces, of which two are in common to the two actions, see
\[\includegraphics[scale=0.7]{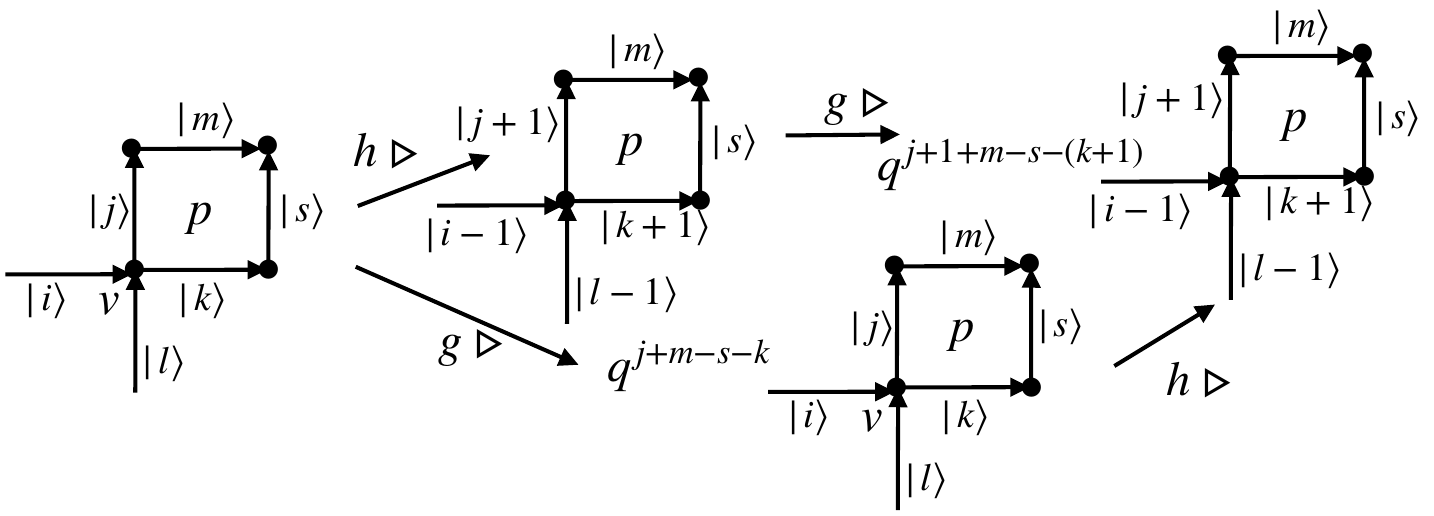}\]
 \endproof

The same applies trivially if $v$ and $p$ are not adjacent (as they have no arrow in common). Thus, we can in fact consider $h\la$ determined by a vertex $v$ and $g\la$ determined by a face $p$ independently. With this in mind, we define 
\[ A(v)=\Lambda\la={1\over n}\sum_m (h\la)^m,\quad B(p)=\Lambda^*\la={1\over n}\sum_m (g\la)^m\]
where now $\Lambda^*=n^{-1}\sum_i g^i$ according to (\ref{Zisom}), and these necessarily commute. In fact it is easy to see that
\[ A(v)^2=A(v),\quad B(p)^2=B(p),\quad [A(v),A(v')]=[B(p),B(p')]=[A(v),B(p)]=0.\]

We then define the Hamiltonian 
\[ H=\sum_v(1-A(v))+\sum_p(1-B(p))=-(\sum_v A(v) + \sum_p B(p))+{\rm const.}\]
and define the set of vacuum states
\[ \CH_{vac}=\{ |\xi\>\in\CH\ |\  A(v)|\xi\>=B(p)|\xi\>=|\xi\>, \forall v,p\}.\]  

Vacuum states are `topologically protected' from errors which are sufficiently local, which we will make precise later.

Next, the irreducible representations of the double in this form are
\[  \pi_{ij}(g)=q^i,\quad \pi_{ij}(h)=q^j,\]
which as all 1-dimensional. We denote these by
\[ 1=\pi_{00},\quad e_i=\pi_{0i},\quad m_i=\pi_{i0},\quad \eps_{ij}=\pi_{ij},\quad i,j\in{1,\cdots,n-1}\]
with braiding
\[\Psi_{1,*}=\Psi_{*,1}=\Psi_{e_i,e_j}=\Psi_{m_i,m_j}=\Psi_{e_i,m_j}=\Psi_{\eps_{jk},m_i}=\Psi_{e_i,\eps_{jk}}=1,\]
\[  \Psi_{m_i,e_j}=q^{ij},\quad \Psi_{m_i,\eps_{jk}}= q^{ik},\quad \Psi_{\eps_{jk},e_i}= q^{ij},\quad \Psi_{\eps_{ij},\eps_{kl}}=q^{il}\]

\bigskip
Next, we define projectors associated to $\pi_{ij}$ namely
\[ P_{ij}={1\over n^2}\sum_{kl}({\rm Tr}_{\pi_{ij}}g^{-k}h^{-l})g^k h^l=P_i^g P_j^h,\quad P_i^g={1\over n}\sum_k q^{-ik}g^k\]
in the group algebra of $\Z_n\times\Z_n$, built from projectors in each $\Z_n$ (here $P_j^h$ is defined in the same way on the other copy). The projectors on one copy obey 
 $P^g_iP^g_j=\delta_{ij}P^g_i$ and $\sum_iP^g_i=1$ and similarly for $P^h_j$, so that 
\[ P_{ij}P_{i'j'}=\delta_{ii'}\delta_{jj'}P_{ij},\quad \sum_{i,j}P_{ij}=1.\]
At every vertex $v$, every face $p$ and every site $(v,p)$, we have specific projection operators on $\CH$ given by
\[ P_i(p)=P_i^g\la,\quad P_j(v)=P_j^h\la,\quad P_{ij}(v,p)=P_{ij}\la\]
for the actions above on the relevant arrows. We consider these projectors as quantum mechanical observables asking, for $i,j\ne 0$:
\begin{itemize}
\item $P_i(p)$  -- {\em is there a quasiparticle of type $m_{i}$  occupying face $p$?} 
\item $P_j(v)$ -- {\em is there a quasiparticle of type $e_{j}$  occupying vertex $v$?}
\item $P_{ij}(v,p)$ -- {\em is there a quasiparticle of type $\eps_{ij}$ occupying site $(v,p)$?}
\end{itemize}
All these projectors commute, so we can ask these questions independently with Boolean logic.  Thus $P_i(p)=\sum_{j=0}^{n-1}P_{ij}(v,p)$ asks if the site is occupied by $i$ at the face and anything, including nothing, at the vertex. Similarly for $P_j(v)$. The remaining case is the complement of the others, so $P_{00}(v,p)$, which asks if there is a trivial representation quasiparticle at $(v,p)$, is equivalent to the absence of the above excitations, i.e. we regard it as a local vacuum. Note also that
\[ P_0(v)=A(v),\quad P_0(p)=B(p),\quad P_{00}(v,p)=A(v)B(p)\]
which gives the meaning of these. This $A(v)$ asks if there is no excitation at the vertex independently of the face, etc. 

\begin{lemma}\label{toricvac} Let $|\psi\>\in\CH$. For all $i,j\in \Z_n$:

\begin{enumerate} 
\item   $P_{i}(p)|\psi\>=|\psi\>$ if and only if $g\la |\psi\> = q^i |\psi\>$ for the four arrows around $p$.
\item   $P_{j}(v)|\psi\>=|\psi\>$ if and only if $h\la |\psi\> = q^j |\psi\>$ for the four arrows around $v$.
\item  $P_{ij}(v,p)|\psi\>=|\psi\>$ if and only if $g\la |\psi\> = q^i |\psi\>$ and $h\la |\psi\> = q^j |\psi\>$ for the six arrows at the site.
\item  $\vac \in \CH_{vac}$ if and only if $P_{ij}(v,p)\vac =0$ for all $(i,j)\ne(0,0)$ and at all sites $(v,p)$. 
\end{enumerate}
On a closed plane, there is a unique vacuum state (up to normalisation):

\[ 
\vac = \prod_{v \in V} A(v) \bigotimes_{E} |0\>
\]
where $0$ is the group identity of $\Z_n$.
\label{lem:toric_vac}
\end{lemma}
\proof (1) $P_{i}(p)$ acts on the four-arrow state $|i_1\>\tens |i_2\>\tens |i_3\>\tens |i_4\>$ in order around the face by ${1\over n}\sum_k q^{-ia}q^{a(i_1+i_2-i_3-i_4)}=\delta_{i,i_1+i_2-i_3-i_4}$. So invariant states are linear combinations of ones with $i_1+i_2-i_3-i_4=i$ going around the face. These are precisely the local states where $g\la |\psi\>=q^i|\psi\>$. 

(2) Linear combinations of $|i_i\>\tens |i_2\>\tens |i_3\>\tens |i_4\>$ in order around the vertex that are invariant under $P_{j}(v)$ are of the form
\[|\psi\>= \sum_{b}q^{-jb}|i_1-b\>\tens |i_2+b\>\tens |i_3+b\>\tens |i_4-b\>\]
and these are also the local states where 
\[ h\la |\psi\>=\sum_{b}q^{-jb}|i_1-b-1\>\tens |i_2+b+1\>\tens |i_3+b+1\>\tens |i_4-b-1\>=q^j|\psi\>\]

(3) Considering the site $(v,p)$ with $p$ to the upper right of $v$ as before, the joint eigenvectors for (1) and (2) are of the form 
\[|\psi\>= \sum_{b}q^{-jb}|i_1-b\>\tens |i_2+b\>\tens |i_3+b\>\tens |i_4-b\>\tens |i_5\>\tens |i_6\>;\quad i_2+i_5-i_6-i_3=i\]
where we take them in order round the vertex then around the face. These are also the local states where $g\la |\psi\>=q^i|\psi\>$ and  $ h\la |\psi\>=q^j|\psi\>$.

(4) We just note that $P_0(v)=A(v)$, $P_0(p)=B(p)$ so $P_{00}(v,p)=A(v)B(p)$. So if $|\psi\>\in \CH_{vac}$ then $P_{00}|\psi\>=|\psi\>$ i.e. there are no excitations, at every site $(v,p)$. Moreover, for $(i,j)\ne (0,0)$, $P_{ij}|\psi\>=P_{ij}P_{00}|\psi\>=0$ by the projector orthogonalilty, again at every site. Conversely, if $P_{ij}|\psi\>=0$ for all $(i,j)\ne (0,0)$ at $(v,p)$ then $P_{00}|\psi\>=|\psi>$ as $\sum_{ij}P_{ij}=1$ while $A(v)|\psi\>=\sum_iP_{i0}|\psi\>=P_{00}|\psi\>=|\psi\>$ and similarly for $B(p)$. If this is true at every site then  $|\psi\>\in H_{vac}$. 

Note that (4) is the same as saying that if the system is in a vacuum state there is no excitation at any site. We can see this directly as $h \la \circ \Lambda \la = \Lambda \la$ at a given vertex. So if $\vac $ is in $\CH_{vac}$ as defined above then $h\la \vac =h\la A(v)\vac =h\Lambda\la\vac =\Lambda\la\vac =\vac $. Similarly for $g\la \vac =\vac $. This agrees with the analysis above. 
The vacuum state $\vac$ may be verified by directly checking the definition of $\CH_{vac}$. We will see later that this state is unique in $\CH_{vac}$ as a special case of Corollary~\ref{cor:plane}.

\subsection{Quasiparticle creation and transportation}\label{sec:quasiparticles}

We now consider concretely how to create quasiparticles on the lattice. Assume the system has state $\vac \in \CH_{vac}$. Consider the arrow between vertices $v_2$ and $v_1$ on the boundary of faces $p_1$ and $p_2$, 
\[
\input{toric_exampleA.tikz}
\]
For some $j\in \Z_n$, consider $Z^{-j}$ acting on $|s\>$, which we denote $Z^{-j}_s$ and takes $|s\> \mapsto q^{-sj}|s\>$: 
\[
\input{toric_exampleB.tikz}
\]
Then, $h \la_{v_1} Z^{-j}_s\vac = q^j Z^{-j}_s h \la_{v_1} \vac = q^{j}Z^{-j}_s\vac$ and similarly $h \la_{v_2} Z^{-j}_s\vac = q^{-j} Z^{-j}_s\vac$, which is easy to check using commutation relations. By Lemma~\ref{toricvac}, all neighbouring sites $(v_1, p_a)$ and $(v_2, p_a)$ are occupied by $m_{j}$ and $m_{-j}$, where $p_a$ is any neighbouring face.
Let $X^{-i}$ further act on $Z^{-j}_s|s\>$ alone, for some $i \in \Z_n$:
\[
\input{toric_exampleC.tikz}
\]
Now, $g \la_{p_1} (X^{-i} Z^{-j})_s\vac = q^i (X^{-i} Z^{-j})_s g \la_{p_1}\vac= q^i (X^{-i} Z^{-j})_s \vac$ and $g \la_{p_2} (X^{-i} Z^{-j})_s \vac = q^{-i} (X^{-i} Z^{-j})_s \vac$. All neighbouring sites $(v_b, p_1)$ and $(v_b, p_2)$ are now occupied by a quasiparticle $e_{i}$ and $e_{-i}$ respectively, where $v_b$ is any neighbouring vertex. In particular, $(v_1, p_1)$ is occupied by $\pi_{i,j}$, while $(v_2, p_2)$ is occupied by $\pi_{-i,-j}$.

Quasiparticles may be moved on the surface by $X$ and $Z$ edge operations. We next apply $X^{i}$ to $|t\>$: 
\[
\input{toric_exampleD.tikz}
\]
Now, $g \la_{p_2} X^{i}_t \otimes (X^{-i} Z^{-j})_s \vac = X^{i}_t \otimes (X^{-i} Z^{-j})_s \vac$  (being careful about edge orientation). Site $(v_2, p_2)$ is now only occupied by $m_{-j}$. However, the previously unoccupied site $(v_3, p_3)$ is now occupied by $e_{-i}$, as $g \la_{p_3} X^{i}_t|\vec p_3\> = q^{i}X^{i}_t|\vec p_3\>$. Now further apply $Z^{-j}$ acting on $|u\>$: 
\[
\input{toric_exampleE.tikz}
\]
$h \la_{v_2} Z^{-j}_u \otimes X^{i}_t \otimes (X^{-i} Z^{-j})_s \vac = Z^{-j}_u \otimes X^{i}_t \otimes (X^{-i} Z^{-j})_s \vac$, and so site $(v_2, p_2)$ is now unoccupied. Site $(v_3, p_3)$ is occupied by $\pi_{-i,-j}$, as $h \la_{v_3} Z^{-j}_u \otimes X^{i}_t \otimes (X^{-i} Z^{-j})_s \vac = q^{-j} Z^{-j}_u \otimes X^{i}_t \otimes (X^{-i} Z^{-j})_s \vac$. This explanation of creation and transport is quite \textit{ad hoc}. In fact, the above operators are specific instances of ribbon operators, which we describe in Section~\ref{secG}. We delay discussing braiding until then, as it is clearer in terms of ribbons.

\section{$D(G)$ models and example of $D(S_3)$}\label{secG}

Let $G$ be a finite group with identity $e\in G$. We recall that the group Hopf algebra $\C G$ base basis $G$ with product extended linearly and $\Delta h=h\tens h$, $\eps h=1$ and $S h=h^{-1}$ for the Hopf algebra structure. Its dual Hopf algebra $\C(G)$ of functions on $G$ has basis of $\delta$-functions $\{\delta_g\}$ with $\Delta\delta_g=\sum_h \delta_h\tens\delta_{h^{-1}g}$, $\eps \delta_g=\delta_{g,e}$ and $S\delta_g=\delta_{g^{-1}}$ for the Hopf algebra structure. The normalised integrals are
\[ \Lambda={1\over |G|}\sum_{h\in G} h\in \C G,\quad \Lambda^*=\delta_e\in \C(G).\]
For the Drinfeld double we have $D(G)=\C(G)\lcross \C G$, see \cite{Ma},  with $\C G$ and $\C(G)$ subalgebras and the cross relations $h\delta_g =\delta_{hgh^{-1}}h$ (a semidirect product). We will often prefer to refer to $D(G)$ explicitly on the tensor product vector space, then for example the cross relation appears explicitly as $(1\tens h)(\delta_g\tens 1)= (\delta_{hgh^{-1}}\tens 1)(1\tens h)=\delta_{hgh^{-1}}\tens h$ and antipode as $S(\delta_g\tens h)=\delta_{h^{-1}g^{-1}h}\tens h^{-1}$. There is also a  quasitriangular structure which in the subalgebra notation is 
\[ \CR=\sum_{h\in G} \delta_h\tens h\in D(G)\tens D(G).\]

More relevant to us is the representation on  Hilbert space $\CH$, which now is a tensor product of $\C G$ at each arrow. As before, this is associated to a pair $(v,p)$  (a `site') where $v$ is a vertex on the boundary of the face $p$. What is different from the Abelian group case such as for toric codes in Section~\ref{secZn} is that now for the $a\la$ action on $\CH$ we have to pay attention to the exact placement of $v$ in relation to $p$ by drawing dashed line (the `cilium') between $v$ and the interior of $p$ and taking the group elements in order around the face according to
\[ \includegraphics[scale=0.7]{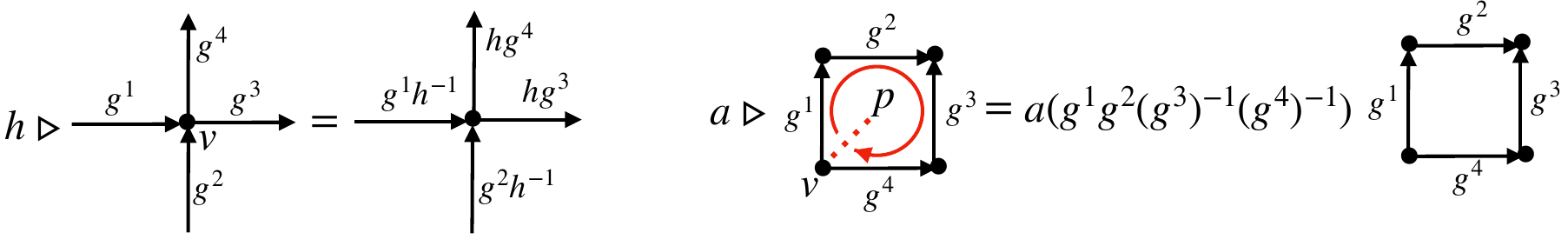}\]

\begin{lemma}\label{lemDGrep} $h\la$ and $a\la$ for all $h\in G$ and $a\in \C(G)$ define a representation of $D(G)$ on $\CH$ associated to each site $(v,p)$. 
\end{lemma}
\proof This follows from the definitions and a check acting on the six affected arrow spaces, see
\[ \includegraphics[scale=0.7]{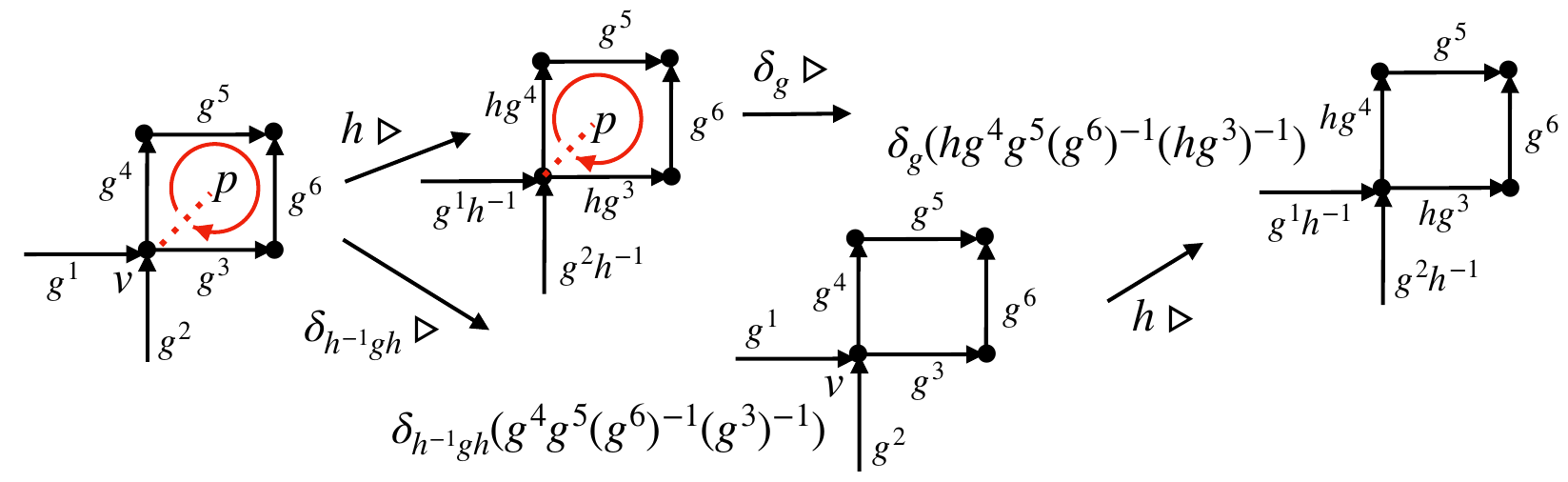}\]
  \endproof 

We next define  
\[ A(v)=\Lambda\la={1\over |G|}\sum_{h\in G}h\la,\quad B(p)=\Lambda^*\la=\delta_e\la\]
where $\delta_{e}(g^1g^2g^3g^4)=1$ iff $g^1g^2g^3g^4=e$ which is iff $(g^4)^{-1}=g^1g^2g^3$ which is iff $g^4g^1g^2g^3=e$. Hence $\delta_{e}(g^1g^2g^3g^4)=\delta_{e}(g^4g^1g^2g^3)$ is invariant under cyclic rotations, hence $\Lambda^*\la$ computed at site $(v,p)$ does not depend on the location of $v$ on the boundary of $p$. Moreover,
\[ A(v)B(p)=|G|^{-1}\sum_h h\delta_e\la=|G|^{-1}\sum_h \delta_{heh^{-1}}h\la=|G|^{-1}\sum_h \delta_{e}h\la=B(p)A(v)\]
 if $v$ is a vertex on the boundary of $p$ by Lemma~\ref{lemDGrep}, and more trivially if not. We also have the rest of
\[ A(v)^2=A(v),\quad B(p)^2=B(p),\quad [A(v),A(v')]=[B(p),B(p')]=[A(v),B(p)]=0\]
for all $v\ne v'$ and $p\ne p'$, as easily checked. We then define
\[ H=\sum_v (1-A(v)) + \sum_p (1-B(p))\]
and the space of vacuum states
\[ \CH_{vac}=\{|\psi\>\in\CH\ |\ A(v)|\psi\>=B(p)|\psi\>=|\psi\>,\quad \forall v,p\}.\]

\subsection{Vacuum space}\label{sec:vac} 

The vacuum space degeneracy depends on the surface topology. Here and throughout the paper, we describe everything very concretely using a square lattice for convenience. While this is obviously possible for a plane, more general surfaces may not admit such a tiling. Precisely, the only 2-dimensional closed orientable surface which admits a (4, 4) tessellation is the torus, which follows from \cite[Thm 1]{tess}. However, the following well-known theorem, and results throughout this paper, apply for other (ciliated, ribbon) graphs embedded into a closed orientable surface. We avoid getting into the weeds on the subject of topological graph theory, but observe that while the lattice will primarily be square, in some places there will have to be irregular faces or vertices. Face and vertex operators generalise straightforwardly to such irregularities.

\begin{theorem}
Let $\Sigma$ be a closed, orientable surface. Then
\[
\dim(\CH_{vac}) = |\mathrm{Hom}(\pi_1(\Sigma), G)/G|.
\]
\label{thm:cui}
where the $G$-action on any $\phi \in \mathrm{Hom}(\pi_1(\Sigma), G)$ is $\phi \mapsto \{h \phi h^{-1}\ |\ h \in G \}$.
\end{theorem}

For completeness we prove this in the Appendix~\ref{app:vacuum}, mostly following \cite{Cui}, and in the process presenting an orthogonal basis for $\CH_{vac}$. This implies, in particular:
\begin{corollary}\label{cor:plane}
Let $\Sigma$ be planar, with no boundaries. Then the vacuum state $\vac$ is unique up to normalisation, and
\[
\vac = \prod_{v \in V} A(v) \bigotimes_{E} e
\]
where $e$ is the group identity of $G$ and $\tens$ is over the arrows.
\end{corollary}
\proof We have assumed that $\pi_1(\Sigma) = \{e\}$ and clearly $\mathrm{Hom}(\{e\}, G) = \{e\}$, $\{e\}/G = \{e\}$. Hence, the vacuum is unique. To find it, define $g := \bigotimes_{E} e$, and observe that $B(p) g = g$ for all $p \in P$, so $g \in S$. Since $B(p)$ commutes with every $A(v)$ commute with, it follows that $B(p)\vac=\vac$. Moreover, applying $A(v)$ for a fixed $v$ to $\vac$, this combines with $A(v)$ in the product to give $A(v)$ again, hence $A(v)\vac=\vac$. Hence, we have constructed the vacuum state. \endproof 

We specify that the plane has no boundaries for Corollary~\ref{cor:plane} because Theorem~\ref{thm:cui} holds only for \textit{closed} surfaces; the `plane' can then be thought of as an infinite sphere. The treatment of boundaries requires adding more algebraic structure to the model, and in general splits vacuum degeneracy \cite{BSW}. It is also obvious that if $\Sigma$ is a closed orientable surface and $G$ is Abelian so that the $G$-action by conjugation is trivial, then
\[
\dim(\CH_{vac}) = |\mathrm{Hom}(\pi_1(\Sigma), G)|.
\]

The Kitaev model may be used to perform fault-tolerant quantum computation -- indeed, the $D(G)$ model corresponds to a class of quantum error-correcting codes in the sense of \cite{KL}, according to \cite{Cui}. If we consider the vacuum to be the logical space of a quantum computer and by following the proof of Theorem~\ref{thm:cui}, we observe that the only non-trivial operators in $\mathrm{End}(\CH_{vac})$ are non-contractible closed loops on the lattice.
Operators which do not form closed paths take the system out of $\CH_{vac}$, and introduce excitations. In particular, considering the quantum computer to be operating in a noisy environment, errors on the lattice which introduce unwanted excitations may be detected using the projectors $A(v), B(p)$ and corrected. Undetectable errors must therefore be sufficiently non-local as to form undetectable non-trivial holonomies; we thus refer to the logical state of the computer as being `topologically protected'.

To run algorithms of practical interest, the model must be capable of supporting a large Hilbert space, but Corollary~\ref{cor:plane} tells us that a boundary-less plane is only capable of supporting a single vacuum state. There are therefore 3 methods of encoding data in Kitaev models:

\begin{enumerate}
\item Build the lattice $\Sigma$ as a torus with $k$ holes, which can encode data in the degenerate vacuum state using $\pi_1(\Sigma)$.
\item Incorporate \textit{gapped boundaries} or \textit{topological defects} into the lattice, which are compatible (in some suitable sense) with the algebra of $D(G)$ and allow for additional vacuum states \cite{BSW, Bom}.
\item Use excited states to encode data. This method requires that $G$ be non-Abelian, as the $D(G)$ model does not admit degenerate excited states on the plane when $G$ is Abelian without the addition of topological features such as boundaries \cite{Kit}.
\end{enumerate}

\subsection{Quasiparticles and projection operators to detect them}

We now return to the underlying algebra of the Kitaev model. The `quasiparticles' in the theory are labelled by irreducible representations of $D(G)$. A couple of standard but not generally  irreducible right representations of $D(G)$ on $\C G$ itself  are
\[ {\rm (i)}\quad g\ra h=gh,\quad  g\ra \delta_h=g \delta_{h,e};\quad {\rm (ii)}\quad  g\ra h=h^{-1}gh,\quad g\ra \delta_h=g \delta_{g,h}.\]
More generally, as a semidirect product, irreducible representations of $D(G)$ are given by standard theory as labelled by pairs $(\CC,\pi)$ consisting of an orbit under the action (i.e. by a conjugacy class $\CC\subset G$ in the present case) and an irrep $\pi$ of the isotropy subgroup $C_G$ of a fixed element $r_{\CC}\in\CC$ (in our case its centraliser i.e. $n\in G$ such that $nr_{\CC}=r_{\CC} n$), the choice of which does not change the group up to isomorphism but does change how it sits inside $G$. Here $\CC$ is called the `magnetic charge' and $\pi$ is called the `electric charge'. Special cases corresponding to $e_i$ and $m_i$ respectively in the toric case are
\[ {\rm chargeons}\quad (\{e\},\pi),\quad \delta_gh\la w=\delta_{g,e}\pi(h)w;\quad  {\rm fluxions}\quad (\CC,1),\quad \delta_g h\la c=\delta_{g,hch^{-1}}hch^{-1}\]
acting on the representation space $V_\pi$ of $\pi$ as an irrep of $G$, and the span $\C\CC$ of the conjugacy class,  respectively. The braiding of two fluxions or a fluxion with a chargeon, for example, are
\[\Psi(f\tens f')=\sum_g g\la f'\tens \delta_g\la f=ff'f^{-1}\tens f,\quad \Psi(f\tens w)=\sum_g g\la w\tens \delta_g\la f=\pi(f)w\tens f.\]
The irrep associated to general $(\CC,\pi)$ can be described as follows\cite{Ma:dg}. First, fix a map
\begin{equation}\label{qC} q:\CC\to G,\quad q_c r_\CC q_{c}^{-1}=c,\quad \forall c\in \CC,\quad \end{equation}
and define from this a `cocycle' $\zeta:\CC\times G\to C_G$ respectively defined and characterised by
\[ \zeta_c(g)=q_{gcg^{-1}}^{-1} g q_{c};\quad \zeta_c(gh)=\zeta_{hch^{-1}}(g)\zeta_c(h)\]
for all $c\in\CC$ and $g,h\in G$. The quantum double action on $\C \CC\tens V_\pi$ is then
\begin{equation}\label{Cpiirrep} \delta_g h\la (c\tens w)= \delta_{g,hch^{-1}}hch^{-1}\tens \pi(\zeta_c(h))w.\end{equation}
This is irreducible and although the formulae depend on the choice of $q$, different choices give isomorphic representations. In particular, we can right multiply $q_c$ by any element  $n_c\in C_G$, and using this freedom we can suppose that
\begin{equation}\label{qsuppl} q_{r_\CC}=e\end{equation}
which, in particular, ensures that $({e},\pi)$ recovers the chargeon representation rather than an equivalent conjugate of it. Also note $G$ is partitioned into the right cosets of $C_G$ with the quotient space $G/C_G$ identified with $\CC$ by its action on $r_\CC$. This implies that every element $g\in G$ can be uniquely factorised as $g=q_c n$ for some $c\in \CC$ and $n\in C_G$. 

We now describe the projectors\cite{Ma:dg} that detect the presence of such quasiparticles, focussing first  on the electric/chargeon sector. Then for each irrep $\pi$,  such quasiparticles will be detected by measuring a projection operators $P_\pi(v)=P_\pi\la v$ for $v\in \CH$ acting at any site, where $P_\pi$ is a central projection element (central idempotent) in the group algebra $\C G$ given by
\begin{equation}\label{Ppi} P_\pi={\dim V_\pi\over |G|}\sum_{g}(\Tr_{\pi}g^{-1})g\end{equation}
These obey $P_\pi P_{\pi'}=\delta_{\pi,\pi'}P_\pi$ by the orthogonality of characters on finite groups, as well as $\sum_\pi P_\pi=1$ and $P_1=\Lambda$. Centrality is immediate by changing the variable $g$ and symmetry of the trace. For reference, the orthogonality relations for characters on any finite group are
\begin{equation}\label{orth1}
\sum_{h\in G}\Tr_\pi(h^{-1})\Tr_{\pi'}(hg)=\delta_{\pi,\pi'}{|G|\over \dim(V_\pi)}\Tr_\pi(g)\end{equation}
\begin{equation}\label{orth2}
\sum_{\pi\in \hat G}\Tr_\pi(g^{-1})\Tr_\pi(h)=\delta_{\CC_g,\CC_h}|C_G(g)|\end{equation}
for all $h,g\in G$ and $\pi,\pi'\in \hat  G$ the set of irreps up to equivalence. Here $\CC_g$ denotes the conjugacy class containing $g$. We likewise have a projection element $\chi_\CC$ in $\C (G)$ defined as the characteristic function of $\CC$ and $P_\CC(v)=\chi_\CC\la v$ for all $v\in \CH$ acting at any site.  The general case is
\begin{equation}\label{PCpi} P_{\CC,\pi}=\sum_{c\in\CC}\delta_c\tens q_cP_\pi q_c^{-1}={\dim V_\pi\over |C_G|} \sum_{c\in\CC}\sum_{n\in C_G} \Tr_\pi(n^{-1})  \delta_c\tens q_c n q_c^{-1}\end{equation}
where $P_\pi\in \C  C_G$ is for $\pi$ as a representation of $C_G$, and associated site projection operators $P_{\CC,\pi}(v)=P_{\CC,\pi}\la v$ for $v\in \CH$ and action at a site. Here $\dim(V_\pi)/|C_G|=\dim(V_{\CC,\pi})/|G|$. Also note  that 
\[  P_{{e},1}=\Lambda^*\tens \Lambda,\quad P_{{e},\pi}=\Lambda^*\tens P_\pi,\quad P_{\CC,1}=\sum_{c\in \CC}\delta_c \tens q_c\Lambda_{C_G}q_c^{-1},\]
the last of which in the Abelian case is $\delta_c\tens \Lambda$ and recovers the chargeon and fluxion projections to the extent possible. We can also define for a fixed $\CC$, 
\[ \sum_{\pi\in \hat C_G} P_{\CC,\pi}=\sum_{c\in\CC}\delta_c\tens q_c(\sum_{\pi\in \hat C_G} P_\pi)q_c^{-1}=\chi_\CC\tens 1.\]
What we can not do in the nonAbelian case is sum over $\CC$ for a fixed nontrivial $\pi$ as these depend on $\CC$, so we do not have a formula like $\sum_\CC P_{\CC,\pi}=1\tens P_\pi$. 

\begin{lemma}\label{lemPCpi} In $D(G)$, the $P_{\CC,\pi}$ are central and form a complete orthogonal set of projections, 
\begin{equation} P_{\CC,\pi} P_{\CC',\pi'}=\delta_{\CC,\CC'}\delta_{\pi,\pi'}P_{\CC,\pi},\quad \sum_{\CC,\pi}P_{\CC,\pi}=1\end{equation}
\end{lemma} 
\proof This is due to \cite{Ma:dg}, but for completeness we now provide more explicit proofs than given there. Thus,
\begin{align*}P_{\CC,\pi}P_{\CC',\pi'}&=\sum_{c\in \CC, d\in \CC'}(\delta_c\tens q_c P_\pi q_c^{-1})(\delta_d\tens q'_dP_{\pi'}q'{}^{-1}_d)\\
&={\dim V_\pi\over |C_G|}\sum_{c\in \CC, d\in \CC'}\delta_c\sum_{n\in C_G}\Tr_\pi(n^{-1})\delta_{c,q_c n q_c^{-1}d q_c n^{-1}q_c^{-1}}\tens q_c n q_c^{-1}q'_dP_{\pi}q'{}^{-1}_d\\
&={\dim V_\pi\over |C_G|}\sum_{c\in \CC, d\in \CC'}\delta_c\sum_{n\in C_G}\Tr_\pi(n^{-1})\delta_{c,d} \tens q_c n q_c^{-1}q'_d P_{\pi'}q'{}^{-1}_d\\
&=\delta_{\CC,\CC'}\sum_{c\in \CC}\delta_c \tens q_c P_{\pi}P_{\pi'} q_c^{-1}=\delta_{\CC,\CC'}\delta_{\pi,\pi'}\sum_{c\in \CC}\delta_c \tens q_c P_{\pi'} q_c^{-1}=\delta_{\CC,\CC'}\delta_{\pi,\pi'}P_{\CC,\pi}
\end{align*}
where $C_G=C_G(r_\CC)$ and $c=q_c n q_c^{-1}d q_c n^{-1}q_c^{-1}$ iff $ d=q_c n^{-1} q_c^{-1} c q_cnq_c^{-1}=q_c n^{-1} r_\CC n q_c^{-1}=q_c r_\CC q_c^{-1}=c$. Note that if $\CC=\CC'$, which is needed for $c=d$,  then $q=q'$ are the same function and we can cancel $q_c q'{}^{-1}_d$ in this case.  We also have
\[ \sum_{\CC,\pi}P_{\CC,\pi}=\sum_{\CC}\sum_{c\in \CC}\delta_c\tens q_c\left(\sum_{\pi}  P_\pi\right) q_c^{-1}=\sum_{\CC}\sum_{c\in \CC}\delta_c\tens 1=\sum_{\CC}\chi_\CC\tens 1=1\tens 1\]
where we sum over irreps $\pi$  of $C_G$ for each $\CC$. For centrality, 
\begin{align*} P_{\CC,\pi}(\delta_h\tens g)&={\dim V_\pi\over |C_G|}\sum_{c\in \CC}\sum_{n\in C_G}\delta_c \Tr_\pi(n^{-1})\delta_{c,q_c n q_c^{-1}hq_cn^{-1}q_c^{-1}}\tens q_c n q_c^{-1}g\\
&={\dim V_\pi\over |C_G|}\delta_h\chi_\CC(h)\tens \sum_{n\in C_G}\Tr_\pi(n^{-1})q_h n q_h^{-1} g=\chi_\CC(h)\delta_h\tens q_h P_\pi q_h^{-1}g\\
(\delta_h\tens g)P_{\CC,\pi}&=(\delta_h\tens g)\sum_{c\in\CC}\delta_c\tens P_\pi q_c^{-1}=\sum_c\delta_h\delta_{gcg^{-1},h}\tens g q_c P_\pi q_c^{-1}=\chi_\CC(h)\delta_h\tens g q_{g^{-1}d g}P_\pi q^{-1}_{g^{-1}d g}\end{align*}
where for the second equality $c=q_c n q_c^{-1}hq_cn^{-1}q_c^{-1}$ iff $c=h$ by the same calculation as above. But $q_h^{-1}g q_{g^{-1}hg}r_\CC q_{g^{-1}hg}^{-1}g^{-1}q_h=q_h^{-1}g g^{-1}h g g^{-1}q_h=r_\CC$ so $q_h^{-1}g q_{g^{-1}hg}\in C_G$ and therefore commutes with $P_\pi$. \endproof

The origin of these projection operators is the Peter-Weyl decomposition which applies to group algebras and other semisimple Hopf algebras including $D(G)$. We look at the group algebra case first in some detail. Thus, for $\C  G$, there is an isomorphism $\C  G\isom \oplus_\pi \End(V_\pi)$ where the map to each component is to send $g \mapsto \pi(g)_{ij} e_i\tens f^j$ where $e_i$ is a basis of $V_\pi$ and $f^j$ is a dual basis. Here, $e_i\tens f^j$ is the elementary matrix with 1 at the $i,j$ row/column if we identify $\End(V_\pi)=M_{\dim(V_\pi)}(\C )$. We check conventions: if $v=v^ie_i$ then $\pi(g)v=v^i \pi_{kj}e_k\<f^j,e_i\>=e_k\pi_{ki}v^i$ so that $\pi(g)$ acts by matrix multiplication on $(v^i)$ as a column vector. In the converse direction we define
\[ \Phi_{\C G}:\oplus_\pi\End(V_\pi)\to \C G,\quad \Phi(e_i\tens f^j)={\dim V_\pi\over |G|}\sum_{g\in G}\pi(g^{-1})_{ji} g\]
which we see obeys $\Phi(e_i\tens f^i)=P_\pi$. One can check that the map $\Phi$ is an isomorphism of bimodules where $\C G$ acts on itself from the left and the right and acts on $\End(V_\pi)=V_\pi\tens V_\pi^*$ on the left by $\pi$ and on the right by its adjoint. Here $h\la e_i=e_k\pi(h)_{ki}$ and $f^j\ra h=\pi(h)_{jk}f^k$  (the dual basis elements transform the same way as vectors) and $\Phi_{\C G}$ is necessarily surjective as the image of $\sum_\pi \sum_i e_i\tens f^i=\sum P_\pi=1$, given that is is a bimodule map. Moreover,  under $\pi'$, the element $\Phi(e_i\tens f^j)$ maps to
\begin{equation}\label{fullorth}  {\dim V_\pi\over |G|}\sum_{g\in G}\pi(g^{-1})_{ji}\pi'(g)_{kl}e_k\tens f^l=\delta_{\pi,\pi'}e_i\tens f^j\end{equation} as required for the inverse in one direction, which proves that $\Phi_{\C G}$ is injective. The  equality (\ref{fullorth}) used here is equivalent to a stronger version of the orthogonality relations for matrix entries of unitary irreducible representations over $\C$, of which (\ref{orth1}) is a consequence.   This also implies that $\pi'(P_\pi)=\id\delta_{\pi,\pi'}$ and hence that 
\begin{equation}\label{Ppief} P_\pi\la e_i=e_k\pi(P_\pi)_{ki}=e_i,\quad f^j\ra P_\pi=\pi(P_\pi)_{jk} f^k=f^j\end{equation}
if $e_i\in V_\pi$ and $f_j\in V_\pi^*$ respectively, or zero if these are in one of the other components. By the equivariance, these actions are equivalent to the projectors $P_\pi$ acting by left or right multiplication, hence $P_\pi \C G= (\C G)P_\pi\isom \End(V_\pi)$ via $\Phi$.

We now similarly let $D(G)$ act on $\End(V_{\CC,\pi})=V_{\CC,\pi}\tens V_{\CC,\pi}^*$ from the left and right by the given left representation and its adjoint as a right one. It also acts on itself by left and right multiplication. 

\begin{theorem}\label{thmPhi} Taking a basis $\{c\tens e_i\}$ of the $D(G)$ representation $V_{\CC,\pi}$, with dual basis $\{\delta_d\tens f^j\}$, the map $\Phi:\oplus_{\CC,\pi}\End(V_{\CC,\pi})\to D(G)$ given on $\End(V_{\CC,\pi})$ by
\[ \Phi(c\tens e_i\tens \delta_d\tens f^j)=\delta_c\tens q_c\Phi_{\C C_G}(e_i\tens f^j)q_d^{-1}={\dim V_\pi\over |C_G|} \sum_{n\in C_G} \pi(n^{-1})_{ji}  \delta_c\tens q_c n q_d^{-1}\]
is an isomorphism of bimodules. 
\end{theorem}
\proof Using the action (\ref{Cpiirrep}) of $D(G)$ on $V_{\CC,\pi}$ in basis terms
\begin{equation}\label{acteu} (\delta_h\tens g)\la (c\tens e_i)=\delta_{h,gcg^{-1}} gcg^{-1}\tens \pi(q^{-1}_{gcg^{-1}}g q_c)_{ki}e_k, \end{equation}
the left module property of $\Phi$ is
\begin{align*}\Phi((\delta_h\tens g)&\la(c\tens e_i)\tens \delta_d\tens f^j)=\Phi(\delta_{h,gcg^{-1}}gcg^{-1}\tens\pi( q^{-1}_{gcg^{-1}}g q_c)_{ki}e_k\tens\delta_d\tens f^j)\\
&=  \delta_{h,gcg^{-1}}{\dim V_\pi\over |C_G|} \sum_{n\in C_G} \pi(n^{-1})_{jk}\pi( q^{-1}_{gcg^{-1}}g q_c)_{ki}  \delta_{gcg^{-1}}\tens q_{gcg^{-1}} n q_d^{-1}\\
&=\delta_{h,gcg^{-1}}{\dim V_\pi\over |C_G|} \sum_{n'\in C_G} \pi(n'{}^{-1})_{ki}  \delta_{gcg^{-1}}\tens  g q_c n' q_d^{-1}\\
&=(\delta_h\tens g)\Phi(c\tens e_i\tens \delta_d\tens f^j)
\end{align*}
where $n'=q_c^{-1}g^{-1}q_{gcg^{-1}}n$. We check that $n'r_\CC n'^{-1}=q_c^{-1}g^{-1}q_{gcg^{-1}}r_\CC q^{-1}_{gcg^{-1}}gq_c=q_c^{-1}g^{-1}(gcg^{-1})g q_c=r_\CC$ so $n'\in C_G$ in our change of variables. For the other side we first use
\begin{align*} \<(\delta_d\tens f^j)&\ra(\delta_h\tens g),c\tens e_i\>:=\<\delta_d\tens f^j,(\delta_h\tens g)\la(c\tens e_i)\\
&=\<\delta_g\tens f^j,\delta_{h,gcg^{-1}}gcg^{-1}\tens \pi(q^{-1}_{gcg^{-1}}g q_c)_{ki}e_k\>=\delta_{d,h}\delta_{d,gcg^{-1}}\pi(q^{-1}_{gcg^{-1}}g q_c)_{ji}\\
&=\delta_{d,h}\<\delta_{g^{-1}dg}\tens \pi(q^{-1}_dgq_{g^{-1}dg})_{jk}f^k,c\tens e_i\>
\end{align*}
for all $c,e_i$, from which we find the dual action
\begin{equation}\label{actfv} (\delta_d\tens f^j)\ra(\delta_h\tens g)=\delta_{h,d}\delta_{g^{-1}dg}\tens \pi(q^{-1}_dgq_{g^{-1}dg})_{jk}f^k\end{equation}
We then proceed similarly for the right module property
\begin{align*} \Phi(c\tens e_i\tens&(\delta_d\tens f^j)\ra (\delta_h\tens g))=\Phi(c\tens e_i\tens\delta_{h,d}\delta_{g^{-1}dg}\tens \pi(q^{-1}_dgq_{g^{-1}dg})_{jk}f^k)\\
&=\delta_{h,d}{\dim V_\pi\over |C_G|}\sum_{n\in C_G}\delta_c\tens \pi(q^{-1}_dgq_{g^{-1}dg})_{jk}\pi(n^{-1})_{ji}q_c n q_{g^{-1}dg}^{-1}\\
&=\delta_{h,d}{\dim V_\pi\over |C_G|}\sum_{n'\in C_G}\delta_c\tens \pi(n'{}^{-1})_{ji}q_c n' q_d^{-1} g\\
&=\Phi(c\tens e_i\tens \delta_d\tens f^j)(\delta_h\tens g)
\end{align*}
where $n'=n q^{-1}_{g^{-1}dg}g^{-1}q_d$ and one can check that this is in $C_G$. For the last line to identify the product in $D(G)$, we need for any $n\in C_G$ that $q_c n q_d^{-1} h q_d n{}^{-1} q_c^{-1}=c$ if and only if $h=d$.

We now check that $\Phi$ is inverse to the composite of the representations $(\CC,\pi)$ as  maps $D(G)\to \End(V_{\CC,\pi})$. It is already surjective as it is a bimodule map and $\Phi(\sum_{\CC,\pi}\sum_{c,i} c\tens e_i\tens \delta_c\tens f^i)=\sum_{\CC,\pi}P_{\CC,\pi}=1\in D(G)$. Therefore it suffices to check that applying the representation (\ref{acteu}) undoes $\Phi$. Focussing on the block $\End(V_{\CC,\pi})$ and acting with its image on $c'\tens e_{i'}\in V_{\CC',\pi'}$, 
\begin{align*} \Phi(c\tens e_i\tens &\delta_d\tens f^j)\la (c'\tens e_{i'})={\dim V_\pi\over |C_G|}\sum_{n\in C_G}\pi(n^{-1})_{ji}(\delta_c\tens q_c n q_d^{-1})\la(c'\tens e_{i'})\\
&={\dim V_\pi\over |C_G|} \sum_{n\in C_G} \pi(n^{-1})_{ji}\delta_{c,q_c n q_d^{-1} c' q_d n^{-1}q_c^{-1}} c\tens \pi'(q^{-1}_c q_c n q_d^{-1}q_{c'})_{j' i'} e_{j'}\\
&=\delta_{\CC,\CC'}\delta_{d,c'}c\tens {\dim V_\pi\over |C_G|} \sum_{n\in C_G} \pi(n^{-1})_{ji}  \pi'(n)_{j' i'} e_{j'}\\
&=\delta_{\CC,\CC'}\delta_{\pi,\pi'}\delta_{d,c'}\delta_{j,i'} c\tens e_i
\end{align*}
as required. Here, $c=q_c nq_d^{-1}c'q_dn^{-1}q_c^{-1}$ iff $n^{-1}r_\CC=q_d^{-1}c' q_d n^{-1}$ which is iff $q_d r_{\CC}q_d^{-1}=c'$ which is iff $c=c'$. This is zero unless $\CC=\CC'$ also. We then used the full orthogonality (\ref{fullorth}) for the group $C_G$. 

By general arguments as in the group case, it follows that $P_{\CC,\pi}$ acts as the identity on $V_{\CC,\pi}$ and $V_{\CC,\pi}^*$ (and zero on other components). One can also check this explicitly, for example, 
\begin{align*} P_{\CC,\pi}\la (c\tens e_i)&={\dim V_\pi\over |C_G|}\sum_{d\in\CC}\sum_{n\in C_G}\Tr_\pi(n^{-1})\delta_{d,q_d n q_d^{-1}c q_d n^{-1}q_d^{-1}}d \tens e_j \pi(\zeta_c(q_d n q_d^{-1})_{ji}\\
&= {\dim V_\pi\over |C_G|}\sum_{n\in C_G}\Tr_\pi(n^{-1})c \tens e_j \pi( n)_{ji}=c\tens e_i \end{align*}
since $d=q_d n q_d^{-1}c q_d n^{-1}q_d^{-1}$ iff $c=q_d n q^{-1}_d d q_d n^{-1}q_d^{-1}=q_d n r_c n^{-1}q_d^{-1}=q_d  r_\CC q_d^{-1}=d$. We used the strong orthogonality relations. Likewise for $f^v\ra P_{\CC,\pi}=f^v$.  \endproof

We see that, while $\Phi$ clearly sends the identity element or `maximally entangled state' of $V_{\CC,\pi}\tens V_{\CC,\pi}^*$ to $P_{\CC,\pi}$, it also implies a basis of all of $D(G)$ broken down into irreps $(\CC,\pi)$ and elements $\Phi(c\tens e_i\tens \delta_d\tens f^j)$ for each block. We will need this result for the discussion of ribbon teleportation.

\begin{figure}
\[ \includegraphics[scale=0.8]{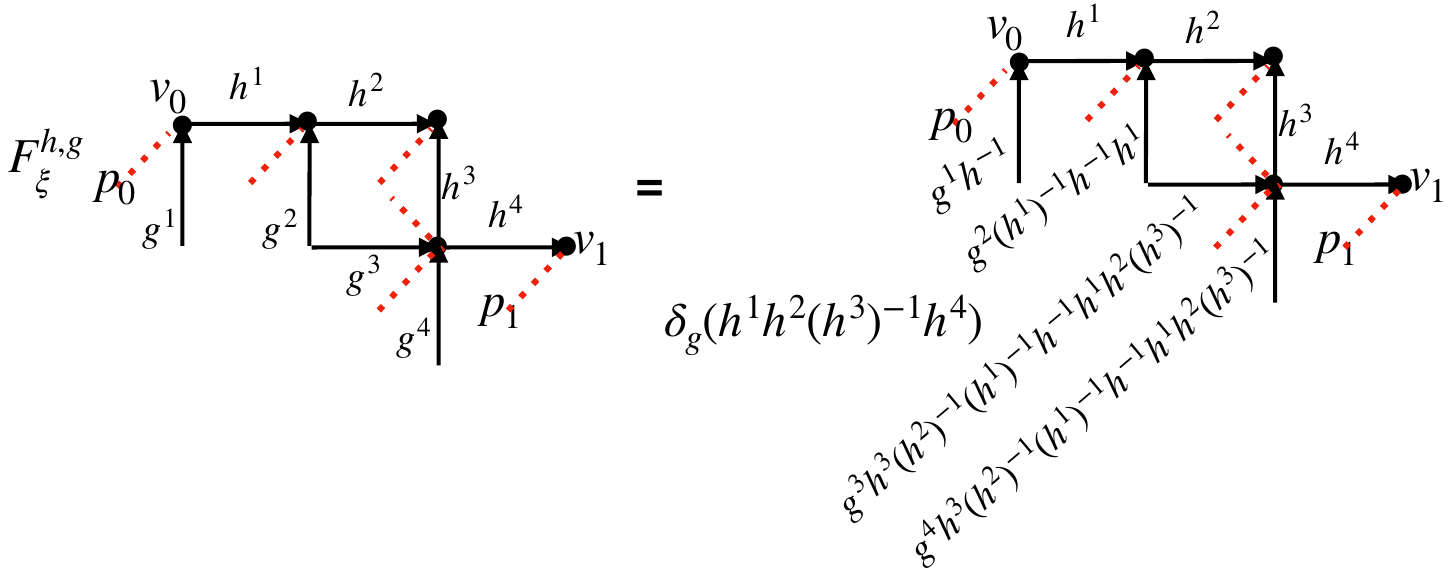}\]
\caption{\label{figribbon} Example of a ribbon operator for a ribbon $\xi$ starting at $s_0=(v_0,p_0)$ to $s_1=(v_1,p_1)$.} 
\end{figure}

\subsection{$D(G)$ ribbon operators}

To discuss the physics further, one needs the notion of a ribbon operator. By definition, a ribbon $\xi$ is a strip of  face width  that  connects two sites $s_0=(v_0,p_0)$ to $s_1=(v_1,p_1)$ 
by a sequence of sites (shown dashed) as for example in Figure~\ref{figribbon}. We call a ribbon \textit{closed} if its endpoints are at the same site, and \textit{open} if the endpoints are at disjoint sites with no intersection.
Note that there exist ribbons which are neither open nor closed, which end at the same vertex but with different faces, say, but we are not concerned with this case. In Figure~\ref{figribbon} we also show an associated ribbon operator $F^{h,g}_\xi$ 
acting on the spaces associated to the participating arrows  and trivially elsewhere. The ribbon has an edge along which we transport $h$ from the initial vertex by conjugation along the path, at 
each vertex of which we apply the conjugated $h$ in the manner of a vertex operation but only to the cross arrow that comes anticlockwise from the dashed site marker. It follows that if we concatenate ribbon $\xi'$ following on from ribbon $\xi$ then we have the first of
\begin{equation}\label{concat} F_{\xi'\circ\xi}^{h,g}=\sum_{f\in G}F_{\xi'}^{f^{-1}hf,f^{-1}g}\circ F_\xi^{h,f};\quad F^{h,g}_\xi \circ F^{h',g'}_\xi=\delta_{g,g'}F_\xi^{hh',g},\end{equation}
where we see the coproduct $\Delta \delta_g$ of $\C(G)$. The latter implies the adjointness
\begin{equation}\label{adjoint}(F_{\xi}^{h,g})^{\dagger} = F_{\xi}^{h^{-1},g}\end{equation}
with respect to the inner product of $\CH$.
 
\begin{example}
\rm Let the state on the l.h.s. of Figure~\ref{figribbon} be $|\psi\>$, and take the inner product with another state:
\[\input{lattice_innerproduct.tikz}\]
\begin{align*}
\<\psi'|(F_{\xi}^{h,g}|\psi\>) &= \delta_g(h^1 h^2 (h^3)^{-1}h^4) \delta_{h'^1}(h^1) \delta_{h'^2}(h^2) \delta_{h'^3}(h^3)\delta_{h'^4}(h^4)\delta_{g'^1}(g^1h^{-1})\\
&\delta_{g'^2}(g^2(h^1)^{-1}h^{-1}h^1)\delta_{g'^3}(g^3h^3(h^2)^{-1}(h^1)^{-1}h^{-1}h^1h^2(h^3)^{-1})\\
&\delta_{g'^4}(g^4h^3(h^2)^{-1}(h^1)^{-1}h^{-1}h^1h^2(h^3)^{-1})\\
&= \delta_g(h'^1 h'^2 (h'^3)^{-1}h'^4) \delta_{h'^1}(h^1) \delta_{h'^2}(h^2) \delta_{h'^3}(h^3)\delta_{h'^4}(h^4)\delta_{g^1}(g'^1h)\\
&\delta_{g^2}(g'^2(h^1)^{-1}hh^1)\delta_{g^3}(g'^3 h^3(h^2)^{-1}(h^1)^{-1}hh^1h^2(h^3)^{-1})\\
&\delta_{g^4}(g'^4h^3(h^2)^{-1}(h^1)^{-1}hh^1h^2(h^3)^{-1})\\
&= (\<\psi'|F_{\xi}^{h^{-1},g}) |\psi\>
\end{align*}
and by (\ref{concat}), $(F^{h, g}_{\xi})^{\dagger}F^{h, g}_{\xi} = F^{h, g}_{\xi}(F^{h, g}_{\xi})^{\dagger} = F^{e, g}_{\xi}$.
\end{example}

Ribbon operators of the form $F^{e,g}_{\xi}$ produce only a scalar $\delta_g(\cdots)$ when applied to a lattice state. It is easy to see that
\begin{equation}
[F^{e,g}_{\xi}, F^{e,g'}_{\xi'}] = 0
\label{eq:delta_ribbons}
\end{equation}
for all $g, g' \in G$ and ribbons $\xi, \xi'$.

\begin{figure}
\[\includegraphics[scale=0.73]{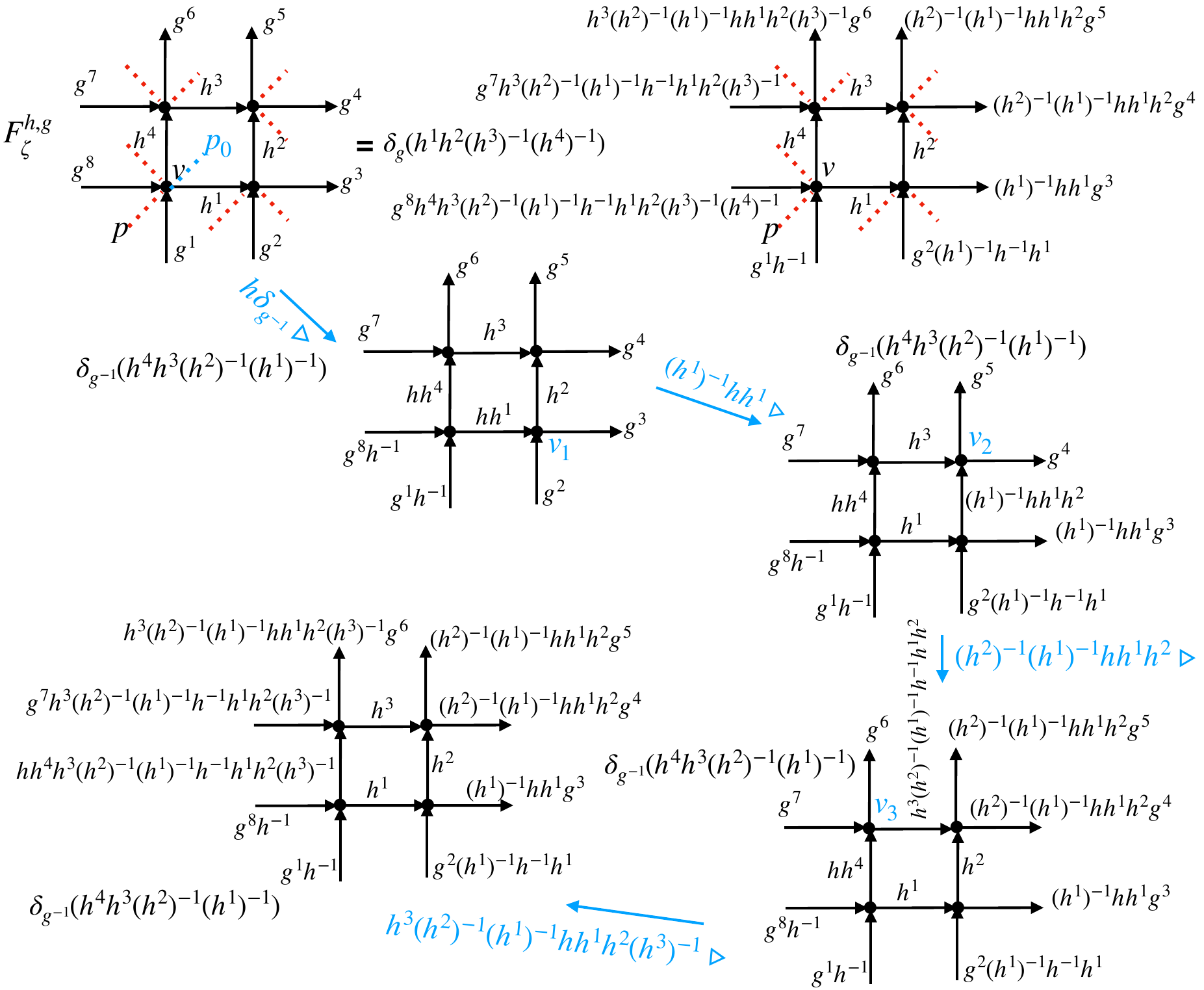}\]
\caption{\label{figcirclerib} (a) Example of a circular ribbon starting at $(v,p)$ and going anticlockwise, and (b) proof that this acts trivially on a vacuum state}
\end{figure}

Another important property of ribbon operators is that closed, contractible ribbons admit a trivial action of the corresponding ribbon operator on a vacuum state.
\begin{example}\rm An example of a closed ribbon operator $F_\zeta^{h,g}$ from site $(v,p)$ going anticlockwise back to itself  is shown in Figure~\ref{figcirclerib}. We compare this with the following sequence of operations (i) $\delta_g\la$ at site $(v,p_0)$, (ii) $h\la$ at $v$, (iii)  $(h^1)^{-1}hh^1\la$ at $v_1$ (iv) $(h^2)^{-1}(h^1)^{-1}hh^1h^2\la$ at $v_2$, and (v) $h^3(h^2)^{-1}(h^1)^{-1}hh^1h^2(h^3)^{-1}\la$ at $v_3$. The final results differ only on the initial arrows $(h^4,g^8)$ where the ribbon sends these to $(h^4, g^8h^{-1}hg^{-1}h^{-1}g)$ (given the $\delta_g$) while the sequence by contrast sends these to $(hg^{-1}h^{-1}gh^4, g^8 h^{-1})$. Thus, the two act the same as long as the state they act on forces $\delta_{g,e}$. This is true for a vacuum state where $\delta_g\la\vac =\delta_g\Lambda^*\la\vac =\delta_{g,e}\Lambda^*\vac =\delta_{g,e}\vac $ and where  $F^{h,g}_\zeta$ can be viewed as starting with $\delta_g\la$, as does our sequence.  We have $h\la\vac =h\Lambda\la\vac =\Lambda\la\vac =\vac $ similarly, hence the action of the sequence (i)-(v) on the vacuum is $\delta_{g,e}\vac $.  We conclude that $F^{h,g}_\zeta\vac =\delta_{g,e}\vac $. \end{example}
		
\begin{lemma}\label{ribcom} Let $\xi$ be a  ribbon between sites $s_0=(v_0,p_0)$ and $s_1=(v_1,p_1)$. Then
\[ [F_\xi^{h,g},f\la_v]=0,\quad [F_\xi^{h,g},\delta_e\la_p]=0,\]
for all $v \notin \{v_0, v_1\}$ and $p \notin \{p_0, p_1\}$.
\[ f\la_{s_0}\circ F_\xi^{h,g}=F_\xi^{fhf^{-1},fg} \circ f\la_{s_0},\quad \delta_f\la_{s_0}\circ F_\xi^{h,g}=F_\xi^{h,g} \circ\delta_{h^{-1}f}\la_{s_0},\]
\[ f\la_{s_1}\circ F_\xi^{h,g}=F_\xi^{h,gf^{-1}} \circ f\la_{s_1},\quad \delta_f\la_{s_1}\circ F_\xi^{h,g}=F_\xi^{h,g}\circ  \delta_{fg^{-1}hg}\la_{s_1}\]
for all ribbons where $s_0,s_1$ are disjoint, i.e. when $s_0$ and $s_1$ share neither vertices or faces. 
\end{lemma}
\proof We refer to the example in Figure~\ref{figribbon} to be concrete, but the arguments are general. (1) Commutation of the ribbon with $f\la$ at sites across from the main path is automatic because the ribbon acts on the  states on the cross arrows ($g^1,\dots,g^4$ in the example) like a vertex operator on the main path, which has an opposite relative orientation to a vertex at the other end of the relevant cross arrow. Hence the two actions are from opposite sides and commute. $f\la$ between $h^1,h^2$  changes these to $h^1f^{-1},fh^2$ in the illustration which does not change the product when it comes to parts of a subsequent ribbon operator at later vertices. It also changes $g^2$ to $g^2f^{-1}$. When we then apply the ribbon operator this changes to $g^2f^{-1}(h^1f^{-1})^{-1}h^{-1}h^1f^{-1}=g^2(h^1)^{-1}h^{-1}h^1f^{-1}$ which is what we get if we apply the ribbon first and then $f\la$ at this vertex. The same cancellation applies at other vertices on the main path other than the endpoints.  

(2) The action of $\delta_f\la$ at a face depends on the cyclic order determined by the vertex part of the site; commutation only holds in general if we chose this correctly or if we restrict to $\delta_e$ as stated (this disagrees with \cite{BSW}). For faces on the other side of the main path $\delta_f\la$ has the form of $\delta_f(... h_i...)$ where the $...$ are states on arrows unaffected by the ribbon. The ribbon op does not change $h_i$ so commutes with $\delta_f\la$. The other relevant faces are those in the body of the ribbon itself and we look at all three in detail. (i) The face bounded by $g^1,h^1,g^2$ and an unkown $x$ has $\delta_f\la=\delta_f(g^1h^1(g^2)^{-1}x^{-1})$ in some cyclic order. But if we apply the ribbon first then $\delta_f\la=\delta_f(g^1h^{-1}h^1 (g^2 (h^1)^{-1}h^{-1}h^1)^{-1}x^{-1})$ in the same cyclic order, which we see is the same unless we started at $g^2$. (ii) The face bounded by $g^2,h^2,h^3,g^3$ has $\delta_f\la=\delta_f(g^2h^2(h^3)^{-1}(g^3)^{-1})$ in some cyclic order. But if we apply the ribbon first then $\delta_f\la=\delta_f(g^2 (h^1)^{-1}h^{-1}h^1h^2h^3(g^3 h^3 (h^2)^{-1} (h^1)^{-1}h^{-1}h^1h^2(h^3)^{-1})^{-1})$ in the same order which again cancels (but only for the order shown). (iii) The face bounded by $g^3,g^4$ and unknowns $x,y$ say has 
$\delta_f\la=\delta_f(g^3(g^4)^{-1}x^{-1}y)$, say, in some cyclic order. If we apply the ribbon first then $g^3$ is replaced by $g^3w$ for a certain expression $w$ but so is $g^4$, so  $\delta_f\la$ is the same as long as we do not start at $g^4$.

(3) We have four remaining cases and again we refer to Figure~\ref{figribbon} to be concrete.  (i) $f\la$ at vertex $v_0$ sends $(h^1,g^1)$ to $(fh^1,g^1 f^{-1})$ (the other two arrows are also changed but this commutes with the ribbon operation). Applying $F^{fhf^{-1},fg}_\xi$ changes this to $(fh^1,g^1f^{-1}(fhf^{-1})^{-1})$ with a factor $\delta_{fg}(fh^1\cdots)$. If we apply $F^{h,g}_\xi$ first then we have $(h^1,g^1h^{-1})$ and a factor $\delta_{g}(h^1\cdots)$ and applying $f\la$ turns the former to $(fh^1,g^1h^{-1}f^{-1})$, which is the same. (ii) Similarly, $f\la$ at vertex $v_1$ sends $h^4$ to $h^4f^{-1}$ (the action on other, unmarked, arrows commutes with the ribbon operator). The ribbon operator $F^{h,gf^{-1}}_\xi$ then gives a factor $\delta_{gf^{-1}}(\cdots h^4 f^{-1})$. If we apply the ribbon first, we have $\delta_g(\cdots h^4)$ and then $f\la$ gives the same as before. (iii) $\delta_{h^{-1}f}\la$ at $v_0$ gives  factor $\delta_{h^{-1}f}((g^1)^{-1}...)$ (for three unmarked arrows around the rest of the face) and the ribbon then gives a factor sends gives a factor $\delta_g(h^1\cdots)$. It also acts on $g^1$.  If we apply the ribbon first, then this gives $\delta_g(h^1\cdots)$ and changes $g^1$ to $g^1h^{-1}$. Then applying $\delta_f\la$ gives a factor $\delta_f((g^1h^{-1})^{-1}\cdots)$, which is the same. (iv) $\delta_{fg^{-1}hg}\la$ at $v_1$ gives a factor $\delta_{fg^{-1}hg}(\cdots g^4h^4)$ for two unmarked arrows at the start of the face). The ribbon then imposes $\delta_{g}(zh^4)$ where $z$ is the product along the ribbon main path up to  $h^4$ (in our case, $h^1h^2(h^3)^{-1}$). If we apply the ribbon first then $g^4$ gets changed to $g^4 z^{-1}h^{-1}z$ and then $\delta_f\la$ gives $\delta_f(\cdots g^4z^{-1}h^{-1}zh^4)$, which is the same, given the  $\delta_{g}(zh^4)$ factor.   \endproof

This means that $F_\xi^{h,g}$ commutes with all terms of the Hamiltonian except those at $s_0,s_1$, where the nontrivial commutation relations will be used to create a quasiparticle at $s_0$ and its antiparticle at $s_1$.
In this sense, a ribbon operator is a generalisation of the creation operators discussed in Section~\ref{secZn}. We briefly consider so-called triangle operators, as they will be useful in future proofs.

\begin{definition}\label{def:triangles}
The direct-triangle and dual-triangle operators $T^g_{\tau}$ and $L^h_{\tau^*}$ respectively are defined by
\[ \includegraphics[scale=0.8]{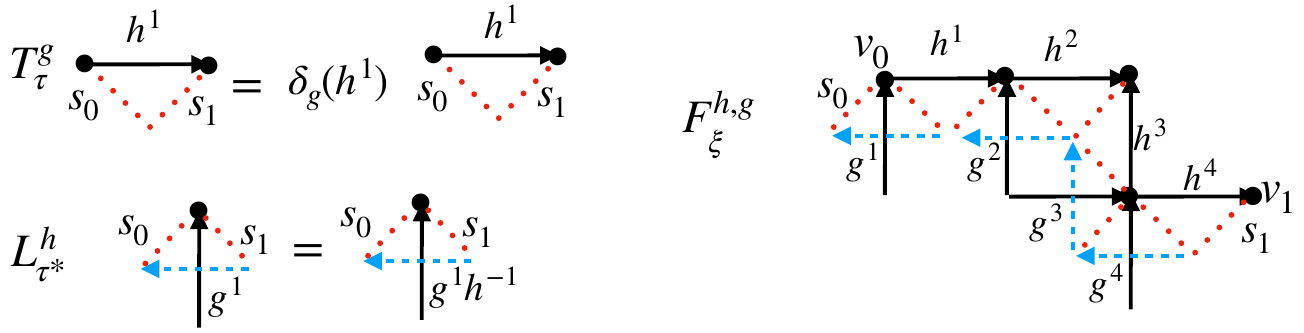}\]
We also show how a ribbon can be built as a sequence of triangle operations. 
\end{definition}
Here the dual lattice inherits an orientation by anticlockwise rotation of the unique arrow that crosses a dual arrow. If the flow around either triangle is clockwise then $h$ or $g$ enters as shown, otherwise their opposite version much as before.
For the dual triangle, this is the same as the arrow pointing inwards towards the vertex. Triangle operations can be viewed as atomic instances of ribbon operators, where
the start and end are adjacent sites, namely the associated ribbons are $F_\tau^{h,g}=T^g_\tau$ and $F_{\tau^*}^{h,g}=\delta_{g,e}L_{\tau^*}^h$ respectively and convolving these via (\ref{concat}) gives the composite $F_\xi^{h,g}$. However, triangle operators are not open ribbons due to $s_0,s_1$ not being disjoint, so they have different commutation relations from those in Lemma~\ref{ribcom}, which we study in full later. It is clear that we have have algebras $\mathcal{A}_{\tau} := \mathrm{span}\{T^g_{\tau}\ |\ g
\in G\}\isom \C(G)$ and $\mathcal{A}_{\tau^*} := \mathrm{span}\{L^h_{\tau^*}\ |\ h
\in G\}\isom \C G$ in view of the composition rules
\[T^g_{\tau} \circ T^{g'}_{\tau} = \delta_{g, g'}T^g_{\tau}, \quad
L^h_{\tau^*} \circ L^{h'}_{\tau^*} = L^{hh'}_{\tau^*}. \]

\begin{proposition}\label{Ls0s1} Let $\vac $ be a vacuum vector on a
plane $\Sigma$. Let $\xi$ be a ribbon between fixed sites $s_0 :=
(v_0, p_0),s_1 := (v_1, p_1)$ and
\[ |\psi^{h,g}\>:=F_\xi^{h,g}\vac .\]

(1) $|\psi^{h,g}\>$ is independent of the choice of ribbon
 between fixed sites $s_0, s_1$.

(2) The space \[ \CL(s_0,s_1):=\{|\psi\>\in \CH\ |\
A(v)|\psi\>=B(p)|\psi\>=|\psi\>,\quad \forall  v \notin \{v_0, v_1\},
p \notin \{p_0, p_1\}\} \] is spanned by $\{|\psi^{h,g}\>\ |\ h, g \in
G \}$.

(3) When sites $s_0$ and $s_1$ are disjoint, $\{|\psi^{h,g}\>\ |\ h, g \in G\}$ is an
orthogonal basis of $\CL(s_0,s_1)$. We call this the `group basis' of $\CL(s_0,s_1)$.

(4) $\CL(s_0,s_1)\subset\CH$ inherits actions at disjoint sites $s_0, s_1$, 
\[ f\la_{s_0}|\psi^{h,g}\>=|\psi^{ fhf^{-1},fg}\>,\quad \delta_f\la_{s_0}|\psi^{h,g}\>=\delta_{f,h}|\psi^{h,g}\>\]
\[ f\la_{s_1}|\psi^{h,g}\>=|\psi^{h,gf^{-1}}\>,\quad \delta_f\la_{s_1}|\psi^{h,g}\>=\delta_{f,g^{-1}h^{-1}g}|\psi^{h,g}\>\]
isomorphic to the left and  right regular representation of $D(G)$ by $|\psi^{h,g}\>\mapsto \delta_hg$.
\end{proposition}

\proof (1) Acting on the vacuum, a contractible, closed ribbon acts
trivially as we have illustrated. Then if $\xi,\xi'$ are two ribbons
between the same sites, we regard the composite of the reverse of
$\xi'$ with $\xi$ as a contractible, closed ribbon, as $\Sigma$ is a
plane. We then use equation~(\ref{concat}).

(2) We leave this proof to Appendix~\ref{app:span}, as it is lengthy and similar in some respects to \cite{Bom}.

(3) This proof can be found in \cite{BSW}, but we include it to clarify that it applies only when $s_0, s_1$ are disjoint. Thus,
\[\<\psi^{h,g}|\psi^{h',g'}\> = \vacket (F^{h, g}_{\xi})^{\dagger}
F^{h', g'}_{\xi}\vac = \vacket F^{h^{-1}, g}_{\xi} F^{h',
g'}_{\xi}\vac = \delta_{g, g'} \vacket F^{h^{-1}h', g}_{\xi}\vac\]
and, if $s_0, s_1$ are disjoint,
\[\vacket F^{h, g}_{\xi} \vac = \vacket (\delta_e \la_{p_1})^{\dagger} F^{h, g}_{\xi} \vac
= \vacket \delta_e \la_{p_1} F^{h, g}_{\xi} \vac = \vacket F^{h,
g}_{\xi} \delta_{g^{-1}hg} \la_{p_1}\vac
\] 
so that $\vacket F^{h, g}_{\xi} \vac = 0$ if $h \ne e$. When $h = e$,
\[\vacket F^{e, g}_{\xi}\vac = \vacket (k\la_{v_1})^{\dagger} F^{e,
g}_{\xi}\vac = \vacket F^{e, gk}_{\xi} k^{-1}\la\vac = \vacket F^{e, gk}_{\xi}
\vac \]
for every $k$, from which we deduce that $\vacket F^{e, g}_{\xi}\vac$ is independent of
$g$. Since $\sum_{g\in G}F^{e, g}_{\xi}= \id$, it follows that $\vacket F^{h, g}_{\xi} \vac = \frac{\delta_{h,e}}{|G|}$
and hence that 
\[ \<\psi^{h,g}|\psi^{h',g'}\> = \delta_{g,g'} \vacket F^{h^{-1}h', g}_{\xi}\vac\ = \frac{1}{|G|} \delta_{h, h'}
\delta_{g, g'}.\]
 Combined with (2), 
$\{|\psi^{h,g}\>\ |\ h,g \in G \}$ is then an orthogonal basis of $\CL(s_0, s_1)$.

If $s_0$, $s_1$ are not disjoint then Lemma~\ref{ribcom} no longer applies, and the commutation relations are different. For example, if $s_0$ and $s_1$ are joined by a direct triangle $\tau$ then $F^{h, g}_{\tau} = T^g_{\tau}$ so $\{|\psi^{h,g}\>\ |\ h,g \in G \}$ are no longer orthogonal.

(4) This follows from the  commutation relations in Lemma~\ref{ribcom} at $s_0$ and $s_1$ using $f\la\vac=\vac$ and $\delta_f\la\vac=\delta_{f,e}\vac$ replacing $f$ as modified by the commutation relations. Making the identification with $D(G)$ we compare the $s_0$ action with the left regular representation $\delta_f\la (\delta_hg)=\delta_f\delta_hg=\delta_{f,h}\delta_h g$ and $f\la(\delta_h g)= f\delta_h g=\delta_{fhf^{-1}}fg$ using the $D(G)$ commutation relations. The right regular representation is made into a left action via the antipode, so $\delta_f\la(\delta_hg)=\delta_h g\delta_{f^{-1}}=\delta_h\delta_{gf^{-1}g^{-1}}g=\delta_{f,g^{-1}h^{-1}g}g$ and $f\la(\delta_h g)=\delta_h g f^{-1}$. These match the stated  $D(G)$ actions at the end sites.  \endproof

\begin{remark}\rm The above Proposition assumes that we begin with a vacuum state $\vac$ on $\Sigma$. It is immediate, however, that the same arguments apply for a state $|\vartheta\>$ which is merely $\textit{locally}$ vacuum -- that is, $B(p)|\vartheta\> = A(v)|\vartheta\> = |\vartheta\>$ for $v, p$ at sites along the ribbon path and in the region between if we change the ribbon path. Thus,  (1) now becomes more precisely that  $|\vartheta^{h,g}\> := F^{h, g}_{\xi}|\vartheta\>$ is invariant under choice of ribbons $\xi$ and $\xi'$ between fixed sites $s_0$, $s_1$ iff the composite of $\xi$ with reversed $\xi'$ forms a closed, contractible ribbon $\xi''$, and where $A(v)|\vartheta\> = B(p)|\vartheta\> = |\vartheta\>$ for all $p$ and $v$ adjacent to $\xi''$ and in the region enclosed by $\xi''$. The intuition is that the ribbons may be smoothly deformed into one another, and thus leave the state invariant by previous arguments. The subspace $\CL'(s_0, s_1)$ is then defined in the natural way, ignoring excitations outwith the local neighbourhood of consideration, and actions are inherited on the sites $s_0$, $s_1$ in the identical manner to $\CL(s_0, s_1)$.
This locality of the Hamiltonian $H$ allows us to create quasiparticles at distance without being concerned about the compounding effects: they may be considered entirely separately. While we don't refer to it explicitly, this remark applies to the corollaries and applications throughout the paper in this context.
\end{remark}

The last part of Proposition~\ref{Ls0s1} implies a new basis of $\CL(s_0,s_1)$ in terms of the quasiparticle content  at the two ends.

\begin{corollary} Let $\xi$ be an open ribbon from $s_0$ to $s_1$. Then $\CL(s_0,s_1)$ has an alternative `quasiparticle basis' consisting for each irrep $\CC,\pi$ of $D(G)$ of the elements
\[  |u,v; \CC,\pi\>= {\dim V_\pi\over |C_G|}F_\xi^{'\CC,\pi;u,v}\vac;\quad F_\xi^{'\CC,\pi;u,v}:= \sum_{n\in C_G}  \pi(n^{-1})_{ji}  F_\xi^{c, q_c n q_d^{-1}}  \]
where  $u=(c,i)$ and $v=(d,j)$ with $c,d\in \CC$ and $i,j=1,\cdots\dim V_\pi$.
\label{cor:particle_basis} 
\end{corollary}
\proof  Here  $|u,v; \CC,\pi\>=\tilde\Phi(e_u\tens f^v)$ by which we mean $\Phi(e_u\tens e^v)$ in Theorem~\ref{thmPhi}, where $e_u=c\tens e^i$ and $f^v=\delta_d\tens f^j$ are basis elements of $V_{\CC,\pi}$ and $V_{\CC,\pi}^*$ respectively, then identified with an element of $\CL(s_0,s_1)$ by the inverse of the last part of Proposition~\ref{Ls0s1}. \endproof

These states behave for the left site action $\la_{s_0}$ on $\CL(s_0,s_1)$ according to a quasiparticle state labelled by basis element $e_u$ and for the right action at $s_1$ according to an anti-quasiparticle state labelled by the dual basis element $f^v$. Recall that we view the left site action $\la_{s_1}$ as a right one via the antipode $S$ of $D(G)$. The ribbon operators $F_\xi^{'\CC,\pi;u,v}$ that create these states from the vacuum are also of interest in their own right and it is claimed in \cite{Bom} that they form a basis of the space of operators that commute with almost all $A(v)$ and $B(p)$ in the same way that $\CL(s_0,s_1)$ is defined.

\begin{corollary}\label{blocktele} If $|\psi\>\in \CL(s_0,s_1)$ and we detect in it a quasiparticle of type $\CC,\pi$ at $s_0$ by nonzero  projection $P_{\CC,\pi}\la_{s_0}|\psi\>$ then 
\[ P_{\CC,\pi}\la_{s_0}|\psi\>=|\psi\>\ra_{s_1}P_{\CC,\pi},\]
hence we also automatically detect it at $s_1$, and vice-versa. In particular, the state
\[ |{\rm Bell}, \xi\>=\sum_{h
\in G}F_\xi^{h,e}\vac\]
 has a nonzero projection $P_{\CC,\pi}\la_{s_0}|{\rm Bell},\xi\>=|{\rm Bell},\xi\>\ra_{s_1}P_{\CC,\pi}\ne 0$ for all $\CC,\pi$. 
\end{corollary}
\proof This is essentially a block version of teleportation. A general state is highly entangled between the different particle types,
\[ |\psi\>=\sum_{\CC,\pi}\sum_{u,v}\phi(\CC,\pi,u,v) \tilde\Phi(e_u\tens f^v)\]
where, as above,  $\tilde\Phi:\End(V_{\CC,\pi})\to \CL(s_0,s_1)$ denotes $\Phi$ combined with the inverse of the identification in Proposition~\ref{Ls0s1}. Here $\{e_u\}$ are a basis  of $V_{\CC,\pi}$ and $\{f^u\}$ a dual basis.  Applying $P_{\CC',\pi'}\la_{s_0}$ and $\ra_{s_1}P_{\CC',\pi'}$ becomes via the bimodule properties
respectively $P_{\CC',\pi'}\la e_u$ and $f^v\ra P_{\CC',\pi'}$. But these projections are zero unless $(\CC',\pi')=(\CC,\pi)$, in which case they act as the identity, as in the proof of Theorem~\ref{thmPhi}. Hence 
\[ P_{\CC,\pi}\la_{s_0}|\psi\>=\sum_{u,v}\phi(\CC,\pi,u,v) \tilde\Phi(e_u\tens f^v)=|\psi\>\ra_{s_1}P_{\CC,\pi}.\]
In particular,
\begin{equation}\label{projuv}P_{\CC', \pi'} \la_{s_0} |u,v;\CC,\pi,\xi\> =|u,v;\CC,\pi,\xi\> \ra_{s_1} P_{\CC', \pi'} = \delta_{\CC', \CC } \delta_{\pi', \pi } |u,v;\CC,\pi,\xi\>.\end{equation}
For the `block Bell state', we consider $1_{D(G)}=\sum_{h\in G} \delta_h \tens e$ which map to $\sum_h |\psi^{h,e}\>$ according to the last part of Proposition~\ref{Ls0s1}. On the other hand, this is $\sum_{\CC,\pi} P_{\CC,\pi}$ by Lemma~\ref{lemPCpi} and hence each term is the image under $\Phi$ of $\sum e_u\tens f^u$ in Theorem~\ref{thmPhi}. Thus,
\[ |{\rm Bell}; \xi\>=\sum_{\CC,\pi}\sum_u\tilde\Phi(e_u\tens f^u)=\sum_{\CC,\pi} \sum_u|u,u;\CC,\pi\>\]
and from (\ref{projuv}) we have
\[ P_{\CC,\pi}\la_{s_0}|{\rm Bell},\xi\>=|{\rm Bell};\CC,\pi,\xi\>=|{\rm Bell},\xi\>\ra_{s_1}P_{\CC,\pi}\ne 0\]
where
\[  |{\rm Bell};\CC,\pi,\xi\>=\sum_u\tilde\Phi(e_u\tens f^u)=\sum_u |u,u;\CC,\pi\>\]
is the claimed nonzero state projected out from $|{\rm Bell};\xi\>$. 
\endproof

We recall that in teleportation one has an entangled `Bell state',  $\sum_i  |v_i\>\tens \<v_i|$ for a basis and dual basis of a Hilbert space, and if we apply from the left a projection $|v_1\>\<v_1|$ say then the state collapses to $|v_1\>\tens \<v_1|$ so that the right factor is an eigenstate if we apply $|v_1\>\<v_1|$ from the right. Equivalently,  if we evaluate against any $\<\psi|$ on the left then the result is $\<\psi|$ in the right factor. We see a similar phenomenon with the block $V_{\CC,\pi}\tens V_{\CC,\pi}^*$ in place of $|v_i\>\tens \<v_i|$. In the toric case discussed below, each block will be 1-dimensional so that we are then a bit closer to the standard case. 

We can also potentially look inside each block, i.e. for each fixed $\CC,\pi$, regard $|{\rm Bell};\CC,\pi,\xi\>\in \CL(s_0,s_1)$ as a `mini Bell state' that can similarly transport a single particle state across the ribbon. We saw in the proof above that this is $\sum_u \Phi(e_u\tens f^u)= P_{\CC,\pi}$ mapped over to this space by the inverse of the identification in the last part of Proposition~\ref{Ls0s1}. We can also write
\begin{equation}\label{WCpi}  |{\rm Bell};\CC,\pi,\xi\>={\dim V_\pi\over |C_G|}W_\xi^{\CC,\pi}\vac;\quad W_\xi^{\CC,\pi}:=\sum_u F_\xi^{'\CC,\pi;u,u}.\end{equation}
so that the `ribbon trace operator' $W_\xi^{\CC,\pi}$ has the physical interpretation of creating a maximally entangled quasiparticle/anti-quasiparticle pair (the mini Bell state) of only the specified type $\CC, \pi$. The issue for teleportation of a single quasiparticle state vector  using a such mini Bell state would be how, in a quantum computer, to create a single particle state or its dual and evaluate it against $e_u$ at $s_0$ or against $f^u$ at $s_1$.  

\begin{lemma} Let $\xi:s_0\to s_1$ and $\xi':s_1\to s_2$ be open ribbons. Then
\[ F_{\xi'\circ\xi}^{'\CC,\pi;u,v}=\sum_w F_{\xi'}^{'\CC,\pi;w,v}\circ F_\xi^{'\CC,\pi;u,w}\]
\end{lemma}
\proof We have using (\ref{concat}), 
\begin{align*}F_{\xi'\circ\xi}^{'\CC,\pi; (c,i),(d,j)}&= \sum _{n\in C_G}  \pi(n^{-1})_{ji} F_{\xi'\circ \xi}^{c , q_c n q_d^{-1}}\\
&=\sum_{f\in G}\sum _{n\in C_G}  \pi(n^{-1})_{ji} F_{\xi'}^{f^{-1}cf,f^{-1}q_c n q_d^{-1}}\circ F_\xi^{c,f}\\
&=\sum_{b\in \CC} \sum_k \sum_{m,n\in C_G}  \pi((m^{-1}n)^{-1})_{jk} \pi(m^{-1})_{ki} F_{\xi'}^{b,q_b m^{-1}n q_d^{-1}} F_\xi^{c, q_c m q_d^{-1}}
\end{align*}
where we uniquely factorised $f^{-1}q_c = q_b m^{-1}$ in terms of some $b\in\CC$ and $m\in C_G$. We then change variables to $n'=m^{-1}n$ and recognise the answer with $w=(b,k)$. 
\endproof
This reflects that $F_\xi'$ are a kind of (nonAbelian) Fourier transform of the original $F_\xi$ with convolution as in (\ref{concat}) becoming multiplication. Invertibility of Fourier transform implies that the space spanned by such operators is the same as the space spanned by the original $F_\xi$, now organised according to the quasiparticle type.
In addition, we have:

\begin{lemma}\label{lemWcomp}
\[W^{e, \pi}_{\xi} \circ W^{e, \pi'}_{\xi} = W^{e, \pi \otimes \pi'}_{\xi}\]
\end{lemma}
\proof
Using (\ref{WCpi}), Corollary~\ref{cor:particle_basis} and (\ref{concat}) in that order,
\begin{align*}
W^{e, \pi}_{\xi} \circ W^{e, \pi'}_{\xi} &= \sum_{n, n' \in G} {\rm Tr}_{\pi}(n^{-1})F^{e, n}_{\xi} {\rm Tr}_{\pi'}(n'^{-1})F^{e, n'}_{\xi}\\
&= \sum_{n, n' \in G} {\rm Tr}_{\pi}(n^{-1}) {\rm Tr}_{\pi'}(n'^{-1}) \delta_{n, n'} F^{e, n}_{\xi}\\
&= \sum_n {\rm Tr}_{\pi \otimes \pi'}(n^{-1}) F^{e, n}_{\xi}
\end{align*}
which we recognise as stated. 
\endproof

We will also need the following. 
\begin{lemma}\label{lemW} Let $\xi:s_0\to s_1$ be an open ribbon. Then $W_\xi^{'\CC,\pi}{}^\dagger=W_\xi^{'\CC^*,\pi^*}$ where $\pi^*$ is the conjugate unitary representation of $C_G$ and $\CC^*=\CC^{-1}$ equipped with $r_{\CC^*}=r_{\CC}^{-1}$ and $q:\CC^{-1}\to G$ given by $q_{c^{-1}}=q_c$ for all $c\in \CC$. 
\end{lemma}
\proof Here
\[ F_\xi^{'\CC,\pi;(c,i),(d,j)}{}^\dagger=\sum_{n\in C_G}  \overline{ \pi(n^{-1})_{ji} } F_\xi^{c^{-1}, q_c n q_d^{-1}}=\sum_{n\in C_G}  \pi^*(n^{-1})_{ij}  F_\xi^{c^{-1}, q_c n q_d^{-1}}=F_\xi^{'\CC^{-1},\pi^*;(c^{-1},j),(d^{-1},i)}\]
where $\CC^*$ is the $\CC^{-1}$ with the basepoint and $q$ function data  as stated and the same $C_G$. We now take the trace by summing over $c=d$ and $i=j$. \endproof

Note that it could be that $\CC=\CC^{-1}$ as a set but is not $\CC^*$ due to a different base point (this happens for the order two conjugacy class of $S_3$). 

The last ingredient we need for applications is a generalisation of the space $\CL(s_0, s_1)$. If $s_0,s_1,\cdots,s_n$ are $n+1$ sites, define the subspace
\[\CL(s_0,s_1,\cdots,s_n) := \{ |\psi\> \in \CH\ |\ A(v)|\psi\> = B(p)|\psi\> = |\psi\>, \forall v \notin \{v_0, v_1,\cdots,v_n \}, p \notin \{p_0,p_1,\cdots,p_n \} \}.\]

\begin{lemma}\label{lem:manysites} Given a $D(G)$ model on a borderless planar lattice $\Sigma$, let $s_0,s_1,\cdots,s_n$ be $n+1$ disjoint sites. Then
\[\dim(\CL(s_0,s_1,\cdots,s_n)) = |G|^{2n}\]
with an orthogonal basis
\[\{|\psi^{\{h^1, h^2,\cdots,h^n\},\{g^1,g^2,\cdots,g^n\}}\>\ |\ h^1,h^2,\cdots,h^n,g^1,g^2,\cdots,g^n \in G \}\]
generalising the group basis of $\CL(s_0,s_1)$. There is another orthogonal basis that is the equivalent generalisation of the quasiparticle basis.
\end{lemma}
\proof
As we saw in the proof of Proposition~\ref{Ls0s1}, the only operations which take the vacuum to states with excitations only at any two sites, say $s_0, s_1$, are ribbon operators along a ribbon $\xi : s_0 \rightarrow s_1$. Now, let $A$ be a complete graph, with sites $s_0,s_1,\cdots,s_n$ as vertices. We then have a contribution to $\dim(\CL(s_0,s_1,\cdots,s_n))$ of $|G|^2 = \dim(\CL(s_0,s_1))$ from an edge in A between $s_0, s_1$, corresponding to some ribbon $\xi:s_0\rightarrow s_1$; we can give this the group basis with labels $\{h^1, g^1\ |\ h^1, g^1 \in G\}$
or the equivalent quasiparticle basis. Edges in $A$ between other vertices/sites contribute similarly, so for example there are another $|G|^2$ orthogonal ribbon operators along the ribbon $\xi':s_1 \rightarrow s_2$, which multiplies with the initial $|G|^2$ from $\xi$. However, by Lemma~\ref{ribcom}, if we have already counted the operators along ribbons $\xi : s_0 \rightarrow s_1$ and $\xi':s_1 \rightarrow s_2$ then any ribbon operator $F^{h, g}_{\xi''}$ for $\xi'':s_0\rightarrow s_2$ has a decomposition into ribbons along $\xi$ and $\xi'$ iff $\xi''$ is
isotopic to $\xi'\circ\xi$. Therefore the only edges which contribute are between vertices which have no alternative path along previously visited edges.
In particular, we define $T$ as any maximally spanning tree on $A$. Then $\dim(\CL(s_0,s_1,\cdots,s_n))$ receives contributions from exactly the $n$ edges in $T$, and we may for example give the group basis with labels $\{h^i, g^i\ |\ h^i, g^i \in G\}$ from each edge.
\endproof

We note that while the dimensions multiply, it is not true that $\CL(s_0,s_1,\cdots,s_n)$ can be presented as $\CL(s_0, s_1) \otimes \cdots \otimes \CL(s_{n-1},s_1)$ where the tensor product is along each edge in $T$. This is because, for example, ribbon operators $F^{h, g}_{\xi}$ and $F^{h', g'}_{\xi'}$ meet at the endpoint $s_1$ and need not commute. On the other hand, if we have some disjoint subsets of $\{s_0,\cdots,s_{n-1}\}$ then the tensor product of the logical spaces associated to each subset form a subspace. For example
\[ \CL(s_0,s_1)\tens \CL(s_2,s_3)\subset \CL(s_0,s_1,s_2,s_3)\]
by sending $F_\xi^{h,g}\vac\tens F_{\xi'}^{h',g'}\vac\mapsto F_\xi^{h,g}\circ F_{\xi'}^{h',g'}\vac$. 

\subsection{Reduction to toric model for $G=\Z_n$}\label{secredZn}

In this section, we verify that everything above reduces correctly to the toric case already covered in Section~\ref{secZn} via the Fourier correspondence (\ref{Zisom}). Here $D(\Z_n)=\C(\Z_n)\tens\C\Z_n\isom \C.\Z_n\times\Z_n=\C[g,h]/\<g^n-1,h^n-1\>$ and we recall that we set $q=e^{2\pi\imath\over n}$. Clearly, at a face
 \[ g\la =\sum_m q^m\delta_m\la= \sum_m q^m\delta_{m, i+j-k-l}=q^{i+j-k-l}\]
  if the state around the face is $|i\>,|j\>,|k\>,|l\>$ with orientations as displayed before. This no longer depends on the starting point. Moreover $h\la$ around a vertex is the action of $1\in \Z_n$ so acts as before. This clearly gives gives $A(v)$ as before and $B(p)=\delta_0\la ={1\over n}\sum_k g^k\la$ as before. 
  
The vacuum degeneracy of the toric model is straightforward to calculate.

\begin{lemma}
Let $\Sigma$ be a closed, orientable surface, and let $G = \Z_n$. Then
\[
\dim(\CH_{vac}) = n^{2k},
\]
where $k$ is the genus of $\Sigma$.
\end{lemma}
\proof
The fundamental group $\pi_1(\Sigma) \cong \Z^{2k}$. $\Z^{2k}$ is a
$2k$-biproduct of $\Z$ in the category of groups, so
$\mathrm{Hom}(\Z^{2k}, \Z_n) \cong \mathrm{Hom}(\Z, \Z_n)^{2k}$. Now,
$|\mathrm{Hom}(\Z, \Z_n)| = n$, so $|\mathrm{Hom}(\Z^{2k}, \Z_n)| =
n^{2k}$. The $G$-action is trivial, so we are done.
\endproof

For representations, the conjugacy classes are singletons $\{i\}$, say, with isotropy group all of $\Z_n$, with irrep $\pi_j$ say. The carrier space is 1-dimensional and the irrep is
\[ g\la \{i\}=\sum_m q^m\delta_{m,i}\{i\}=q^i\{i\},\quad h\la \{i\}=q^j\{i\}\]
as employed before. Projectors simplify to those from Section~\ref{secZn}. Thus,  
\begin{align*}P_j &={\dim V_\pi\over |G|}\sum_{g}(\Tr_{j}g^{-1})g = \frac{1}{n} \sum_{k} q^{-jk} g^k\\
P_{\{i\},j}&=\delta_i\tens P_j =\delta_i \tens {1 \over n} \sum_{l \in G} q^{-jl} g^l\cong P_i \tens P_j = {1\over n^2}\sum_{k,l}q^{-(ik+jl)}h^k g^l=P_{ij}
\end{align*}
by  Fourier isomorphism between $\C \Z_n$ and $\C(\Z_n)$.

The ribbon operators are now labelled as $F_\xi^{a,b}$ say, where $a,b\in\Z_n$ and have a simpler form. For example, 
\[\includegraphics[scale=0.73]{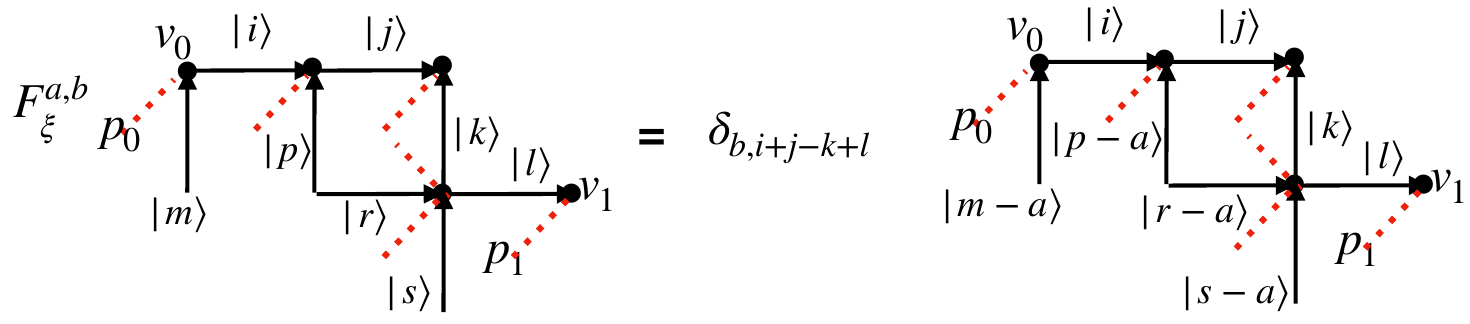}\]
for the ribbon in Figure~\ref{figribbon}. Concatenation of ribbon operators simplifies to:
\begin{equation}\label{concatZn}F_{\xi'\circ\xi}^{a,b}=\sum_{f\in G}F_{\xi'}^{a,b-f}\circ F_\xi^{a,f}.\end{equation}
The commutation relations in Lemma~\ref{ribcom} simplify to 
\[ h\la_{s_0}\circ F^{a,b}_\xi=F_\xi^{a,b+1}\circ h\la_{s_0},\quad g\la_{s_0}F^{a,b}_\xi=q^a F^{a,b}_\xi g\la_{s_0}\]
\[  h\la_{s_1}\circ F^{a,b}_\xi=F_\xi^{a,b-1}\circ h\la_{s_0},\quad g\la_{s_0}F^{a,b}_\xi=q^{-a} F^{a,b}_\xi g\la_{s_0}\]
i.e. these `q-commute'. For example, 
\[g\la_{s_0}\circ F^{a,b}_\xi=\sum_m q^m \delta_m\la_{s_0}\circ F^{a,b}_\xi=\sum_mq^m F^{a,b}\delta_{m-a}\la_{s_0}={1\over n}\sum_{m,k}q^m F^{a,b}_\xi q^{-(m-q)k}g^k\la_{s_0}.\]
The sum over $m$ forces $k=1$ which then gives the answer stated. Consequently, on states $|\psi^{a,b}\>=F^{a,b}_\xi\vac$ we just have that 
\[ h\la_{s_0}|\psi^{a,b}\>=|\psi^{a,b+1}\>,\quad g\la_{s_0}|\psi^{a,b}\>=q^a |\psi^{a,b}\>\]
which commute, and similarly at $s_1$.

For the quasiparticle basis, the relevant ribbon operator and its adjoint are
\[ F_\xi^{'i,j}:= \sum_{k} q^{-jk}  F_\xi^{i, k},\quad F_\xi^{'i,j}{}^\dagger=F_\xi^{'-i,-j}
\]
where we omit $u, v$ as these are trivial and the $i,j$ play the role of $\CC,\pi$ respectively in the construction of $P_{ij}$ above. We see that $F'_\xi$ is just a Fourier transform in the second argument, which  takes convolution to multiplication so that (\ref{concatZn}) becomes
\begin{equation}
F^{'i,j}_{\xi' \circ \xi} =
\sum_k q^{-j(k-l)-jl} \sum_l F^{i, k-l}_{\xi'} \circ F^{i, l}_{\xi}=F'^{i,j}_{\xi'} \circ F'^{i,j}_{\xi}.
\label{concatZn_2}
\end{equation}
Also, since there are no $u,v$ indices, $W_{\xi}^{i, j} = F^{'i,j}_{\xi}$. The following three subsections show how the above might be used in practice to perform operations relevant to quantum computation.

\subsubsection{Toric Bell state and teleportation} According to our general theory and our calculations above, the state which is maximally entangled between the particle types is 
\[ |{\rm Bell};\xi\>=\sum_{i, j} {1\over n}F_\xi^{'i,j}\vac=\frac{1}{n} \sum_{i, j, k} q^{-jk} |\psi^{i,k}\>=\sum_i F^{i, 0}_{\xi}\vac = \sum_i|\psi^{i,0}\>, \]
where the second-to-last step is the Fourier transform. Here $\<{\rm Bell};\xi|{\rm Bell};\xi\>=n$. For a concrete example, consider 
\[\input{toric_maxentangle_v3.tikz}\]
for a ribbon $\xi : s_0 \rightarrow s_1$, where $s_0 = (v_0, p_0), s_1 = (v_1, p_1)$ and a generic term in $\vac$ with relevant arrow values $|a\>\tens\cdots\tens|p\>$ as shown. We  also know that 
\[ |i,j\>:=|{\rm Bell};i,j,\xi\>=P_{ij}\la_{s_0}|{\rm Bell};\xi\>= |{\rm Bell};\xi\>\ra_{s_1} P_{ij}={1\over n}W_\xi^{i,j}\vac\]
are the `mini-Bell' states associated to each $i,j$. In our case (as there are no $u,v$ indices) these have no internal substructure as an entangled sum of internal states and we just regard them as a basis of $\CL(s_0,s_1)$ as we vary $i,j$. To illustrate how this goes explicitly, consider  $P_{ij}= {1\over n^2}\sum_{x,y}q^{-i x-jy} g^x h^y$ as above, acting at $s_0$, say. Then, renaming the dummy index $i$ in $|{\rm Bell};\xi\>$ as $z$, 
\begin{align*} P_{ij}&\la_{s_0}|{\rm Bell};\xi\>={1\over n^2}\sum_{x,y,z} q^{-i x-jy}\delta_{0,k+l+m}q^{x(f+z+y-c-b+a-y)}\\
&\qquad\qquad\qquad |f+z+y\>\tens |k+y\>\tens |d-y\>\tens |a-y\>\tens|p-z\> \\
&={1\over n}\sum_{y}q^{-jy}\delta_{0,k+l+m} |i+c+b-a+y\>\tens |k+y\>\tens |d-y\>\tens |a-y\>\tens|p-i+f-c-b+a\>\\
&={1\over n}\sum_{y}q^{-jy}\delta_{0,k+l+m} |i+f+y\>\tens |k+y\>\tens |d-y\>\tens |a-y\>\tens|p-i\>\\
&={1\over n}\sum_y q^{-jy}\delta_{y,k+l+m}|f+i\>\tens |k\>\tens |d\>\tens |a\>\tens|p-i\>\\
&={1\over n}\sum_y q^{-jy}F_\xi^{i,y}\vac={1\over n}W_\xi^{i,j}\vac\end{align*}
for the affected arrows located in line with the previous diagram. The sum over $x$ forced the value $z=i-f+c+b-a$ for the second equality. We then used that $\delta_0\la_{s_0}$ acts as the identity on the vacuum so that $f+c+b-a=0$ around the face $p_0$ for the third equality. Likewise, we use that $h^{-y}\la_{s_0}$ is the identity on the vacuum for the fourth (this shifts the original values $f,k,d,a$ around the vertex $v_0$ by $\mp y$). We then recognise the action of $W_\xi^{i,j}$ as expected. Similarly for $\ra_{s_1}P_{ij}$.

We now explain our teleportation point of view in this toric case. Here $\{|i,j\>\}$ are a basis of $\CL(s_0,s_1)$ and we have seen that  $P_{ij}(s_0)$ applied to $|{\rm Bell};\xi\>$ collapses the ribbon state to $|i,j\>$, and a quasi-particle of type $(i,j)$ now occupies $s_0$. This is a local operation at $s_0$ but the resulting state when measured at $P_{-i,-j}(s_1)=\ra_{s_1}P_{ij}$ is also an eigenstate and detects a quasiparticle of type $(-i,-j)$ locally at $s_1$. Here the right action of $P_{ij}$ is the left action of $S(P_{ij})$, where $S$ is the antipode of the group algebra of $\Z_n\times\Z_n$. Although the details are not the same as usual quantum teleportation, we follow the same principle of using a maximally entangled state to transfer information along the length of an extended object, our case the ribbon.  We can use this to transmit any state vector in the vector space spanned by the particle types, i.e. a vector $\vec \psi=(\psi_{ij})$. We set up a state $|{\rm Bell};\xi\>$ and apply the operator $\sum_{i,j}\psi_{ij} P_{ij}(s_0)$ to it locally at $s_0$. This results in 
\[ |\psi\>:=\sum_{i,j}\psi_{ij}P_{ij}(s_0)|{\rm Bell};\xi\>=\sum_{i,j}\psi_{ij}|i,j\>\in \CL(s_0,s_1)\]
 as a ribbon state that encodes our vector $\vec\psi$. We can then read off the latter at the other end by applying the operator $P_{-i,-j}(s_1)$ locally at $s_1$, where
\begin{align*}P_{-i,-j}(s_1)|\psi\>&=|\psi\>\ra_{s_1}P_{ij}=P_{ij}\la_{s_0}|\psi\>=\sum_{k,l}\psi_{kl}P_{ij}(s_0)|k,l\>=\psi_{ij}P_{i,j}(s_0)|{\rm Bell};\xi\>\end{align*}
This is the component of $|\psi\>$ that contains the $\psi_{ij}$ coefficient. It is also equal to $\psi_{ij}P_{-i,-j}(s_1)|{\rm Bell};\xi\>$ making it clear that we can extract the coefficient by local operations at $s_1$. 

While such a teleportation scheme is possible when the projectors $P_{ij}$ can be applied to the lattice, in reality such projectors can only be applied probabilistically, by performing measurements.
In particular, assuming that application of projectors can be performed deterministically in general has grave complexity-theoretic consequences, such as allowing NP-complete problems to always be solved in polynomial time on a quantum computer \cite{Aaronson}.
This means that, while the above scheme is illustrative of the entanglement between sites $s_0$ and $s_1$, it is unclear how to leverage this property to be computationally useful when the superposition may only be collapsed by a measurement.

\subsubsection{Quasiparticle creation and transportation redux}\label{sec:quasi_redux}

Next, we show how to create and transport quasiparticles using 
the $W^{i, j}_{\xi}$ operators, which are equal to $F^{'i, j}_{\xi}$ in the $\Z_n$ case, and relate this to the \textit{ad hoc} description of Section~\ref{sec:quasiparticles}.   This pertains to the following lattice in the vacuum state:
\[
\input{toric_creation.tikz}
\]
with $\xi : s_0 \rightarrow s_1$ as shown. We apply 
\[
\input{toric_creation2.tikz}
\]
We see that only $|s\>$ is affected and $\sum_{k}q^{-jk}\delta_k(s) = q^{-js}$. In terms of our $X,Z$ operations, we have
$W^{i, j}_{\xi}|s\> =X^{-i} Z^{-j}|s\>$. Recall from Section~\ref{sec:quasiparticles} that the effect of this is that particles $\pi_{i, j}$ and $\pi_{-i,-j}$ appear at sites $s_0$ and $s_1$ respectively, which we tested using projectors.
In other words, we have the mini-Bell state $|\mathrm{Bell}; i, j, \xi\>$. 

Next, we consider a further site $s_2$
\[
\input{toric_transport.tikz}
\]
then the further effect of a operator $W^{i,j}_{\xi'}$ for an open ribbon  $\xi':s_1\to s_2$ is
\[
\input{toric_transport2.tikz}
\]
We see that $W^{i,j}_{\xi'} : |t\> \otimes |u\> \mapsto
X^{i}|t\> \otimes Z^{-j}|u\>$ while leaving the other states unchanged. We saw in Section~\ref{sec:quasiparticles} that quasiparticles
$\pi_{i, j}$ and $\pi_{-i,-j}$ now occupy sites $s_0$ and $s_2$
respectively. So the effect of this second ribbon operator is to transport the $\pi_{-i,-j}$ excitation from
$s_1$ to $s_2$. We also know from (\ref{concatZn_2}) that these two ribbon operations compose to $W^{i,j}_{\xi' \circ \xi}$ along the composite ribbon, so we create the state $|\mathrm{Bell}; i, j, \xi' \circ \xi\>$. In other words, creation at sites $s_0$ and $s_1$ followed by transport from
$s_1$ to $s_2$ is equal to creation at sites $s_0$ and $s_2$. The combined operation is
\[
\input{toric_concatenation_v2.tikz}
\]
which we see affects only the states $|s\>\tens|t\>\tens|u\>$ along the ribbon and has the particle content at the ends
as previously analysed. 

\subsubsection{Quasiparticle braiding}

This section gives an
example of braiding on the lattice, and relates it to the braiding of irreducible representations of $D(G)$ given at the start of Section~\ref{secZn}. We do not prove explicitly that all such lattice braidings correspond to braids in the representation category,
but the broad arguments are easy to see. Let $\xi:s_0\to s_1$ be the following ribbon acting on  a vacuum state $\vac$,
\[\input{braiding1.tikz}\]
where we have labelled the relevant edges $|k\>$ to $|q\>$ as shown. $W^{0,
-j}_{\xi}$ creates quasiparticles $e_{-j}$, $e_j$ at $s_0$, $s_1$,
and takes the vacuum to
\[\input{braiding2.tikz}\]
Also consider another ribbon operator $W^{-i, 0}_{\xi'}$ for $\xi':s_2\to s_3$, creating $m_{-i}$, $m_i$
quasiparticles at $s_2,s_3$ according to
\[\input{braiding3.tikz}\]
The combined effect of these is the state $|\psi\> := W^{-i, 0}_{\xi'} \otimes W^{0,-j}_{\xi}\vac$
\[\input{braiding4.tikz}\]
Now let $\xi''$ be a ribbon rotating anti-clockwise from
$s_1$ back to $s_1$ around the face of $s_3$, according to
\[\input{braiding5.tikz}\]
Acting on $|\psi\>$, we move the $e_j$ quasiparticle at $s_1$ around the $m_i$ at $s_3$ using  $W^{0, -j}_{\xi''}$ and resulting in
\[\input{braiding6.tikz}\]
Now use  $Z^jX^i  = q^{ij}X^i Z^j$ on $|q\>$ so that the $Z^{\pm j}$ operators that make up 
$W^{0, -j}_{\xi''}$ act on $\vac$. But the latter is a face operator 
$g^{-j} \la$ around the face of $s_3$ and acts trivially on the vacuum. Hence the effect of  $W^{0,-j}_{\xi''}$ on $|\psi\>$ is to send 
\[|\psi\> \mapsto q^{ij}|\psi\>
\]
as expected for the braiding of $m_i$ with $e_j$. To visualise this braiding, we should think in terms
of worldlines to take account of the  temporal aspect: we first create the
quasiparticles, and then transport one around the other. We identify
the map above with 
\[\Psi_{m_i, e_j}\circ \Psi_{e_j, m_i} : e_j \otimes
m_i \rightarrow e_j \otimes m_i\] 
and have braided the worldline of the
$e_j$ quasiparticle around that of the $m_i$ quasiparticle and back
again.

While this is only one instance of braiding, any ribbon operator on the plane which forms a closed loop around another occupied site
will admit a similar braiding, as the same argument from above applies but taking a product of vertex and face operators, rather than just $g^{-j} \la$ in this example. We assert that the state at the occupied site will always admit commutation relations such that
the appropriate phase factor is produced.

\subsection{Details for $D(S_3)$ and applications}\label{secS3}

$S_3$ is the smallest nonAbelian group. We let $S_3$ be generated by $u=(12)$, $v=(23)$ with relations $u^2=v^2=e$ and $uvu=vuv$ ($=w=(13)$). This has three irreducible representations:
\[ 1, \quad \sigma={\rm sign},\quad \tau;\quad \sigma\tens\sigma=1,\quad \sigma\tens\tau=\tau\tens\sigma=\tau,\quad  \tau\tens\tau=1\oplus\sigma\oplus\tau\]
where $\tau$ is the only 2-dimensional one and $\sign=-1$ on $u,v,w$ and $+1$ otherwise. The irreps of $D(S_3)$ are given by pairs $(\mathcal{C}, \pi)$, where $\mathcal{C}$
is a conjugacy class in $S_3$ and $\pi$ is an irrep of the centraliser
of a distinguished element $r_{\CC}$ in $\mathcal{C}$, i.e. an isotropy subgroup, and we also need to fix $q_c$ for each $c\in \CC$ such that $c=q_cr_\CC q_c^{-1}$. We take these as follows:

\begin{enumerate}
\item The trivial conjugacy class  $\CC=\{e\}$, $r_\CC=q_c=e$ and $C_G=S_3$, giving exactly 3  chargeons $(\{e\}, 1)$, $(\{e\}, \sigma)$ and $(\{e\}, \tau)$ as $D(S_3)$ irreps.

\item $\CC= \{u,
v, w\}$, $r_\CC=u$, $q_u=e, q_v=w, q_w=v$ and $C_G=\Z_2=\{e,u\}$, giving $(\{u,v,w\},1)$ and $(\{u,v,w\},-1)$ as 2 irreps of $D(S_3)$, where we indicate the representation $\pi_{-1}(u)=-1$ of $C_G$.

\item $\CC= \{uv, vu\}$, with $r_\CC=uv$, $q_{uv}=e$, $q_{vu}=v$ and $C_G=\Z_3=\{e,uv,vu\}$, giving $(\{uv,vu\},1)$,   $(\{uv,vu\},\omega)$, $(\{uv,vu\},\omega^*)$  as 3 irreps of $D(S_3)$, where  $\omega=e^{2\pi\imath\over 3}$ and we indicate irreps $\pi_\omega(uv)=\omega$ and $\pi_{\omega^*}(uv)=\omega^{-1}$ of $C_G$.
\end{enumerate}

Thus, there are  8 irreps of $D(S_3)$. To describe the projectors, we denote the conjugacy class $\mathcal{C}$ by its chosen element $r_{\CC}$ as shorthand, for example $P_{u, \pi} := P_{\{u, v, w\}, \pi}$.  The chargeons have projectors
\[P_1 = \frac{1}{6}\sum_g g, \quad P_{\sigma} =
\frac{1}{6}(e-u-v-w+uv+vu), \quad P_{\tau} = \frac{1}{6}(2e-uv-vu)
\]
in $\C S_3$ with actual $D(S_3)$ projectors $P_{e,1}=\delta_e \otimes P_1, P_{e,\sigma}=\delta_e \otimes P_\sigma, P_{e,\tau}=\delta_e \otimes P_\tau$. The fluxion projectors  are
\begin{align*}P_{u, 1} &= \sum_{c\in \CC}\delta_c \tens q_c\Lambda_{C_G}q_c^{-1} = \frac{1}{2}(\delta_u \otimes (e + u) + \delta_v \otimes (e + v) + \delta_w \otimes (e + w))\\
P_{uv, 1}& = \frac{1}{3}(\delta_{uv}+\delta_{vu})(e + uv + vu) \end{align*}
along with $P_{e, 1}$ from before which can be viewed as either. The remaining projectors after a short computation are
\begin{align*}
P_{u, -1} &= \frac{1}{2}(\delta_u \otimes (e-u) + \delta_v \otimes (e-v) + \delta_w \otimes (e-w)) \\
P_{uv, \omega} 
&= \frac{1}{3}(\delta_{uv} \otimes (e + \omega^{-1}  uv + \omega vu) + \delta_{vu} \otimes (e + \omega uv + \omega^{-1} vu))\\
P_{uv, \omega^{-1}} &= \frac{1}{3}(\delta_{uv} \otimes (e + \omega  uv + \omega^{-1} vu) + \delta_{vu} \otimes (e + \omega^{-1} uv + \omega vu))
\end{align*}

On a lattice $\Sigma$ where each edge has an associated state in $\mathbb{C}S_3$, $\CL(s_0, s_1)$ has the quasiparticle basis $|u, v; \CC, \pi\>$ from Corollary~\ref{cor:particle_basis}, where unlike the $\Z_n$ case
$u = (c, i), v = (d, j)$ can have different $i,j$ as  not all irreps are 1-dimensional. To avoid a clash with group elements of $S_3$, we will refer to the pairs $(c, i), (d, j)$ directly. We again refer to $\CC$ by its representative. Then in our case, the  ribbon operators required to create these bases from vacuum for each chargeon are
\begin{align*}
&F^{'e, 1}_{\xi} = \sum_{n \in S_3} F^{e, n}_{\xi} = \id,\quad F^{'e, \sigma}_{\xi} = \sum_{n \in S_3} \sign(n) F^{e, n}_{\xi},\quad F^{'e, \tau; i, j}_{\xi} = \sum_{n \in S_3} \tau(n^{-1})_{ji} F^{e, n}_{\xi} 
\end{align*}
The last of these is the only case with $i,j$ indices as the other $\pi$ are 1-dimensional. Similarly, for fluxions:
\begin{align*}
&F^{'u, 1; c, d}_{\xi} = F^{c, q_cq_d^{-1}}_{\xi} + F^{c, q_cuq_d^{-1}}_{\xi},\quad F^{'uv, 1; c, d}_{\xi} = F^{c, q_cq_d^{-1}}_{\xi} + F^{c, q_cuvq_d^{-1}}_{\xi} + F^{c, q_cvuq_d^{-1}}_{\xi}
\end{align*}
where in the first case have indices $c,d\in\{u,v,w\}$ and in the second case $c,d\in\{uv,vu\}$.  The remaining quasiparticle basis operators are
\begin{align*}
F^{'u, -1; c, d}_{\xi} &= F^{c, q_cq_d^{-1}}_{\xi} - F^{c, q_cuq_d^{-1}}_{\xi} \\
F^{'uv, \omega; c,d}_{\xi} &= F^{c, q_cq_d^{-1}}_{\xi} + \omega^{-1} F^{c, q_c uv q_d^{-1}}_{\xi} + \omega F^{c, q_c vu q_d^{-1}}_{\xi}\\
F^{'uv, \omega^{*}; c,d}_{\xi} & = F^{c, q_cq_d^{-1}}_{\xi} + \omega F^{c, q_c uv q_d^{-1}}_{\xi} + \omega^{-1} F^{c, q_c vu q_d^{-1}}_{\xi}
\end{align*}
with corresponding indices as before. 

We will mainly need the traces $W_\xi^{\CC,\pi}$ of these defined in (\ref{WCpi}). Up to normalisation, these are just the  $P_{\CC,\pi}$ already computed but converted to ribbon operators according to the last part of Proposition~\ref{Ls0s1}.  For chargeons these come out as 
\[W^{e, 1}_{\xi} = \id, \quad W^{e, \sigma}_{\xi} = \sum_{n \in S_3} \sign(n) F^{e, n}_{\xi}, \quad W^{e, \tau}_{\xi} = \sum_{j} F^{'e, \tau; j, j}_{\xi} = 2F^{e, e}_{\xi} - F^{e, uv}_{\xi} - F^{e, vu}_{\xi}.\]
For fluxions we have
\begin{align*}
&W^{u, 1}_{\xi} = \sum_{c \in \{u,v,w \}} F^{c, e}_{\xi} + F^{c, c}_{\xi},\quad W^{uv,1}_{\xi} = \sum_{c \in \{uv, vu\}} F^{c, e}_{\xi} + F^{c, uv}_{\xi} + F^{c, vu}_{\xi} \end{align*}
and the other ones are
\begin{align*}
&W^{u, -1}_{\xi} = \sum_{c \in \{u,v,w \}} F^{c, e}_{\xi} - F^{c, c}_{\xi}\\
&W^{uv, \omega}_{\xi} = F^{uv, e}_{\xi}+F^{vu, e}_{\xi}+ \omega (F^{uv, vu}_{\xi}+ F^{vu, uv}_{\xi})+\omega^{-1}(F^{uv, uv}_{\xi}+F^{vu, vu}_{\xi})\\
&W^{uv, \omega^*}_{\xi} = F^{uv, e}_{\xi}+F^{vu, e}_{\xi}+\omega (F^{uv, uv}_{\xi}+F^{vu, vu}_{\xi}) + \omega^{-1} (F^{uv, vu}_{\xi}+F^{vu, uv}_{\xi})
\end{align*}
Note that the $\CC=\{uv,vu\}$ class is self-inverse but its elements are not self-inverse, so $\CC^*$ is the same class $\CC$ but with $r_{\CC^*}=vu$ and $q^*_{uv}=q_{vu}=v, q^*_{vu}=q_{uv}=e$. Hence Lemma~\ref{lemW} says that
\[ W^{uv, \omega}_{\xi}{}^\dagger=W^{vu, \omega^*}_{\xi}=\sum_{n}\pi_{\omega^*}(n^{-1})(F^{uv,vnv^{-1}}+F^{vu,n})=W^{uv, \omega}_{\xi}\]
so this works out as self-adjoint (as one can also check directly). Similarly for $W^{uv, \omega^*}_{\xi}$, and more obviously for the other cases. 

The maximally entangled state is then
\begin{align*}|{\rm Bell};\xi\>&=\Big({1\over 6}(W_\xi^{e,1}+W_\xi^{e,\sigma}+2 W_\xi^{e,\tau})+{1\over 2}(W_\xi^{u,1}+W_\xi^{u,-1})\\
&\qquad\qquad\qquad+ {1\over 3}(W_\xi^{uv,1}+ W_\xi^{uv,\omega}+W_\xi^{uv,\omega^*})\Big)\vac \\
&=\sum_{h\in S_3} F_\xi^{h,e}\vac
\end{align*}
as required by Corollary~\ref{blocktele}, the first expression being as a sum of 8 mini Bell states. 

\subsubsection{Protected qubit system using $S_3$ ribbons}

Here, we provide a concrete construction of a protected logical qubit within the
$D(S_3)$ Kitaev model, elaborating on ideas in \cite{Woot}. Let $\Sigma$ be a lattice in the vacuum state.
Let $\xi$ be a ribbon between sites $s_0 := (v_0, p_0)$ and $s_1 := (v_1, p_1)$.
\[\input{ds3_1.tikz}\]
This particular choice of ribbon and sites is just for illustrative purposes; any open ribbon will do. We focus initially on the chargeon sector with $W_\xi^\tau:=W_{\xi}^{e,\tau}$. If we apply this to the vacuum the lattice is now occupied by quasiparticles $\pi$ and $\pi^*$ at
sites $s_0$ and $s_1$. Next, let $\xi':s_2\to s_3$ be another ribbon and apply the ribbon operator $W^{\tau}_{\xi'}$
\[\input{ds3_2.tikz}\]
We call this state $|0_L\>:=W_{\xi'}^\tau\circ W_\xi^\tau\vac$ for reasons which will become clear. $|0_L\> \in \CL(s_0,s_1,s_2,s_3)$, and now $\tau$ quasiparticles occupy the lattice at sites $s_0, s_1, s_2, s_3$, which is obvious as
$P_{\tau} \la_{s_i} W_{\xi'}^\tau\circ W_\xi^\tau\vac = W_{\xi'}^\tau\circ W_\xi^\tau\vac$ for all $i$.

Next, let $\xi'':s_0\to s_2$ connect across as
\[\input{ds3_3.tikz}\]
and apply the ribbon operator $W^{\sigma}_{\xi''}$ to $|0_L\>$, defining $|1_L\> := W^{\sigma}_{\xi''}|0_L\>$. We claim that $1_L\>$ still has only $\tau$ excitations at $s_0, s_1, s_2, s_3$. We check this by expanding $P_{\tau}$ and $W^{\sigma}_{\xi''}$:
\begin{align*}
P_{\tau} \la_{s_0} W^{\sigma}_{\xi''} &= \frac{1}{6}(2e\la_{s_0} - uv\la_{s_0}-vu\la_{s_0})(F^{e, e}_{\xi''}-F^{e, u}_{\xi''}-F^{e, v}_{\xi''}-F^{e, w}_{\xi''}+F^{e, uv}_{\xi''}+F^{e, vu}_{\xi''})\\
&= \frac{1}{6}(2(F^{e, e}_{\xi''}-F^{e, u}_{\xi''}-F^{e, v}_{\xi''}-F^{e, w}_{\xi''}+F^{e, uv}_{\xi''}+F^{e, vu}_{\xi''})e\la_{s_0}\\
&-(F^{e, vu}_{\xi''}-F^{e, w}_{\xi''}-F^{e, u}_{\xi''}-F^{e, v}_{\xi''}+F^{e, e}_{\xi''}+F^{e, vu}_{\xi''})uv\la_{s_0}\\
&-(F^{e, uv}_{\xi''}-F^{e, v}_{\xi''}-F^{e, w}_{\xi''}-F^{e, u}_{\xi''}+F^{e, vu}_{\xi''}+F^{e, e}_{\xi''})vu\la_{s_0})\\
&= (F^{e, e}_{\xi''}-F^{e, u}_{\xi''}-F^{e, v}_{\xi''}-F^{e, w}_{\xi''}+F^{e, uv}_{\xi''}+F^{e, vu}_{\xi''})\frac{1}{6}(2e\la_{s_0} - uv\la_{s_0}-vu\la_{s_0})\\
&= W^{\sigma}_{\xi''}\circ P_{\tau} \la_{s_0}
\end{align*}
by Lemma~\ref{ribcom}. Therefore
\begin{align*}
P_{\tau} \la_{s_0}|1_L\> &= P_{\tau} \la_{s_0}W^{\sigma}_{\xi''}\circ W^{\tau}_{\xi'}\circ W^{\tau}_{\xi}\vac\\
&=W^{\sigma}_{\xi''}\circ P_{\tau} \la_{s_0} W^{\tau}_{\xi'}\circ W^{\tau}_{\xi}\vac\\
&= W^{\sigma}_{\xi''}\circ P_{\tau} \la_{s_0}|0_L\> = W^{\sigma}_{\xi''}|0_L\> = |1_L\>
\end{align*}
and an identical calculation applies at $s_1$.

The states $|0_L\>$ and $|1_L\>$ are therefore indistinguishable by local projectors, as the orthogonality of projectors shown in Lemma~\ref{lemPCpi} means that for all $P_{\CC, \pi}$, $P_{\CC, \pi} \la_{s_i} |0_L\> = P_{\CC, \pi} \la_{s_i}|1_L\> = 0$, $\forall s_i$ iff $\CC, \pi \neq e, \tau$, and 
$P_{e, \tau} \la_{s_i} |0_L\> = |0_L\>$, $P_{e, \tau} \la_{s_i} |1_L\> = |1_L\>$.
A physical explanation is that the $\sigma$ quasiparticles generated by $W^{\sigma}_{\xi''}$ at sites $s_0, s_2$ `fuse' with the extant $\tau$ quasiparticles, as we have $\sigma \otimes \tau = \tau$.
Now, $|0_L\>$ and $|1_L\>$ are orthogonal since
\begin{align*}
\<0_L|1_L\> &= \vacket W^{\tau}_{\xi}{}^\dagger \circ W^{\tau}_{\xi'}{}^\dagger  \circ W^{\sigma}_{\xi''} \circ  W^{\tau}_{\xi'} \circ  W^{\tau}_{\xi} \vac\\
&= \vacket W^{\sigma}_{\xi''} \circ W^{\tau}_{\xi} \circ  W^{\tau}_{\xi'} \circ W^{\tau}_{\xi'} \circ  W^{\tau}_{\xi} \vac\\
&= \vacket W^{\sigma}_{\xi''}  \circ W^{\tau \otimes \tau}_{\xi'} \circ W^{\tau \otimes \tau}_{\xi} \vac
\end{align*}
by (\ref{eq:delta_ribbons}), Lemma~\ref{lemWcomp} and Lemma~\ref{lemW}. By the arguments of Lemma~\ref{lem:manysites}, $W^{\tau \otimes \tau}_{\xi'}W^{\tau \otimes \tau}_{\xi} \vac$ has no support in $\CL(s_0, s_2)$, while $W^{\sigma}_{\xi''}\vac$ has no support in $\CL(s_0, s_1)$ or $\CL(s_2, s_3)$, and so 
\[\<0_L|1_L\> = 0.\]
Thus, $\CH_L := \mathrm{span}(\{ |0_L\>, |1_L\> \})$ is a 2-dimensional subspace of $\CL(s_0,s_1,s_2,s_3)$ that is degenerate under $H$. We call $\CH_L$ a \textit{logical qubit} on the lattice.
By similar arguments as for the vacuum in Section~\ref{sec:vac}, any state in $\CH_L$ is `topologically protected'; local errors leave the state unaffected. In this case, the two types of errors which
are undetectable and affect $\CH_L$ are (a) loops enclosing at least one occupied site and (b) ribbon operators extending between occupied sites. Therefore, quasiparticles should be placed at distant locations
to minimise errors.

We then identify $W^{\sigma}_{\xi''}$ with $X_L$, the logical $X$ gate, which is justified as 
\[W^{\sigma}_{\xi''}\circ W^{\sigma}_{\xi''} = W^{\sigma \otimes \sigma}_{\xi''} = W^{1}_{\xi''} = \id\]
by Lemma~\ref{lemWcomp}. Therefore, $X_L$ is involutive as desired for any implementation of a qubit computation within the model, for example by ZX-calculus based on $\C\Z_2$ as a quasiFrobenius algebra. Clearly, we can obtain any $X_L$ basis rotation on the logical qubit by exponentiation. In \cite{Woot} it is argued that we can in fact acquire universal quantum computation by an implementation of the logical Hadamard, entangling gates and measurements. For completeness, we outline some aspects of these further steps in  Appendix~\ref{app:universal_comp}.

\section{Aspects of general $D(H)$ models}\label{secH}

The Kitaev model is known to generalise with $\C G$ replaced by any finite-dimensional Hopf algebra $H$ with antipode $S$ obeying $S^2=\id$  (which over $\C$ or another field of characteristic zero is equivalent to $H$ semisimple or cosemisimple). Although less well studied, that one can obtain topological invariants as a version of the Turaev-Viro invariant was shown in \cite{Kir,Meu}. That one has an action of the Drinfeld quantum double $D(H)$ \cite{Dri} at each site is more immediate and was first noted in \cite{BMCA}. We just replace the group action $g\la$ by $h\la$ acting in the tensor product representation with factors in order going around the vertex as in Figure~\ref{figHact}, which now depends on where $p$ is located. We likewise replace the action of $\delta_g$ by  $a\la$ for $a\in H^*$ and likewise just take the tensor product action around the face in the order depending on where $v$ is located. We use the Hopf algebra regular and coregular representations 
\begin{equation}\label{Hverfacactions} h\la g=hg\quad {\rm or}\quad h\la g=gS h;\quad a\la g=a(g_1)g_2\quad{\rm or}\quad a\la g=a(Sg_2)g_1\end{equation}
with the first choice if the arrow is outbound for the vertex /in the same direction as the rotation around the face.  Here $\Delta g=g_1\tens g_2$ (sum understood) denotes the coproduct $\Delta:H\to H\tens H$ and $a\la$ is a right action of $H^*$ viewed as a left action of $H^{*op}$. The antipode $S:H\to H$ is characterised by $(Sh_1)h_2=h_1Sh_2=1\eps(h)$ for all $h\in H$, where $\eps\in H^*$ is the counit. We refer to \cite{Ma} for more details.

We have also used better conventions for $D(H)$,  namely the double cross product construction introduced by the 2nd author in \cite{Ma:phy}. Here  $D(H)\isom H^{*op}\dcross H$, where $H$ left acts on $H^*$ and $H^*$ left acts on $H$ by the coadjoint actions
\[ h\la a=a_2\<h, (Sa_1)a_3)\>,\quad  h\ra a=h_2\<a, (Sh_1)h_3\>\]
with the left action of $H^*$ viewed as a right action of $H^{*op}$. The numerical suffices denote iterated coproducts (sums understood) and $\<\ ,\ \>$ is the duality pairing or evaluation. These then form a matched pair of Hopf algebras\cite{Ma:phy} and give the Drinfeld double explicitly as  \cite[Thm~7.1.1]{Ma},
\[ (a\tens h)(b\tens g)=b_2 a\tens h_2 g\<Sh_1,b_1\>\<h_3,b_3\>,\quad \Delta (a\tens h)=a_1\tens h_1\tens a_2\tens h_2.\]
\[ S(a\tens h)=S^{-1}a_2\tens S h_2\<h_1,a_1\>\<Sh_3,a_3\>,\quad \CR=\sum_a f^a\tens 1\tens 1\tens e_a,\]
where we also give the factorisable quasitriangular structure. Here $\{e_a\}$ is a basis of $H$ and $\{f^a\}$ is a dual basis. We will also sometimes employ a subalgebra notation where $h,a$ are viewed in $D(H)$ with cross relations $ha=a_2 h_2  \<Sh_1,a_1\>\<h_3,a_3\>$ and $\CR=\sum_a f^a\tens e_a$. 
While this much is clear, explicit properties of ribbon operators have not been much studied as far as we can tell even for $S^2=\id$, and we do so here. Moreover, we will explore how much can be done without this semisimplicity assumption. 

From Hopf algebra theory, we will particularly need that every finite-dimensional Hopf algebra $H$ has, uniquely up to normalisation, a left integral element $\Lambda\in H$  such that $h\Lambda =\eps(h)\Lambda$ and a right-invariant integral map $\int\in H^*$ such that $(\int h_1)h_2=1\int h$ for all $h$. Ditto with left-right swapped. In the semisimple case in characteristic zero the integrals can be normalised so that $\eps(\Lambda)=\int 1=1$, are both left and right integrals at the same time, and obey  $\int hg=\int gh$ and $\Delta\Lambda={\rm flip}\Delta\Lambda$ (here $\int =\Tr_H/\dim H$ is the normalised trace in the left regular representation), see \cite{Sch} for an account (the general theory underlying this goes back to the work of Larson and Radford). If we denote irreps of $H$ by $(\pi,V_\pi)$ then analogously to the group case, one has a complete orthogonal set of central idempotents $P_\pi$ given by
\begin{equation}\label{Phopf} P_\pi=\dim(V_\pi) \Lambda_1 \Tr_\pi(S\Lambda_2)\end{equation}
whereby $P_\pi H= H P_\pi\isom \End V_\pi$. Note that $\sum_\pi P_\pi=\Lambda_1\int S\Lambda_2\dim H=1$ as part of the Frobenius structure where $\Lambda$ is currently normalised so that $\int\Lambda=1/\dim H$ compared to usual normalisation in \cite{Sch,Ma:fro}. We omit the proof but part of the theory is the orthogonality relation $\Tr_{H^*}(\chi_\pi\chi_{\pi'})=\delta_{\pi,\pi'}\dim H$ for normalised characters $\chi_\pi=\Tr_\pi/\dim V_\pi$.  Moreover, in this case of $H$ semisimple, $D(H)$ is also, with integrals $\Lambda_D=\int\tens\Lambda$ and $\int_D=\Lambda\tens\int$. Hence the same result applied to $D(H)$ tells us that $D(H)\isom \oplus_{\tilde\pi} \End(V_{\tilde\pi})$ now for irreps $(\tilde\pi,V_{\tilde\pi})$ of $D(H)$. Hence our ideas about Bell states and ribbon teleportation still apply in this case, with quasiparticles detected by projectors $P_{\tilde\pi}$. 

Also note that a representation of $D(H)$ can also be described as a $H$-crossed \cite{Ma,Ma:pri} or `Drinfeld-Yetter' module consisting of a left action or representation $\pi$ of $H$ and a compatible {\em left coaction} of $H$ $\Delta_L$ (this is equivalent to a compatible right action of $H^*$  or left action of $H^{*op}$ on the same vector space, these being two subalgebras from which $D(H)$ is built).  If $V_{\tilde\pi}$ has basis $\{e_i\}$ then  the structures for a crossed module are $\Delta_L e_i=\rho_{ij}\tens e_j$ where $a\la e_i=\<a,\rho_{ij}\>e_j$ is the corresponding action, and $h\la e_i=\pi(h)_{ki}e_k$ as usual. We sum over the repeated $k$ and $\rho_{ij}\in H$ is required to obey
\[ \Delta \rho_{ij}=\rho_{ik}\tens\rho_{kj},\quad \eps\rho_{ij}=\delta_{ij}, \quad h_1\rho_{ik}\pi(h_2)_{jk}  =\pi(h_1)_{ki}\rho_{kj} h_2 \]
for all $h\in H$,  again summing over $k$.  

\begin{lemma} Let $S^2=\id$ and  $\tilde\pi$ an irrep of $D(H)$ with $\pi,\rho$ its associated crossed module data with respect to a basis $\{e_i\}$ as above. Then 
\[ P_{\tilde\pi}=\dim(V_{\tilde\pi})\sum_{a,i,j}f^a\tens \Lambda_1 \pi(S\Lambda_2)_{ij} (\int e_a S\rho_{ij}).\]
Moreover, when specialised to $D(G)$, we recover the projectors $P_{\CC,\pi}$ in (\ref{PCpi}). 
\end{lemma}
\proof The general formula for the tensor product integral becomes
\begin{align*} P_{\tilde\pi}&=\dim(V_{\tilde\pi}) \int_1\tens \Lambda_1\Tr_{\tilde\pi}((S\int_2)S\Lambda_2)=\dim(V_{\tilde\pi})f^a\int_1 e_a\tens \Lambda_1\<S\int_2,\rho_{kj}\> \<f^i, e_j\>\pi(S\Lambda_2)_{ki} \end{align*}
which becomes as stated using Hopf algebra duality. In the $D(G)$ case, we let $\tilde\pi=(\CC,\pi)$ as before and go back to (\ref{Phopf}). Then
\begin{align*} P_{\tilde\pi}&={\dim(V_{\CC,\pi})\over |G|}\sum_{h,g\in G}(\delta_h\tens g)\Tr_{\tilde\pi}(S(\delta_{h^{-1}}\tens g))\\
&={\dim V_\pi\over |C_G|}\sum_{h,g\in G}\sum_{c\in \CC,i}(\delta_h\tens g)\<\delta_c\tens f^i,(\delta_{g^{-1}hg}\tens g^{-1})\la (c\tens e_i)\>\\ &={\dim V_\pi\over |C_G|}\sum_{h,g\in G}\sum_{c\in \CC}(\delta_h \tens g) \delta_{g^{-1}hg,g^{-1}cg} \delta_{c,g^{-1}cg}\Tr_\pi(q^{-1}_{g^{-1}cg}g^{-1}q_c)=P_{\CC,\pi} \end{align*}
where $\la$ is the action (\ref{acteu}). We view the restrictions setting $h=c$ and  $g\in C_G(c)$. Changing variables to $n=q_cgq_c^{-1}$, this is equivalent to a sum over $n\in C_G(r_\CC)$ as usual and $n^{-1}$ in the trace. \endproof
   
How exactly to construct and classify irreps $\tilde \pi$, however, depends on the structure of $D(H)$, which is no longer generally a semidirect product. This therefore has to be handled on a case by case basis before one can do practical quantum computations. 

\begin{example}\rm  Let $G=G_+.G_-$ be a finite group that factorises into two subgroups $G_\pm$, neither of which need be normal and $H=\C(G_-)\bicross \C G_+$  the associated bicrossproduct (or `bismash product') quantum group\cite{Tak,Ma:phy,Ma}, which is semisimple. It is shown in \cite{BGM} that $D(H)\isom D(G)_F$, where the latter is  a Drinfeld twist of $D(G)$ by a 2-cocycle 
\[ F=\sum_{g\in G}1\tens g_-\tens\delta_{g^{-1}}\tens 1\in D(G)\tens D(G)\]
in the sense of \cite{Ma}. We write $g=g_+g_-$ for the unique factorisation of any element of $G$. Explicitly, $D(G)_F$ has the same algebra as $D(G)$ but a conjugated coproduct 
\[ \Delta(\delta_g\tens h)=F(\Delta_{D(G)}(\delta_g\tens h))F^{-1}=\sum_{f\in G}\delta_{f_- g f f_-^{-1}}\tens f_- h (h^{-1}fh)_-^{-1}\tens \delta_{f^{-1}}\tens h\]
after a short computation. The nontrivial isomorphism with $D(H)$ in \cite{BGM} is needed to identify the $H$ and $H^{*op}$ subalgebras but where this is not required, we can work directly with this twisted description. In particular, irreps of $D(G)_F$ are the same as those of $D(G)$ (since the algebra is not changed) and can be identified with irreps of $D(H)$ by the isomorphism. The braided tensor category is different from but monoidally equivalent to that of $D(G)$. 
\end{example}

We  will be concerned more with the formalism with explicit models, such as based on this construction,  deferred to a sequel.  We see, however, that there are plenty of examples. Note that $G\rcross G$ by Ad is an example with one subgroup normal, so $H=D(G)$ is covered by this analysis and $D(D(G))\isom D(G\rcross G)_F$.  

\subsection{$D(H)$ site operators.} 
\begin{figure}\[ \includegraphics[scale=0.7]{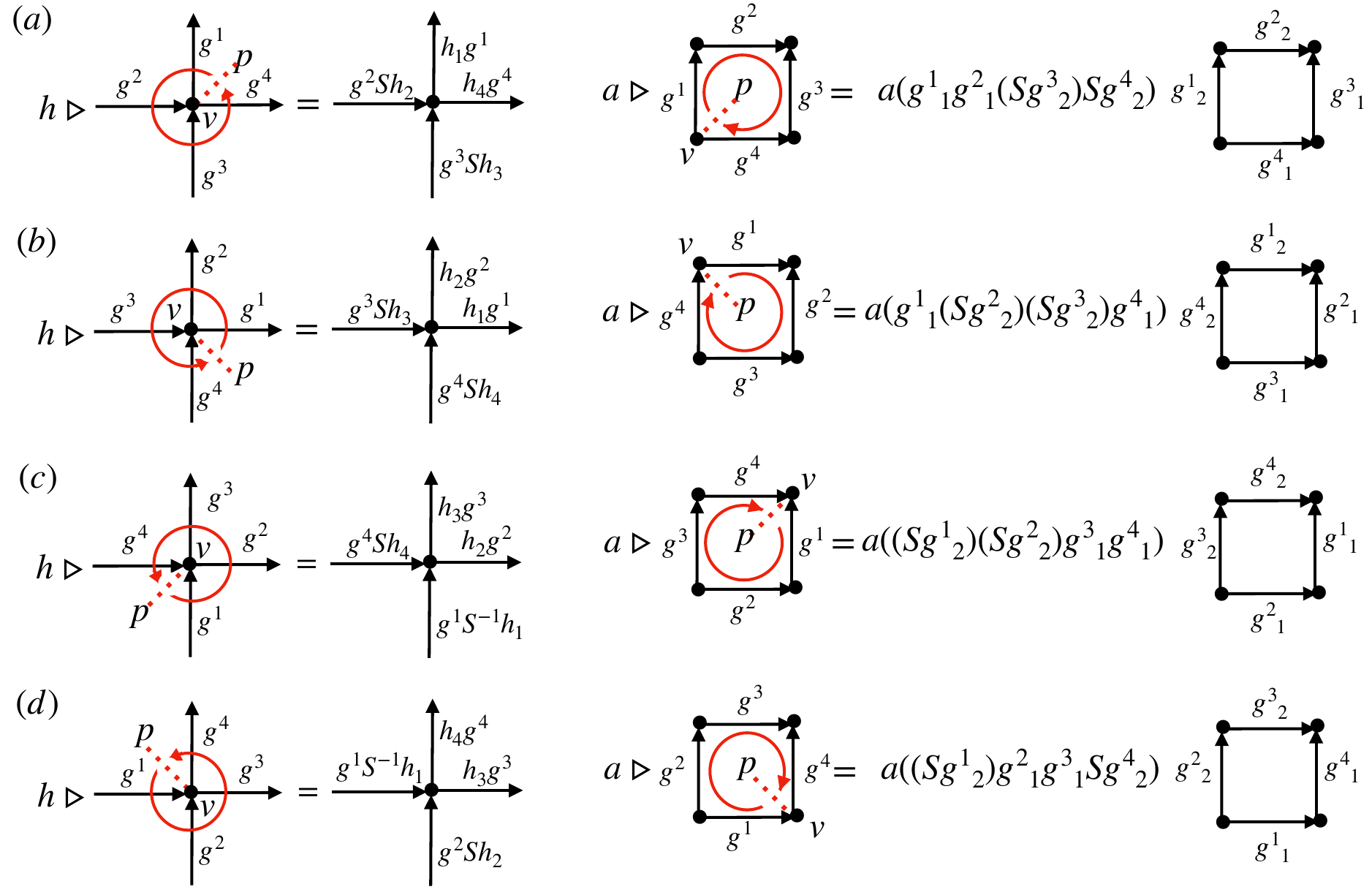}\]
\caption{\label{figHact} Kitaev model representation of $D(H)$ at a site $(v,p)$ for general not necessarily semisimple Hopf algebras. Most instances of the antipode $S$ can be equally $S^{-1}$ but if we use $S$ in all the $a\la$ then we have to use $S^{-1}$ in $h\la$ if this occurs in the first arrow encountered in going around the vertex.}
\end{figure}

By working with the above cleaner form of the Drinfeld double, our modest new observation in this section is that the same format for the Kitaev model works in the general case without assuming $S^2=\id$ provided we use additional information from the lattice geometry to distinguish the four cases (a)-(d) in Figure~\ref{figHact} which follow the same rules as above but sometimes specify to use $S^{-1}$. We focus on the case of a square lattice with its standard orientation as this is most relevant to computer science, rather than on a general ciliated ribbon graph.

\begin{figure}\[ \includegraphics[scale=0.7]{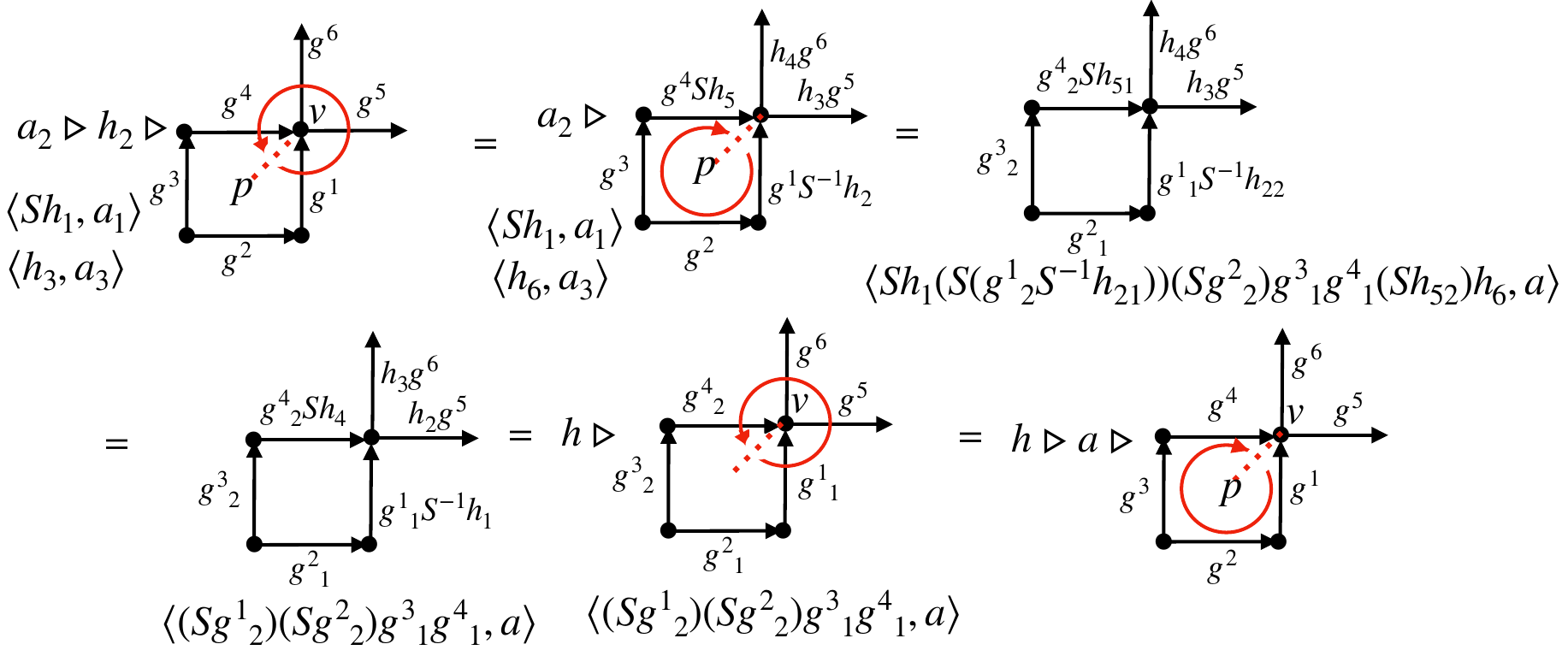}\]
\caption{\label{figpfHact} Proof that case (c) of Figure~\ref{figHact} works in Theorem~\ref{thmDact}.}
\end{figure}

\begin{theorem}\label{thmDact} If $(v,p)$ is a site in the lattice then the actions for the form in Figure~\ref{figHact}, where we act as shown and by the identity on other arrows, is a representation of $D(H)$ provided we use (as shown) $S^{-1}$ if the first arrow going around the vertex is inward and $S$ if the last arrow is inward. We can freely choose $S$ or $S^{-1}$ if the inward arrow is in one of the intermediate places. 
\end{theorem}
\proof We have to check the relations $a_2h_2\<Sh_1,a_1\>\<h_3,a_3\>=ha$ for all $h\in H$ and $a\in H^*$. The proof of the hardest case (c) is in Figure~\ref{figpfHact} and for the cancellation for the 3rd equality, we see that we need $S^{-1}$ when the first vertex going around is inward and $S$ when the last vertex is, as is the case here. The other cases are similar but slightly easier as unconstrained on the choice of $S$ where there is no inward arrow in one or both of these  positions.  \endproof

We could equally well decide to always use $S$ for  $h\la$ and use $S^{-1}$ in $a\la$ if the contraflowing is in first position going around the face and $S$ if it is in last position. (This is the same as above in the dual lattice and faces and arrows interchanged and with the roles of $H,H^*$ interchanged.) {\em We see that in the non-involutive $S$ case there is still some freedom in the choice of $S$ or $S^{-1}$, which we need to fix by what we want to do with these $D(H)$-representations.} 

We also know that our finite dimensional Hopf algebra has up to scale a unique right integral  $\Lambda\in H$ and a unique right-invariant integral $ \int:H \to k$ so we can proceed to define operators \[
A(v,p)=\Lambda\la,\quad B(v,p)=\int\la \]
on $\CH$. It is striking that exactly this integral data is also key to a Frobenius Hopf algebra interacting pair for  ZX calculus based on $H$, see \cite{CoD,Ma:fro} at this level of generality. Clearly 
\[ A(v,p)^2=\eps(\Lambda)A(v,p),\quad B(v,p)^2=(\int 1)B(v,p)\]
but without further assumptions, both operators depend on both parts of the site. One can also check that
\[ [A(v,p),A(v',p')]=0,\quad [B(v,p),B(v',p')]=0\]
for all $v,v',p,p'$ with in the first case $v\ne v'$ and in the second case $p\ne p'$. The first is because if, in the worst case, the vertices are adjacent then the common arrow is pointing in for one vertex and out for the other, hence the element $g$ in the middle gets multiplied by something on the left and something on the right, which does not depend on the order by associativity. Similarly for two faces with an arrow in common. We do {\em not} in general have that $[A(v,p),B(v,p)]=0$.

For the Hamiltonian, there are two possible approaches. (i) we could we fix the vertex of all site to be at the bottom left of the face (Case (a) in Figure~\ref{figHact}). Thus if $v$ is a vertex, we define $p_v$ as the face to its upper right. Then
\[ H=\sum_v (1-A(v,p_v)+1- B(v,p_v))\]
makes sense. (ii) Alternatively, motivated by \cite{Meu} we can define
\[ H_K=\prod_{(v,p)}A(v,p)B(v,p)\]
The $D(G)$ model admits a Hamiltonian which is necessarily frustration-free, meaning that any vacuum state is also the lowest energy state of any given local term. This condition is broken by general $D(H)$ models. Let $A(v_1, p_1)$ be a local term. First, consider the Hamiltonian from (i), ignoring the additive constant:
\begin{align*}
    A(v_1, &p_{v_1})\vac = - A(v_1, p_{v_1}) \sum_v (A(v,p_v)+ B(v,p_v))\vac\\ 
    &= -(\eps(\Lambda)A(v_1, p_{v_1})+A(v_1, p_{v_1})B(v_1, p_{v_1}) + \sum_{v \neq v_1} (A(v,p_v)+ B(v,p_v)))\vac
\end{align*}
So in general, $A(v_1, p_{v_1})\vac \neq \vac$. Next, consider the Hamiltonian from (ii):
\begin{align*}
    A(v_1, p_{v_1})\vac &= A(v_1, p_{v_1}) \prod_{(v,p)}A(v,p)B(v,p)\vac = \eps(\Lambda)\vac
\end{align*}
The fact that the integral actions are no longer idempotent also breaks the interpretation of these actions as `check operators' to be measured and detect unwanted excitations. In these more general models, it is unclear what error-correcting capabilities still exist on the lattice, or whether there are alternative methods of preserving fault-tolerance. They don't appear to fit under the umbrella of `surface codes' in the usual sense.
We still preserve some locality as a feature of the model, in the sense that a locally vacuum state can be defined for example in case (ii) as being the image of $\prod_{(v \in V_R,p \in P_R)}A(v,p)B(v,p)$, where $V_R$ and $P_R$ are sets of vertices and faces in some region $R$.

In the semisimple case where $S^2=\id$, we have already noted that $\Delta\Lambda={\rm flip}\Delta\Lambda$  and that the integrals can be normalised so that $\eps(\Lambda)=\int 1=1$. This implies  that $A(v,p)=A(v)$ independently of $p$ and $B(v,p)=B(p)$ independently of $v$, are projectors and, using the commutation relations of $D(H)$ and Theorem~\ref{thmDact} that $[A(v),B(p)]=0$. Therefore, we recover both the frustration-free property of $H$ and the interpretation of the lattice as a fault-tolerant quantum memory. In this case, it is claimed in \cite{Meu} that
\[ H_K=\prod_v A(v)\circ\prod_p B(p),\quad \CH_{vac}={\rm image}(H_K)\]
results in the latter `protected space' being a topological invariant of the surface obtained by gluing discs on the faces of a ribbon graph. This motivates the definition above. It is also claimed in \cite{Meu} that a particle at $(v,p)$ corresponds to a defect where we leave out the site $(v,p)$ in the product. 

While the $D(G)$ model on a lattice $\Sigma$ allows for the convenient expression in Theorem~\ref{thm:cui} for $\dim(\CH_{vac})$ in terms of the fundamental group $\pi_1(\Sigma)$, the proof of this relies on the invertibility of group elements and the invariance under orientation of $\delta_e \la$. The topological content for $D(H)$ models can similarly be related to holonomy as in \cite{Meu} but is more complicated. The topological content in the $D(H)$ model in the non-semisimple case is less clear and will be more indirect. For example,  reversal of orientation cannot be expressed simply via the antipode as this no longer squares to the identity. 

\subsection{$D(H)$ triangle and ribbon operators}

Canonical representations of $D(H)$ that we will need are left and right actions of $D(H)$ on $H$, \cite[Ex.~7.1.8]{Ma}
\begin{equation}\label{DHactH} h\la g=h_1 g Sh_2,\quad a\la g=a(g_1)g_2;\quad g\ra h=(Sh_1)gh_2,\quad g\ra a= g_1 a(g_2)\end{equation}
and left and right actions of $D(H)$ on $H^*$,
\begin{equation}\label{DHactH*} h\la b=\<Sh,b_1\>b_2,\quad a\la b=(S^{-1}a_2) b a_1;\quad b\ra h=b_1\<Sh,b_2\>,\quad b\ra a=a_2 b S^{-1} a_1\end{equation}
which is essentially the same construction with the roles of $H,H^{*op}$ swapped. Moreover, in the quasitriangular case, there is a braided monoidal functor ${}_H\CM\to {}_{D(H)}\CM$, see \cite{Ma,Ma:pri}, which can be used to obtain a class of nice representations of $D(H)$ from irreps of $H$. 

Also note that if $D$ is any Hopf algebra, for example $D=D(H)$, it acts on its dual as a module algebra by the left and right coregular representations
\begin{equation}\label{DactD*} d\la \phi=\<Sd,\phi_1\>\phi_2,\quad \phi\ra d=\phi_1 \<Sd,\phi_2\>\end{equation}
for all $d\in D$ and $\phi\in D^*$.  These already feature in (\ref{DHactH*}) for the action of $H$. Also, if $D$ acts from the left (say) on a vector space $\CH$ by an action $\la$ then it acts on the linear operators $\End(\CH)$ as a module algebra from both the left and the right \cite{Ma}.
\begin{equation}\label{DactL} (d\la L)(\psi)=d_1 \la L(Sd_2\la \psi),\quad  (L\ra d)(\psi)=Sd_1 \la L(d_2\la\psi)\end{equation}
for all $d$ in the Hopf algebra and $L\in \End(\CH)$.  We will use  (\ref{DactL}) with different site actions of $D(H)$ in Theorem~\ref{thmDact} for example $\la_{s_0}$ and $\la_{s_1}$ for the two halves. These commute if $s_0,s_1$ are far enough apart. In the case of $D(H)$, its dual is $H\tens H^*$ as an algebra and has the coproduct
\begin{equation}\label{DeltaD*} \Delta_{D(H)^*}(h\tens a)=\sum_{a,b} h_2\tens f^a a_1 f^b\tens Se_a h_1 e_b \tens a_2.\end{equation}

Next, we define Hopf algebra triangle and ribbon operators at least in the case $S^2=\id$. 
Before attempting this, we need to better understand the $D(G)$ case, both ribbon covariance properties and the construction of a ribbon a sequence of triangle and dual triangle ops as defined in Definition~\ref{def:triangles}. 

\begin{lemma} For $D(G)$ and with left and right actions on $\End(\CH)$ induced as in (\ref{DactL}) by the initial and final site actions:

\begin{enumerate}
\item If $\tau^*$ is a dual triangle, the triangle operator $L_{\tau*}^h=\sum_gF_{\tau^*}^{h\tens \delta_g}$  is a left and right module map $L_{\tau^*}:\C G\to \End(\CH)$, where $D(G)$ acts as in (\ref{DHactH})  by 
\[ (\delta_a\tens b)\la h=\delta_{a,bhb^{-1}}bhb^{-1},\quad h\ra(\delta_a\tens b)=b^{-1}hb\, \delta_{a,h}.\] 
\item If $\tau$ is a direct triangle, the triangle operator $T_\tau^{\delta_g}=F_\tau^{h\tens\delta_g}$ for any $h$ is a left and right module map $T_\tau:\C(G)\to \End(\CH)$, where $D(G)$ acts as in (\ref{DHactH*})  by                                                                                                                                              
\[( \delta_a\tens b)\la\delta_g=\delta_{a,e}\delta_{bg},\quad \delta_g\ra (\delta_a\tens b)=\delta_{a,e}\delta_{gb}.\]
\item If $\xi$ is an open ribbon then $\tilde F_\xi^{h\tens\delta_g}:= F_\xi^{h^{-1},g}$ is a left and right module map $\tilde F:D(G)^*\to \End(\CH)$, where $D(G)$ acts by (\ref{DactD*}). Moreover, if $\xi,\xi'$ are composeable ribbons then 
\[ \tilde F_{\xi'\circ\xi}^{h\tens\delta_g}= \tilde F_{\xi'}^{(h\tens\delta_g)_2}\circ \tilde F_{\xi}^{(h\tens\delta_g)_1}\]
using the coproduct (\ref{DeltaD*}) of $D(G)^*$. This also applies to $F_\xi^{h\tens\delta_g}=F_\xi^{h,g}$. 
\end{enumerate}
\end{lemma}
\proof (1) The relations we find for dual triangles are
\[ (\delta_a\tens b)\la_{s_0}\circ\tilde F^{b^{-1}hb\tens\delta_g}=\tilde F^{h\tens\delta_g}\circ(\delta_{ha}\tens b)\la_{s_0},\quad (\delta_{ah}\tens b)\la_{s_1}\circ\tilde F^{b^{-1}hb\tens\delta_g}=\tilde F^{h\tens\delta_g}\circ(\delta_{a}\tens b)\la_{s_1}\]
which we interpret as stated. 

(2) The relations we find for direct triangles are
\[ (\delta_a\tens b)\la_{s_0}\circ \tilde F^{h\tens\delta_g}=\tilde F^{h\tens\delta_{bg}}\circ (\delta_{a}\tens b)\la_{s_0},\quad \tilde F^{h\tens\delta_g}\circ (\delta_a\tens b)\la_{s_1}=(\delta_{a}\tens b)\la_{s_1}\circ \tilde F^{h\tens\delta_{gb}}\]
which we interpret as stated. These same commutation rules hold for the action $\la_t$ at any site $t$ that has the same vertex as $s_0,s_1$ respectively, while the action at other sites commutes with the triangle operator. 

(3) Here $D(G)^*=\C G\tens\C(G)$ as an algebra while its coproduct dual to the product of $D(G)$ is
\[ \Delta_{D(G)^*}(h\tens\delta_g)=\sum_{f\in G}h\tens\delta_{f}\tens f^{-1}hf\tens \delta_{f^{-1}g}\]
Then the composition rule in equation (\ref{concat}) for $F_\xi$ is already in the form stated. The same then applies  $\tilde F_\xi$ as $S\tens\id$ is clearly a coalgebra map. For equivariance, we have from Lemma~\ref{ribcom},
\begin{align*} \<S(\delta_a\tens b)_1, &(h\tens\delta_g)_1\>\tilde F_\xi^{(h\tens\delta_g)\t}\circ (\delta_a\tens b)_1\la_{s_0} \\
&=\sum_{x,f}\<S(\delta_{x^{-1}}\tens b), h\tens\delta_f\>\tilde F_\xi^{f^{-1}hf\tens\delta_{f^{-1}g}}\circ (\delta_{xa}\tens b)_1\la_{s_0} \\
&=\sum_{x,f}\<\delta_{b^{-1}xb}\tens b^{-1}, h\tens\delta_f\>\tilde F_\xi^{f^{-1}hf\tens\delta_{f^{-1}g}}\circ (\delta_{xa}\tens b)_1\la_{s_0}\\
&=\tilde F_\xi^{bhb^{-1}\tens\delta_{fg}}\circ (\delta_{bhb^{-1}a}\tens b)\la_{s_0}=\delta_a\la \tilde F_\xi^{bhb^{-1}\tens\delta_{bg}}\circ b\la_{s_0}=(\delta_a\tens b)\la_{s_0} \tilde F_\xi^{h\tens\delta_g}
\end{align*}
where  $f=b^{-1}$ and $x=bhb^{-1}$. We used the commutation relations from Lemma~\ref{ribcom}. Similarly for the other side,
\begin{align*}(\delta_a\tens b)&_1\la_{s_1}\circ \tilde F^{(h\tens\delta_g)_1}_\xi \<S (\delta_a\tens b)\t, (h\tens \delta_g)\t\>\\
&=\sum_{x,f}(\delta_{ax}\tens b)\la_{s_1}\circ \tilde F^{h\tens\delta_{f}}_\xi \<S(\delta_{x^{-1}}\tens b), f^{-1}hf\tens \delta_{f^{-1}g}\>\\
&=\sum_{x,f}(\delta_{ax}\tens b)\la_{s_1}\circ \tilde F^{h\tens\delta_{f}}_\xi \<\delta_{b^{-1}x b}\tens b^{-1}, f^{-1}hf\tens \delta_{f^{-1}g}\>\\
&=(\delta_{ag^{-1}hg}\tens b)\la_{s_1}\circ \tilde F^{h\tens\delta_{gb}}_\xi=\delta_{ag^{-1}hg}\la_{s_1}\circ\tilde F^{h\tens \delta_g}_\xi\circ b\la_{s_1}=\tilde F_\xi^{h\tens \delta_g}\circ(\delta_{a}\tens b)\la_{s_1}
\end{align*}
where $gb=f$ and $x=bf^{-1}hfb^{-1}=g^{-1}hg$. 
\endproof

The additional commutation relations for $T_\tau$ mentioned in the proof can best be said as the action on it as an operator in $\End(\CH)$,
\[ (\delta_a\tens b)\la_t(T^{\delta_g})=T^{(\delta_a\tens b)\la_t\delta_g}; \quad (\delta_a\tens b)\la_t\delta_g=\delta_{a,e}\begin{cases}\delta_{bg} \\ \delta_{g b^{-1}}\\ \delta_g \end{cases}\]
where we act as per $L^b$ at vertex $t$ on $g$ regarded effectively as living on the arrow of the direct triangle, i.e. $bg$ if the arrow in relation to the vertex of $t$ is outgoing, $gb^{-1}$ if incoming and $g$ otherwise. We can then derive the left and right module properties for ribbon operators by iterating those for triangle operators. To illustrate this, consider 
\[ \tilde F^{h\tens\delta_g}_{\tau_2\circ\tau_1^*}=T_{\tau_2}^{\delta_g}\circ L_{\tau_1^*}^{h^{-1}}=\sum_f \tilde F_{\tau_2}^{f^{-1}hf\tens\delta_{f^{-1}g}}\circ \tilde F_{\tau_1^*}^{h\tens\delta_f}=(\tilde F_{\tau_2}\circ \tilde F_{\tau_1^*})(h\tens\delta_g)\]
 where $\tau_1^*: s_0\to s_1$ and $\tau_2:s_1\to s_2$ and $\tilde F_{\tau}^{h\tens\delta_g}=T_\tau^{\delta_g}$ and $\tilde F_{\tau^*}^{h\tens\delta_g}=\delta_{g,e}L_{\tau^*}^{h^{-1}}$ are the associated ribbon operators which we convolve as in part (3) of the lemma. Using the first expression and the triangle operator left module properties
 \begin{align*}(\delta_a\tens b)\la_{s_0}(\tilde F_{\tau_2\circ\tau_1^*}^{h\tens\delta_g})&=(\delta_a\tens b)_1\la (T_{\tau_2}^{\delta_g})\circ (\delta_a\tens b)_2\la(L_{\tau_1^*}^{h^{-1}})\\
 &=\sum_x T_{\tau_2}^{(\delta_{ax^{-1}}\tens b)\la_{s_0}\delta_g}\circ L_{\tau_1^*}^{(\delta_x\tens b)\la h^{-1}}\\
 &=T_{\tau_2}^{\delta_{bg}}\circ L_{\tau_1^*}^{\delta_{a,bh^{-1}b^{-1}} bh^{-1}b^{-1}}=\delta_{a^{-1},bhb^{-1}}\tilde F_{\tau_2\circ\tau_1^*}^{bhb^{-1}\tens\delta_{bg}}=\tilde F_{\tau_2\circ\tau_1^*}^{(\delta_a\tens b)\la(h\tens\delta_g)}\end{align*}
 where from (\ref{DactD*}),
 \[ (\delta_a\tens b)\la(h\tens\delta_g)=\sum_f\<\delta_{b^{-1}a^{-1}b}\tens b^{-1},h\tens \delta_f\>f^{-1}hf\tens \delta_{f^{-1}g}=\delta_{h,b^{-1}a^{-1}b}bhb^{-1}\tens\delta_{bg}\]
The similar calculation for $(\tilde F_{\tau_2\circ\tau_1^*}^{h\tens\delta_g})\ra_{s_2}(\delta_a\tens b)$ is not so easy as $\ra_{s_2}$ does not enjoy simple commutation relations with $L^{h^{-1}}$. There is a similar story for
\[ \tilde F^{h\tens\delta_g}_{\tau_2^*\circ\tau_1}=L_{\tau_2^*}^{g^{-1}h^{-1}g}\circ T_{\tau_1}^{\delta_g}=\sum_f \tilde F_{\tau_2^*}^{f^{-1}hf\tens\delta_{f^{-1}g}}\circ \tilde F_{\tau_1}^{h\tens\delta_f}=(\tilde F_{\tau_2^*}\circ \tilde F_{\tau_1})(h\tens\delta_g)\]
as the other smallest open ribbon.

Now proceeding to the Hopf algebra case, we define triangle operations in the obvious manner as partial vertex and face operators, with left multiplication by $h$ or right multiplication by $S^{\pm1}h$ for dual triangles, depending on orientation,
\[ \includegraphics[scale=0.5]{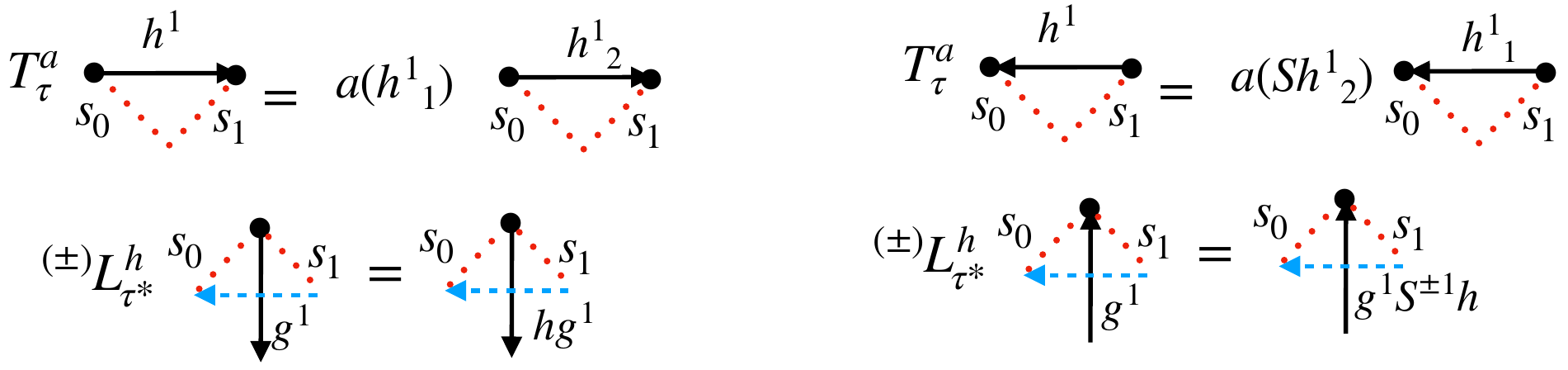}
\]
Recall that we have chosen to use $S$ throughout the face operations but $S^{-1}$ if the first arrow was inward in a vertex operation. As a result, we need both versions ${}^{(\pm)}L^h_{\tau^*}$ to express both left and right covariance. We could equally well have put this complication on the $T^a_\tau$ side. 

\begin{figure}
\[ \includegraphics[scale=0.7]{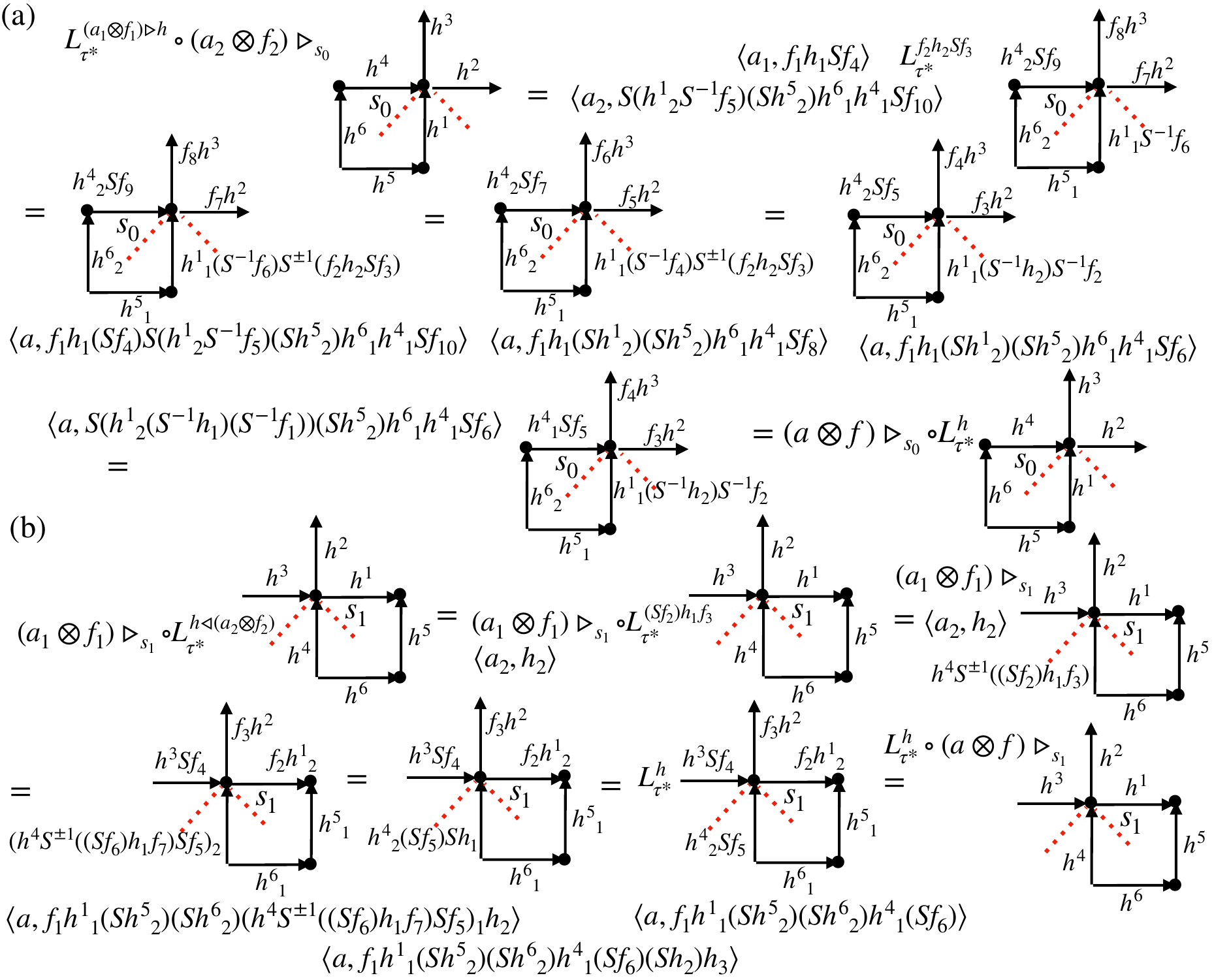}\]
\caption{\label{figHcotri} (a) Proof of left covariance of dual triangle operator needing the $(-)$ version for the 3rd equality. (b) Proof of right covariance needing the $(+)$ version. They coincide when $S^2=\id$.}
\end{figure}

\begin{figure}
\[ \includegraphics[scale=0.5]{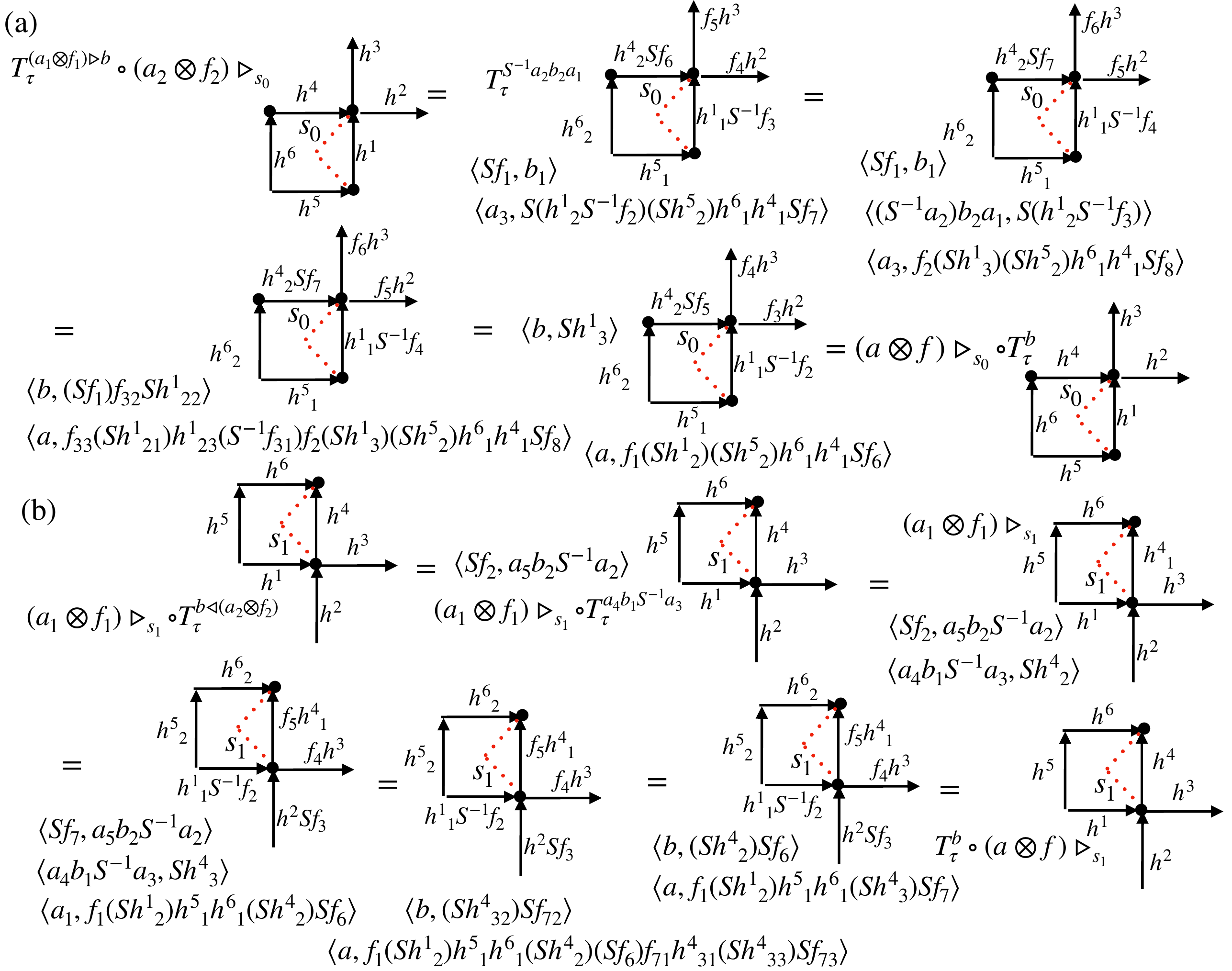}\]
\caption{\label{figHtri} Proof of  (a) left covariance and (b) right covariance of direct triangle operator.}
\end{figure}

\begin{lemma} Let $H$ be a finite-dimensional Hopf algebra, $D(H)$ act on $H$, $H^*$ as in (\ref{DHactH}) and (\ref{DHactH*}) and act on $\End(\CH)$ as in (\ref{DactL})  from the left induced by $\la_{s_0}$  and from the right induced by $\la_{s_1}$. 
\begin{enumerate}
\item For dual triangles,  ${}^{(-)}L_{\tau^*}:H\to \End(\CH)$ is a left module map and  ${}^{(+)}L_{\tau^*}:H\to \End(\CH)$ is right module map. 

\item The direct triangle operator $T_\tau: H^*\to \End(\CH)$  is a left and right module map. 
\end{enumerate}
\end{lemma}
\proof This is shown in Figure~\ref{figHcotri} and Figure~\ref{figHtri}  for sample orientations where $S^\pm$ appears in the dual triangle operation. The other orientations are similar with less work. 
We use the definitions and the actions of $D(H)$ on $H$ and $H^*$.   
 \endproof

Next, we define ribbon operators $F^{h\tens a}_\xi$ associated to a ribbon $\xi$ by convolution-composition of triangle operations, where $h\in H$ and $a\in H^*$.  They are a special case of the `holonomy' maps defined in \cite{Meu} but even so, it is nontrivial to write them out explicitly in our case and in our notations. The first step it to view triangle operators as ribbon operators by
\begin{equation}\label{elerib} \tilde F_\tau^{h\tens a}=\eps(h) T_\tau^a,\quad {}^{(\pm)} \tilde F_{\tau^*}^{h\tens a}=\eps(a){}^{(\pm)}L_{\tau^*}^{S^{-1}h}.\end{equation}
Next, the ribbon operators for two composeable ribbons can be convolution-composed by
\begin{equation}\label{convolH} \tilde F_{\xi'\circ\xi}^\phi= \tilde F_{\xi'}^{\phi_2}\circ \tilde F_{\xi}^{\phi_1}\end{equation}
where now we use the coproduct (\ref{DeltaD*}) on $\phi\in D(H)^*$. This is an associative operation, so starting with a triangle operation viewed as a ribbon operator and extending to a ribbon by composing a series of these,  we correspondingly define the associated ribbon operator by iterating this formula. Because $\eps\tens\id$ and $\id\tens\eps$ are coalgebra maps from $D(H)^*$ to $H^*, H^{cop}$ respectively, the convolution of direct triangle operators viewed as ribbons is the same as convolution of $T$s via the coproduct of $H^*$ and (due to the $S^{-1}$) the convolution of dual triangle operators as ribbons is the same as convolution of $L$s via the coproduct of $H$. It follows that the $D(H)$ site actions in Theorem~\ref{thmDact} can be viewed as ribbon operators, where  for our default conventions $a\la= T^{a_4}\circ\cdots \circ T^{a_1}$ going clockwise around a face and $h\la={}^{(+)} L^{h_4}\circ \cdots {}^{(+)}L^{h_2}\circ{} ^{(-)}L^{h_1}$ going anticlockwise around a vertex. The sign refers to the use of $S^{\pm 1}$ if applicable and as noted, we can also have different patterns of signs, including in the $T$'s, and still get an action of $D(H)$. 

In particular, this means that we no longer have a clear route to topological invariance as we can construct a contractible, closed ribbon equal to $A(v, p)$ (or $B(v, p)$). $A(v, p)\vac \neq \vac$ in general, so ribbon operators are no longer invariant up to isotopy. As a consequence, it is not clear that $\dim(\CH_{vac})$ is a topological invariant in general, albeit it is known to be one in the semisimple case. 

Next, we wish to prove a generalisation of Lemma~\ref{ribcom} from Section~\ref{secG}. However, we could previously rely on topological invariance to justify claims in our proofs by bending ribbons into a convenient shape and eliminating contractible loops.
In the non-semisimple case, we need to specify a new class of ribbon for which our generalisation applies and where these steps are not needed.

\begin{definition}
\rm Recall that a ribbon is a sequence of triangles between sites. Let us represent it as a list of sites in order $[s_0, s_1, \cdots, s_n] := [(v_0, p_0), (v_1, p_1), \cdots, (v_n, p_n) ]$, which must change either the vertex or face between each site. 
A \textit{strongly open} ribbon $\xi$ is an open ribbon which satisfies the following condition: any two sites $s_i := (v_i, p_i)$, $s_j := (v_j, p_j)$ inside $\xi$ may have $v_i = v_j$ only if every site in the sequence of sites $[s_{i+1},\cdots,s_{j-1}]$ between $s_i, s_j$ also has 
$v_{i+1} = \cdots = v_{j-1}$. Similarly, $p_i = p_j$ is only allowed if $p_{i+1} = \cdots = p_{j-1}$.
\end{definition}

The intuition behind this is that the ribbon $\xi$ does not bend `too quickly' or get `too close' to itself; equivalently, it is saying that all subribbons of $\xi$ are either on a single vertex/face or are themselves open.

\begin{figure}%
\centering
\subfloat[]{\input{stronglyopen.tikz}\label{ribbonA}}\hspace{5mm}
\subfloat[]{\input{NOTstronglyopen1.tikz}\label{ribbonB}}\hspace{5mm}
\subfloat[]{\input{NOTstronglyopen2.tikz}\label{ribbonC}}%
\caption{\label{fig:stronglyopen} (A) is a strongly open ribbon. (B) and (C) are open, but not strongly open.}
\end{figure}

\begin{example}
\rm Figure~\ref{ribbonA} shows a strongly open ribbon. While it has rotations at a vertex and a face, it never returns to previously seen vertices or faces. However, Figure~\ref{ribbonB} is an open ribbon which is not strongly open: at the self-crossing of the ribbon, there are sites $s_2$, $s_3$ such that $s_2 = s_3$ but they are not sequentially adjacent in the ribbon.
Similarly, Figure~\ref{ribbonC} is an open ribbon which does not cross itself but gets `too close' -- sites $s_4$, $s_5$ intersect at $p_4$.
\end{example}

\begin{figure}
\[ \includegraphics[scale=0.7]{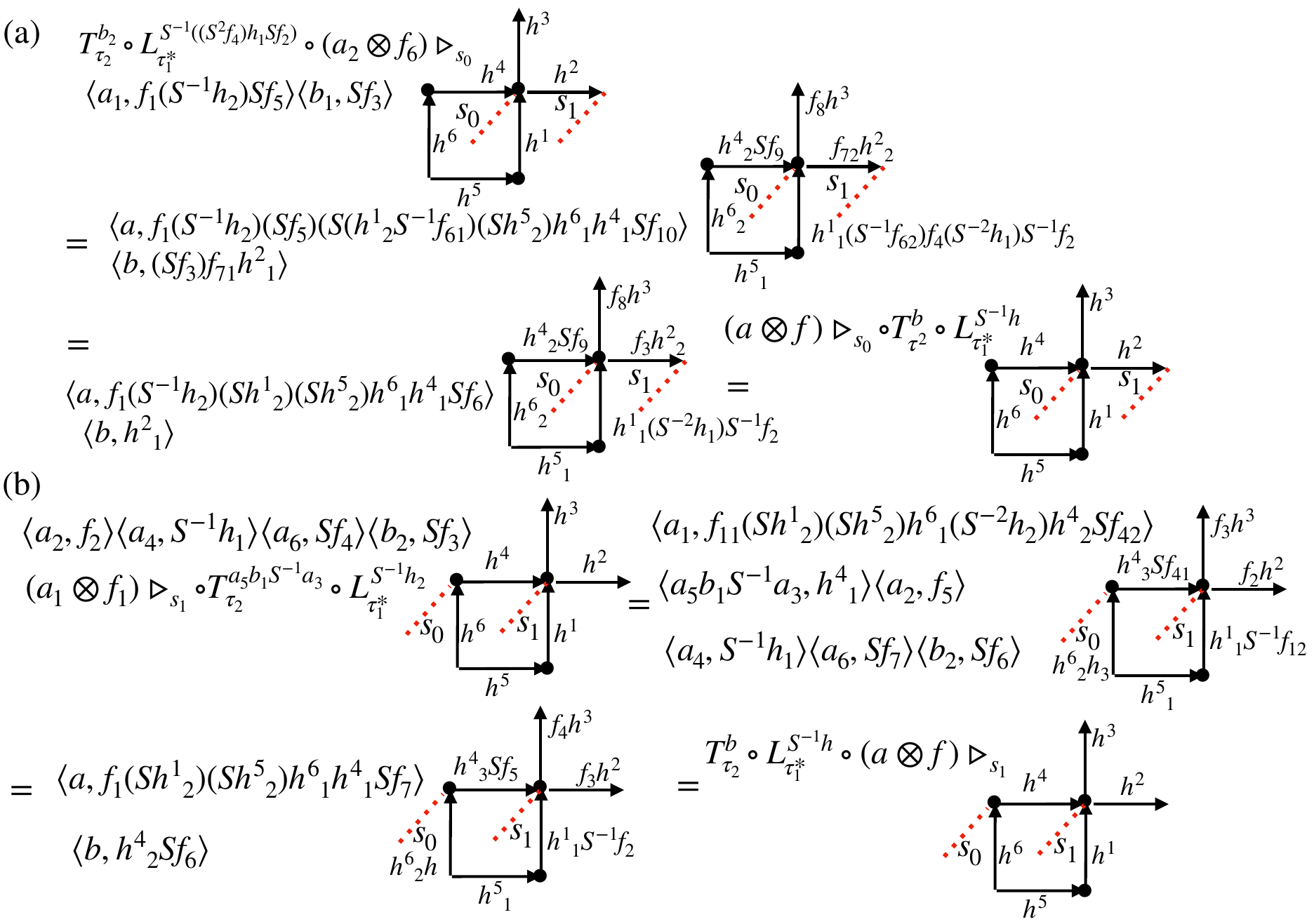}
\]
\caption{\label{figTLpf} Proof of covariance of the 1st elementary open ribbon (a) from the left  using ${}^{(-)}L$  (b) from the right using ${}^{(+)}L$.}
\end{figure}

\begin{figure}
\[ \includegraphics[scale=0.73]{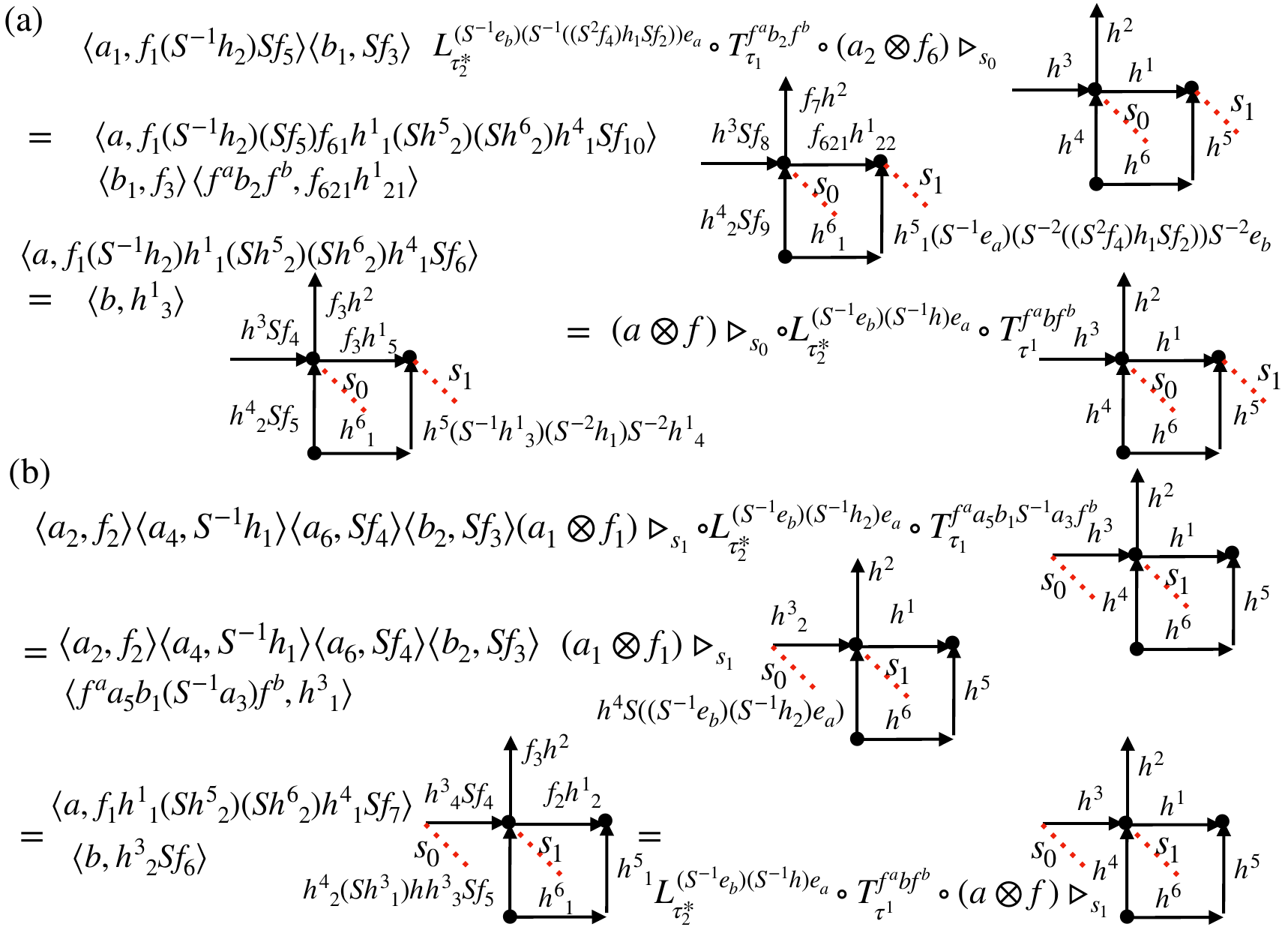}\]
\caption{\label{figLTpf} Proof of covariance of the 2nd elementary open ribbon (a) from the left  using ${}^{(-)}L$  (b) from the right using ${}^{(+)}L$.}
\end{figure}

\begin{proposition}\label{propHrib} Let $\xi$ be a strongly open ribbon from site $s_0$ to site $s_1$. Then 
 $\tilde F_\xi:D(H)^*\to \End(\CH)$ defined iteratively is a left and right module map under $D(H)$, where $D(H)$ acts on itself by (\ref{DHactH}). The actions on $\End(\CH)$ are as before according to $\la_{s_0}$ and $\la_{s_1}$. Moreover, 
if $\Lambda$ and $\Lambda^*$ are cocommutative then $\tilde F_\xi$ commutes with $A(t)$ and $B(t)$ for all sites $t$ disjoint from $s_0,s_1$.
\end{proposition}
\proof The left and right module map properties to be proven are equivalent to
\begin{equation}\label{Dbimod} d\la_{s_0} \circ \tilde F^\phi_\xi=\<S d_1, \phi_1\> \tilde F^{\phi_2}_\xi \circ d_2 \la_{s_0},\quad \tilde F^\phi_\xi \circ d\la_{s_1}= d_1\la_{s_1}\circ \tilde F^{\phi_1}_\xi \<S d_2, \phi_2\>\end{equation}
for $d\in D(H), \phi\in D(H)^*$, which using (\ref{DeltaD*}) and the antipode of $D(H)$ come down to
\[ (a\tens f)\la_{s_0}\circ \tilde F^{h\tens b}=\<a_1, f_1(S^{-1}h_2) Sf_5\>\<b_1,Sf_3\>\tilde F_\xi^{(S^2 f_4) h_1 Sf_2\tens b_2} \circ (a_2\tens f_6)\la_{s_0} \]
\[   \tilde F_\xi^{h\tens b} \circ (a\tens f)\la_{s_1}= \<a_2,f_2\>\<a_4,S^{-1}h_1\>\<a_6,Sf_4\>\<b_2,Sf_3\> (a_1\tens f_1)\la_{s_1}\circ \tilde F_\xi^{h_2\tens a_5 b_1 S^{-1}a_3}  \]

(i) The elementary open ribbon operators 
\[ {}^{(\pm)}\tilde F_{\tau_2\circ\tau_1^*}^{h\tens b}=(\tilde F_{\tau_2}\circ\tilde F_{\tau_1^*})(h\tens b)=T_{\tau_2}^b\circ {}^{(\pm)}L_{\tau_1^*}^{S^{-1}h}\]
\[ {}^{(\pm)}\tilde F_{\tau_2^*\circ\tau_1}^{h\tens b}=(\tilde F_{\tau_2^*}\circ\tilde F_{\tau_1})(h\tens b)=\sum_{a,b}{}^{(\pm)}L_{\tau_2^*}^{(S^{-1}e_b)(S^{-1}h)e_a}\circ T_{\tau_1}^{f^ab f^b}\]
obey the $s_0$ (left) module condition for the $(-)$ case and the $s_1$ (right) module condition for the $(+)$ case. 
These are shown for a sample orientation in Figures~\ref{figTLpf} and~\ref{figLTpf}. In the latter, we use the Hopf algebra duality axioms to identify $\<f^a,\ \>$ $\<f^b,\ \>$ and transfer the other sides to $e_a,e_b$ respectively, then apply cancellations. 

While these elementary ribbons are the smallest open ribbons, not every open ribbon can be generated iteratively starting from one of these elementary ribbons. Any open ribbon can be generated beginning from a rotation around a vertex or face, followed by 
extension to further sites. However, we can just replace the first ${}^{\pm} L_{\tau^*}^{S^{-1}h}$ in the equation above with the appropriate convolution of $L$ operators, and the same for the $T$ case. The left and right module properties will be preserved for the $(+)$ and $(-)$ cases respectively.

(ii) We proceed by induction. Let $\xi : s_0 \to s_2$ be a strongly open ribbon. First observe that if $\xi':s_0\to s_1$ and $\xi'':s_1\to s_2$ and $\tilde F_{\xi'}$ is a left module map with respect to its start then 
\begin{align*} (\tilde F_{\xi''}\circ\tilde F_{\xi'})^{d\la\phi}&=\<Sd,\phi_1\>\tilde F_{\xi''}^{\phi_{22}}\circ \tilde F_{\xi'}^{\phi_{21}}=\<Sd,\phi_{11}\>\tilde F_{\xi''}^{\phi_{2}}\circ \tilde F_{\xi'}^{\phi_{12}}=\tilde F_{\xi''}^{\phi_2}\circ d_1\la_{s_0}\circ\tilde F_{\xi'}^{\phi_1}\circ Sd_2\la_{s_0}.\end{align*}
Now, $\xi''$ is disjoint from $s_0$ as $\xi$ is strongly open. Hence, $\tilde F_{\xi''}$ commutes with $\la_{s_0}$, and so $\tilde F_\xi$ is a left module map with respect to its start. 

Similarly, if $\tilde F_{\xi''}$ is a right module map with respect to its end then 
\begin{align*} (\tilde F_{\xi''}\circ\tilde F_{\xi'})^{\phi\ra d}&=\<Sd,\phi_2\>\tilde F_{\xi''}^{\phi_{12}}\circ \tilde F_{\xi'}^{\phi_{11}}=\<Sd,\phi_{22}\>\tilde F_{\xi''}^{\phi_{21}}\circ \tilde F_{\xi'}^{\phi_{1}}=Sd_1\la_{s_2}\tilde F_{\xi''}^{\phi_2}\circ d_2\la_{s_2}\circ\tilde F_{\xi'}^{\phi_1}.\end{align*}
As before, $\xi'$ is disjoint from $s_2$, as $\xi$ is strongly open. Hence, $\tilde F_{\xi'}$ commutes with $\la_{s_2}$ and $\tilde F_\xi$ is a right module map with respect to its start. 
 
Now suppose that the left and right module property holds for strongly open ribbons up to some number of triangles in length. Let $\xi$ be a strongly open ribbon that is not an elementary one from part (i). Then (a) we write  $\xi''\circ\xi'$ where $\xi''=\tau$ or $\tau^*$ and  $\xi'$ is also a strongly open ribbon. In that case our first observation applies and $\tilde F_\xi$ is a left module map. And (b) we can write it as $\xi''\circ\xi'$ where $\xi'=\tau$ or $\tau^*$ and now $\xi''$ is a strongly open ribbon. Then $\tilde F_\xi$ is also a right module map, hence a left and right module map. 

(iii) We also note that if $\tilde F_{\xi''}$ is a left module map and $\tilde F_{\xi'}$ a right one then 
\begin{align*}
d\la_{s_1}\circ (\tilde F_{\xi''}\circ \tilde F_{\xi'})^{\phi}&=d\la_{s_1}\circ \tilde F_{\xi''}^{\phi_{2}}\circ \tilde F_{\xi'}^{\phi_{1}}= \<Sd_1,\phi_{21}\> \tilde F_{\xi''}^{\phi_{22}}\circ d_2\la_{s_1}\circ\tilde F^{\phi_{1}}\\ &=\tilde F_{\xi''}^{\phi_{2}}\circ \<Sd_1,\phi_{12}\>d_2\la_{s_1}\circ\tilde F^{\phi_{11}}=\tilde F_{\xi''}^{\phi_{2}}\circ \<Sd_2,\phi_{12}\>d_1\la_{s_1}\circ\tilde F^{\phi_{11}}\\
&= \tilde F_{\xi''}^{\phi_2}\circ\tilde F_{\xi'}^{\phi_1}\circ d\la_{s_1}=(\tilde F_{\xi''}\circ \tilde F_{\xi'})^{\phi}\circ d\la_{s_1}
\end{align*}
 provided for the 4th equality we have $d$ cocommutative in the sense $d_1\tens d_2=d_2\tens d_1$. As $\xi''$ and $\xi'$ are both strongly open, this implies that the action by such elements $d$ at interior sites commute with $\tilde F_\xi$.
\endproof 

Observe that this argument holds because the ribbon $\xi$ is strongly open: all subribbons of $\xi$ are either themselves (strongly) open or are rotations around a vertex/face. A ribbon which is open but not strongly open may have subribbons which are not open but have interior sites disjoint from the endpoints, and therefore the above inductive argument breaks down.  In the semisimple case, the condition of \textit{strongly} open in Proposition~\ref{propHrib} can be relaxed to just open, as we have topological invariance and so any subribbons which are not open may be smoothly deformed to their shortest path, which will be a rotation with no interior sites disjoint from the endpoints. We also know in the semisimple case that $\Lambda,\Lambda^*$ are cocommutative, so the last part of the proposition applies. 

The left and right module property and convolution of strongly open ribbons can also be viewed as follows in terms of $D=D(H)$. Recall that two left actions of $D$ on $\CH$, as was the case above using the site actions at the ends of open ribbons, induce left and right $D$-module structures on $A=\End(\CH)$ as in (\ref{DactL}) which are compatible with the product (i.e. $A$ is a left and right $D$-module algebra). They commute (i.e. make $A$ into bimodule) if the end sites are far enough apart. Also recall that $D$ acts on $D^*$ by (\ref{DactD*}) and on itself by the product, so that $D^*,D$ are $D$-bimodules. We will use a compact notation where $\check F=\check F^1\tens\check F^2$ (summation understood) denotes an element of  a tensor product over the field and $\check F_{21}=\check F^2\tens\check F^1$. 

\begin{lemma}\label{lemFbimod} Let $D$ be a finite-dimensional Hopf algebra and $A$ a left and right $D$-module algebra with actions denoted by dot.  We let $\{e_\alpha\}$ be a basis of $D$ and $\{f^\alpha\}$ a dual basis. The following are equivalent.

\begin{enumerate} \item $\tilde F:D^*\to A$ is left and right $D$-module map.
\item  $\check{F}:= \check F^1\tens\check F^2=\sum_\alpha S^{-1}e_\alpha \tens \tilde F^{f^\alpha}\in D\tens A$ obeys   $d\check F=\check F.d$ and $d.\check F_{21}=\check F_{21}d$ for all $d\in D$, using the  product or  action of $D$ on the adjacent factor.  
\item $(S\tens\id)\check F$ is invariant under the left and right tensor product $D$-actions.
\end{enumerate}
If $A$ is an algebra and $\tilde F'',\tilde F':D^*\to A$ are linear maps, their convolution product $(\tilde F''\circ \tilde F')^\phi=\tilde F''{}^{\phi_2}\circ\tilde F'{}^{\phi_1}$ is equivalent to the product of the corresponding $\check F'',\check F'$ in the tensor product algebra.   If $A$ is a bimodule and $\tilde F$ a bimodule map then $f=\check F^1.\check F^2, g=\check F^2.\check F^1\in A$  are in the bimodule centre. 
\end{lemma}
\proof This is elementary but we give some details for completeness. First, as linear maps it is obvious that $\tilde F:D^*\to A$ is equivalent to an element $e_\alpha\tens \tilde F^{\alpha} \in D\tens A$ (summation over repeated labels understood). It is also clear that if $\tilde F'',\tilde F'$ are two such linear maps then $e_\alpha\tens (\tilde F''\circ \tilde F')^{f^\alpha}=e_\alpha\tens \tilde F''{}^{f^\alpha{}_2} F'{}^{f^\alpha{}_1}=e_\beta e_\alpha \tens \tilde F^{f^\alpha} F^{f^\beta}$ which is the product in $D^{op}\tens A$. The $S^{-1}$ means that the corresponding $\check F'',\check F'$ multiply in $D\tens A$. 

Moreover, suppose $\tilde F$ is a left and write $D$-module map. We denote the left and right tensor product actions of $D$ on $D\tens A$ by $\la$, $\ra$. Then
\begin{align*} d\la(e_\alpha\tens \tilde F^{f^\alpha})&=d_1 e_a\tens d_2.\tilde F^{f^a}=d_1 e_a\tens \tilde F^{d_2\la f^a}=d_1 e_\alpha\tens \<Sd_1,f^\alpha{}_1\>\tilde F^{f^\alpha{}_2}\\&=d_1 e_\alpha e_\beta \tens \<Sd_1,f^\alpha\>\tilde F^{f^\beta}=d_1 (S d_2) e_\beta\tens \tilde F^{e^\beta}=\eps(d)(e_\alpha\tens F^{f^\alpha})\end{align*}
and similarly from the other side for the right action $\ra$. This argument can be reversed to prove equivalence of (1) and (3). Similarly, 
\begin{align*}S^{-1}e_\alpha\tens d. \tilde F^{\alpha}&=S^{-1}e_\alpha\tens\<Sd,f^\alpha{}_1\>\tilde F^{f^\alpha{}_2}=(S^{-1} e_\beta)(S^{-1}e_\alpha)\tens \<Sd,f^\alpha\>\tilde F^{f^\beta}=(S^{-1}e_\beta)d\tens  \tilde F^{\beta}\\
S^{-1}e_\alpha\tens \tilde F^{\alpha}.d&=S^{-1}e_\alpha\tens\<Sd,f^\alpha{}_2\>\tilde F^{f^\alpha{}_1}=(S^{-1} e_\beta)(S^{-1}e_\alpha)\tens \<Sd,f^\beta\>\tilde F^{f^\alpha}=d(S^{-1}e_\alpha)\tens  \tilde F^{\alpha}\end{align*}
which can be reversed for the equivalence of (1) and (2). Then in the bimodule case,  $d. f=d.(\check F^1.\check F^2)=\check F^1.(\check F^2.d)=f.d$ and $g.d=(\check F^2.\check F^1).d=(d.\check F^2).\check F^1=d.g$ for all $d\in D$.   \endproof

This is relevant to us in the case where $\tilde F=\tilde F_\xi$ for an open ribbon from $s_0$ to $s_1$ and  $A=\End(\CH)$ with left and right module structures induced by the site actions on $\CH$ at $s_0,s_1$ of $D=D(H)$. We write ${}_{s_0} A_{s_1}$ for the algebra $A$ with this left and right $D$-module structure. For example, the associated $f=f_\xi$ and $g=g_\xi$ are in here. Moreover, if  $\xi=\xi''\circ\xi'$ then $\check F'\in D\tens {}_{s_0}A_{s_1}$ and $\check F''\in D\tens {}_{s_1}A_{s_2}$ while $\check F''\check F'\in D\tens{}_{s_0}A_{s_2}$. This gives a functor from the `ribbon path groupoid' to the category of $D$-modules  $\CH$ with morphisms given by elements of $D\tens A$. The composition of morphisms is given by the tensor product algebra $D\tens A$ plus an assignment of the left and right $D$-module structures on $A$ for the result. This is such that the product of ${}_{s_1} A_{s_2}$ and ${}_{s_0} A_{s_1}$ is deemed to lie in ${}_{s_0} A_{s_2}$. The morphisms that arise from open ribbons also obey the centrality properties (2). This is in the spirit of the `holonomy' point of view in \cite{Meu}.

\subsection{Quasiparticle spaces for $D(H)$ ribbons} Finally, we fix a vaccum state $\vac$ and consider quasiparticle spaces  
\[ \CL_\xi(s_0,s_1)=\{\tilde F^{\phi}_\xi\vac\ |\ \phi\in D(H)^*\}\subset \CH\]
much as before, where $\xi:s_0\to s_1$ is a fixed strongly open ribbon. We make  $\CH$ a left and right $D(H)$-module where $d$ acts from the left by $d\la_{s_0}\psi$ and from the right by $\psi\ra_{s_1}d:=Sd\la_{s_1}\psi$. These commute on $\CH$ so that we have a bimodule when $s_0,s_1$ are sufficiently far apart, meaning that $p_0$ and $p_1$ do not share an edge, and neither do $v_0$ and $v_1$. However the next proposition shows that they always commute when we restrict to $\psi\in \CL_\xi(s_0,s_1)$. We moreover dualise the left and right actions to respectively right and left actions on $\CL_\xi(s_0,s_1)^*$ which then also form a bimodule. Recall that $D(H)^*$ is always a $D(H)$-bimodule by (\ref{DactD*}) and $D(H)$ a  $D(H)$-bimodule by left and right multiplication. 

\begin{proposition}\label{propHLs0s1} Let $\xi:s_0\to s_1$ be a strongly open ribbon and $\CL_\xi(s_0,s_1)$ as above. This is a bimodule and
\begin{enumerate} \item $D(H)^*\twoheadrightarrow \CL_\xi(s_0,s_1)$ sending $\phi \mapsto \tilde F_\xi^\phi\vac$ is a bimodule map. \item $\CL_\xi(s_0,s_1)^*\hookrightarrow D(H)$ sending $\<\Phi|\mapsto \check F^1 \<\Phi|\check F^2\vac$ is a bimodule map. 
\end{enumerate}
\end{proposition}
\proof If $\Lambda\in H$ and $\Lambda^*\in H^*$ are integral elements then $\Lambda_D:=\Lambda^*\tens\Lambda\in D(H)$ is an integral element in $D(H)$ and if $\vac\in \CH_{vac}$ then $\Lambda_D\la\vac=\vac$ at any site. It follows that if $d\in D=D(H)$ then $d\la\vac=d\Lambda_D\la\vac=\eps(d)\Lambda_D\la\vac=\eps(d)\vac$ as we have seen before. Then
\begin{align*} d\la_{s_0} \circ \tilde F^\phi_\xi\vac&=\<S d_1, \phi_1\> \tilde F^{\phi_2}_\xi \circ d_2 \la_{s_0}\vac=\<Sd,\phi_1\>\tilde F^{\phi_2}\vac=\tilde F^{d\la\phi}\vac\\
 Sd\la_{s_1}\circ \tilde F^\phi\vac&=Sd_1\la_{s_1}\tilde F^\phi_\xi \circ d_2\la_{s_1}\vac= (Sd_1) d_2\la_{s_1}\circ \tilde F^{\phi_1}_\xi\vac \<S d_3, \phi_2\>=\tilde F^{\phi\ra d}_\xi\vac\end{align*}
which implies that $\CL_\xi(s_0,s_1)$ is a bimodule and proves (1).  Moreover, we can unpack the centrality in Lemma~\ref{lemFbimod} explicitly and apply it as
\[d\check F_\xi\vac=\check F^1\tens (\check F^2)\ra_{s_1} d\vac=\check F^1\tens Sd\o\la_{s_1}\circ \check F^2\circ d_2\la_{s_1}\vac=\check F^1\tens Sd \la_{s_1}\circ \check F^2\vac.\]
\[\check F^1d\tens \check F^2\vac=\check F^1\tens d\la(\check F^2)\vac=\check F^1\tens d_1\la_{s_0}\circ \check F^2\circ Sd_2\la_{s_0}\vac=\check F^1\tens d \la_{s_0}\circ \check F^2\vac\]
so that
\[ d\check F^1\<\Phi|\check F^2\vac=\check F^1\<\Phi|Sd\la_{s_1}\circ\check F^2\vac=\check F^1\<d\la \Phi | \check F^2\vac \]
\[ \check F^1\<\Phi|\check F^2\vac d=\check F^1\<\Phi|d\la_{s_0}\circ\check F^2\vac=\check F^1\<\Phi\ra d| \check F^2\vac  \]
which is (2). Here the left action on $\CL_\xi(s_0,s_1)$ dualises to the right action  $(\Phi\ra d)(\psi)=\Phi(d\la_{s_0} \psi)$ and the right action on $\CL_\xi(s_0,s_1)$ dualises to the left action $(d\la\Phi)(\psi)=\Phi(\psi\ra_{s_1}d)=\Phi(Sd\la_{s_1}\psi)$.  \endproof

The maps in the proposition are expected to be  isomorphisms in line with  Proposition~\ref{Ls0s1} for the $D(G)$ case, but this requires more proof. For example, this follows if  $\tilde F_\xi^\phi\vac=0$ implies that $\phi=0$, which is expected to follow from unitarity properties with respect to a $*$-structure. Likewise, it is expected that $\CL_\xi(s_0,s_1)$ is independent of $\xi$ at least in the $H$ semisimple case and characterised in terms of $A(t),B(t)$ in the manner that was done in Proposition~\ref{Ls0s1}.

\section{Concluding remarks}\label{secrem}

We have given a self-contained treatment of the Kitaev model for a finite group $G$, focussed on the quasiparticle content and ribbon equivariance properties expressed in terms of the quantum double $D(G)$. This was largely avoided in works such as \cite{Kit,Bom}, while \cite{BSW} starts to take a quantum double view, and we built on this. As well as a systematic treatment of the core of the theory, we have then demonstrated how quasiparticles could be created and manipulated in practice, with details in the case of $D(S_3)$ of the construction of logical operations and gates. We also showed the existence of a `Bell state' that exists in the ribbon space $\CL(s_0,s_1)$ created by ribbon operations for an open ribbon between $s_0,s_1$ and which can be used to teleport quasiparticle information between the endpoints. We also illustrated these ideas for the toric case of $D(\Z_n)$. 

Beyond this practical side, we also looked closely as the obstruction to generalising such models to the `quantum case' where the group algebra $\C G$ is replaced by a finite-dimensional Hopf algebra $H$. That this works when $S^2=\id$ (e.g. the Hopf algebra is semisimple and we work over a field of characteristic zero such as $\C$) is well known as are its link to topological invariants \cite{Kir,Meu} such as the Turaev-Viro invariant and the Kuperberg invariant\cite{Kup}. As far as we can tell, ribbon operators at this level have not been studied very explicitly, athough contained in principle in \cite{Meu} as part of a theory of `holonomy', following the work of \cite{BMCA} for the site operations. As well as our own work we have noted \cite{Chen}. We provided a self-contained treatment of the core properties in the $S^2=\id$ case but we could also see by giving direct proofs what is involved in the general case. We found that site operation work perfectly well but we must use $S^{-1}$ in certain key places. To be concrete, we put this on the vertex side but this complication can be put in different places leading in fact to a set of possible site operations all forming representations of $D(H)$. Dual triangle and ribbon equivariance properties then become more complicated with ${}^{-}L$ needed in some places for good behaviour with respect to the initial site action $s_0$. We also noted that the Peter-Weyl decomposition whereby $D(H)$ is a direct sum of endomorphism spaces for the irreps holds when $S^2=\id$. More generally, one will have some blocks associated to irreps but these will not be the whole story. Hence our ideas on ribbon teleportation will be more complicated in general. Likewise, the actions of the integrals $A(v,p)$ and $B(v,p)$ are no longer projectors in the nonsemisimple case, but square to zero, which considerably changes how the physics should be approached and requires further work. It will also be necessary to look at $*$-structures needed to formulate unitarity at this level, possibly using the notion of  flip Hopf $*$-algebras as recently initiated for ZX calculus in \cite{Ma:fro}.

Nevertheless, there are good reasons to persist with the general case, namely in order to link up with 2+1 quantum gravity and the Turaev-Viro invariant of 3-manifolds in a graph version. In quantum gravity, the relevant 3-manifold would be $\R\times\Sigma $ where $\R$ is time and $\Sigma$ is a surface with marked points, but we would make a discrete approximation of the latter by a (ciliated, ribbon) graph, or in the simplest case a square lattice as here. Since the Turaev-Viro invariant is based on $D(u_q(sl_2))$, the goal would be to have a more Kitaev model point of view in contrast to current Hamiltonian constructions\cite{AA}. Going the other way, it would be interesting to try to regard the $D(S_3)$ model as leading to a baby version of quantum gravity in the context of discrete noncommutative geometry\cite{Ma:dg}. 

Another longer term motivation for the current work is the need for some kind of compiler or `functor' from surface code models such as the Kitaev one to ZX-calculus\cite{CD} as more widely used in quantum computing. The Z,X here are Fourier dual and it would be useful to understand even in the toric case of $D(\Z_n)\cong \C \Z_n^2$ how the surface code theory relates to ZX calculus based on $\C \Z_n$ as a quasispecial Frobenius algebra. There are current ideas about this but they appear to require a notion of boundary defects. This and the notion of condensates will both need to be studied more systematically by the methods in the present work, building on current literature such as \cite{BSW}. We also note that in topology, the Jones invariant and its underlying Chern-Simons theory are based on the quantum group $u_q(sl_2)$ and such invariants are related via surgery on the knot to the Turaev-Viro invariant based on $D(u_q(sl_2))$, suggesting the possibility of a general link between  $D(H)$ surface code theory and ZX calculus on $H$. The latter on general Hopf algebras and braided-Hopf algebras was recently studied in \cite{CoD,Ma:fro}. These are some directions for further work.

\appendix

\section{The vacuum space of $D(G)$ models}\label{app:vacuum}

This appendix finds expressions for an orthogonal basis of $\CH_{vac}$ in the $D(G)$ models, following \cite{Cui}. This is included for completeness in order to have a self-contained account of the theory. Let $g := \bigotimes_{l \in E} g^l$ be the state in $\CH$ with a group element $g^l$ viewed in $\C G$ at each edge. Let $\gamma$ be an oriented path in the lattice.
Now, define
\[
\gamma (g) := \prod_{l \in \gamma} (g^l)^{\eps}
\]
where $\eps= 1$ if the path orientation agrees with the lattice orientation, and $-1$ otherwise. For example, given a segment of the lattice segment  with an oriented path $\gamma$,
\[
\input{border_example2.tikz}
\]

we would have $\gamma (g) := (g^{12})^{-1} g^9 (g^6)^{-1} g^3 g^1$. (We choose arrow composition to be this way round, rather than $\gamma (g) := g^1 g^3 (g^6)^{-1} g^9 (g^{12})^{-1}$, as it is convenient for the proof of Theorem~\ref{thmcui} below.)

Observe that for $B(p)$ at a given face $p$, the condition $B(p) g = g$ is equivalent to $\del p (g) = e$, where $\del p$ is the boundary of $p$ interpreted as a clockwise-oriented path, and $e$ is the identity element of $G$.
Note that the choice of basepoint of this path is immaterial, as the product is still $e$ under cyclic rotations. Now, consider two adjacent boundaries $\del p_1$ (red) and $\del p_2$ (cyan) such that $\del p_1 (g) =\del p_2 (g) = e$. Then 
$\del p_{1,2} (g)=\del p_1 (g)  \del p_2 (g) = e$ for the boundary of the combined face,
\[
\input{border_composition.tikz}
\]
It follows that the subspace $\{\psi\ |\ B(p)\psi = \psi\ {\rm for\ all\ }p \}$  is spanned by the following set:
\[
S = \{ g\ |\ \del p(g) = e\  {\rm for\ all\ } p\} = \{ g\ |\ \gamma (g) = e\ {\rm for\ all\ contractible\ closed}\ \gamma\}.
\]
Clearly, $S$ is invariant under change of orientation of $\gamma$. Next, we define an equivalence relation on S. We say that $g \sim g'$ if $g' = \bigotimes_{v \in V} h_v \la_v g$ for some collecton $\{h_v\in G\}$.
In other words, there is some sequence of vertex operators that takes $g$ to $g'$. The set of equivalence classes is $[S]$, and a given class is called $[g]$. Define
\[
\kappa_{[g]} := \sum_{g' \in [g]} g' \in \CH
\]
For any two tensor products states $g = \bigotimes_{l \in E} g^l$ and $g' = \bigotimes_{l \in E} g'^l $ in $\CH$, define the inner product $(g, g') =\prod_{l\in E} \delta_{g^l, g'{}^l}$.
\begin{lemma}
$\{\kappa_{[g]}\ |\ [g] \in [S]\}$ forms an orthogonal basis of $\CH_{vac}$.
\label{lem:vacuum_basis}
\end{lemma}
\proof
Clearly, $h \la \kappa_{[g]} = \kappa_{[g]}$. Therefore, $\kappa_{[g]} = A(v) \kappa_{[g]}, \forall v \in V$ and we also know that $\kappa_{[g]} = B(p) \kappa_{[g]}, \forall p \in P$, so  $\kappa_{[g]} \in \CH_{vac}$. In addition, for any two $\kappa_{[g]}$ and $\kappa_{[g']}$, either $\kappa_{[g]} = \kappa_{[g']}$ or they have no overlapping terms, by definition of the equivalence relation.
Therefore, $(\kappa_{[g]}, \kappa_{[g']}) = |[g]|\delta_{[g], [g']}$, where $|[g]|$ is the cardinality of $[g]$. Thus all $\kappa_{[g]}$ are orthogonal.

Next we prove that $\kappa_{[g]}$ span $\CH_{vac}$. For any state $\psi \in \CH_{vac}$, write $\psi = \sum_{g \in S} \alpha_g g$, where $g = \bigotimes_{l \in E} g^l $. Now, choose a vertex $v$. We know that $h \la_v \psi = \psi$, $\forall h \in G$.
Given some $g$, consider the set of states $\{g'\}$ such that $g'$ agrees with $g$ everywhere except at $v$, where $g' = h^{'} \la_v g$ for some $h^{'} \in G$. For any such $g' $, $h \la_v g' \in \{g'\}$, so by definition $h \la_v$ permutes through the set.
Therefore, as all $g$ are orthogonal and $h\la_v \sum_{g \in S} \alpha_g g = \sum_{g \in S} \alpha_g g$, each element in $\{g'\}$ must appear with the same weight. Repeating for all vertices, it is clear that $\psi = \sum_{[g]\in [S]} \beta_{[g]} \kappa_{[g]}$, for some coefficients $\{\beta_{[g]}\}$, and hence that $\{\kappa_{[g]}\ |\ [g] \in [S]\}$ spans $\CH_{vac}$.
\endproof

\begin{theorem}\cite{Cui}\label{thmcui}
Let $\Sigma$ be a closed, orientable surface. Then
\[
\dim(\CH_{vac}) = |\mathrm{Hom}(\pi_1(\Sigma), G)/G|.
\]
where the $G$-action on any $\phi \in \mathrm{Hom}(\pi_1(\Sigma), G)$ is $\phi \mapsto \{h \phi h^{-1}\ |\ h \in G \}$.
\end{theorem}
\proof We define an equivalence relation between closed, but not necessarily contractible, paths acting on the ground state, by $\gamma \sim \gamma^{'}$ if $\gamma =  \gamma^{'} \prod_{p \in I} \del p$, for some set of faces $I \subseteq P$.
Denoting the set of all closed paths $K$, the equivalence relation defines a homotopy class of $\Sigma$. By taking the obvious group composition we identify $[K]$, the set of equivalence classes, with $\pi_1(\Sigma)$. We now define a map
\[
\Theta : S \rightarrow \mathrm{Hom}(\pi_1(\Sigma), G),\quad \Theta(g)([\gamma]) := \gamma(g), 
\]
where $\gamma$ is any closed path in $[\gamma]$. The choice of $\gamma$ is immaterial, as $\del p (g) = e, \forall p \in P$. For any $g \in S$, let $[\gamma]_0$ be the class of contractible, closed paths, i.e. the identity of $\pi_1(\Sigma)$. By definition, $\gamma_0(g) = e \in G$, for any $\gamma_0 \in [\gamma]_0$.
Now, again for any $g \in S$, let $[\gamma]_a$ and $[\gamma]_b$ be two classes of closed paths. Let $\gamma_a(g) = g_a$ and $\gamma_b(g) = g_b$. Observe that $(\gamma_a \circ \gamma_b)(g) = g_a g_b$.
Therefore the image of $\Theta$ is indeed in the set $\mathrm{Hom}(\pi_1(\Sigma), G)$ of group homomorphisms.

Next, we show that $\Theta$ is surjective. For any group homomorphism $\phi : \pi_1(\Sigma) \rightarrow G$, consider a maximum spanning tree $T$ on $\Sigma$, with root $r$. By definition, $T$ has $m := |V|-1$ edges. For any edge $\epsilon$ outwith the tree, let $u_{\epsilon}$ and $v_{\epsilon}$ be the end vertices of $\epsilon$.
$u_{\epsilon}$ and $v_{\epsilon}$ are in $T$. There is now a unique path $\gamma_u$ through $T$ from $r$ to $u_{\epsilon}$ and $\gamma_v$ from $r$ to $v_{\epsilon}$. Therefore, we may define a closed path:
\[
\gamma_{\epsilon} = \gamma_u \circ \epsilon \circ \gamma^{-1}_v
\]
where $\gamma^{-1}_v$ is the reverse path of $\gamma_v$. By construction of $T$, the group element $\epsilon (g)$ associated to $\epsilon$ is uniquely fixed by the group elements $\gamma_u(g)$ and $\gamma_v(g)$.
Conversely, given edge $\epsilon$ each group element $\epsilon (g)$ may be acquired by $|G|^m$ choices of group elements for edges in $T$. Applying the same logic for any edge outwith $T$, $\Theta$ is therefore a $|G|^m$-to-$1$ map.
This is invariant under choice of root $r$ and maximum spanning tree $T$.

The proof is now completed by setting up a bijection between $[S]$ and orbits of $\mathrm{Hom}(\pi_1(\Sigma), G)$ under the $G$-action. By definition, any closed path through a given vertex $v \neq r$ will have exactly one incoming arrow and one outgoing arrow. The product along this path is invariant under $h \la_v$, so $\Theta(g) = \Theta(h\la_v g)$.
However, $\Theta(h \la_r g) = h \Theta(g) h^{-1}$, as $r$ is the endpoint of the path. Therefore, the preimage of any $\phi \in \mathrm{Hom}(\pi_1(\Sigma), G)$ is exactly the set of elements of $S$ which agree on edges adjacent to $r$, but are just related by some family $h_v \la_v$ at other vertices. Additionally, if $\Theta(g) = \phi$, then $\Theta([g]) = G \la \phi = \{h \phi h^{-1}\ |\ h \in G\}$. Therefore, $[S] \cong \mathrm{Hom}(\pi_1(\Sigma), G)/G$, 
where the $G$-action is conjugacy as above. Using Lemma~\ref{lem:vacuum_basis}, $\dim(\CH_{vac}) = | \mathrm{Hom}(\pi_1(\Sigma), G)/G|$.
\endproof

\section{Proof of part (2) of Proposition~\ref{Ls0s1}}\label{app:span}

That $|\psi^{h,g}\>\in \CL(s_0,s_1)$ follows from the commutation
relations with operators at sites $t \neq s_0, s_1$ in Lemma~\ref{ribcom}. In this Appendix, we 
show these states span $\CL(s_0,s_1)$. Note that there is a stronger claim in  \cite[Prop~7]{Bom} for their  `ribbon algebra' $\mathcal{F}_{\rho}$ but we have not been able to reproduce the  proof there at a number of points, specifically  (B58), (B59), (B62) and (B63) appear to assume that certain projectors are right-cancellable, which in general  is not possible. 

Our proof by induction will involve 3 series of cases:  (i) the base cases, when $s_0$ and $s_1$ are separated only be a single edge (direct or dual);  (ii) the case when $s_0$, $s_1$ are distance 2 away, i.e. the smallest ribbon connecting them has exactly 2 triangle operatorions; (iii) distance 3 or greater.

(i) The two base cases occur when $s_0$ and $s_1$ are adjacent, so the minimal ribbon $\xi$ required is of length 1, either a direct or dual triangle. 
We start with a direct triangle, for example
\[\input{direct_basecase.tikz}\]
where $s_0 = (v_0, p_0)$ and $s_1 = (v_1, p_0)$. Consider a state $|\Psi\> \in \CL(s_0,s_1)$ and all operators $O$ on $\CL(s_0,s_1)$ such that  $O\vac = |\Psi\>$. Since the conditions on $|\Psi\>$ away from the end sites are the same as for a vacuum, we can assume that $O$ acts trivially on $\vac$ around all vertices $v \notin \{v_0, v_1\}$ and $p \neq p_0$ and hence that $O$ can be chosen to act only on the edge
 shared by $(v_0, p_0)$ and $(v_1, p_0)$, which has state $g^1$ as in the diagram. A fuller explanation requires arguments similar to those for the vacuum in Appendix~\ref{app:vacuum}.

Note next that $\End(\C G)\isom\C(G)\lcross\C G\isom \C G\rcross\C(G)$, which is to say any operator
acting on $g^1\in \C G$ is a sum of terms factorising as $\C G$ acting by multiplication (one can fix the side to be from the left or the right) and $\C(G)$ acting by evaluation against the coproduct, the second of these being the action of a direct triangle operator. 
But any contributions from a nontrivial part of $\C G$ in $O$ will cease to satisfy $B(p_2)O\vac = O\vac$ from the conditions for $\CL(s_0,s_1)$, so $O\in\C(G)$ acts like a direct triangle operator. Hence $\{T^g_{\xi}\vac\ |\ g \in G \}$ span $\CL(s_0,s_1)$, and $T^g_{\xi}=F^{e, g}_{\xi} $, (or any $h$ in place of $e$) so $\CL(s_0,s_1)$ is spanned by $\{\psi^{e, g}\ | g \in G \}$, and therefore also by $\{\psi^{h, g}\ | h, g \in G \}$.

For the dual-triangle, we similarly consider
\[\input{dual_basecase.tikz}\]
Now, $|\Psi\>$ can be characterised by an operator $O$ which acts only on $g^2$ and similarly factorises as $\C G$ and $\C(G)$ acting as before, where the first is the action of a dual triangle operator.
But any contributions from a nontrivial part of $\C(G)$ in $O$ will cease to satisfy $A(v_2)O\vac = O\vac$, so $O\in \C G$ acts like a dual triangle operator. Hence $\{L^gh{\xi}\vac\ |\ h \in G \}$ span $\CL(s_0,s_1)$, and $L^h_{\xi} = F^{h, e}_{\xi}$, so $\CL(s_0,s_1)$ is spanned by $\{\psi^{h, e}\ | h \in G \}$, and therefore also by $\{\psi^{h, g}\ | h, g \in G \}$.
This concludes the base cases.

(ii) There are four distance 2 cases, which can all be calculated. If $s_0, s_1$ occupy positions as in
\[\input{B1_ii_a.tikz}\]
where the smallest ribbon has 1 direct and 1 dual triangle such that $\xi = \tau \circ \tau^*$, then by the same arguments as above $|\Psi\> \in \CL(s_0, s_1)$ can be characterised by an operator $O$ such that $|\Psi\> = O\vac$, where
$O$ acts only on the edges $g^1, g^2$, and we can see by considering $A(v)$ and $B(p)$ acting at $v$ and $p$ adjacent to $g^1, g^2$ that $O$ must be a sum of terms $T^g_{\tau}\circ L^h_{\tau^*}$. We can then set $F^{h, g}_{\tau_2 \circ \tau_1} = T^g_{\tau_2}\circ L^h_{\tau_1}$. 
That $\{F^{h, g}_{\tau \circ \tau^*}\vac\ |\ h, g \in G\}$ spans $\CL(s_0, s1)$ is then immediate. The same argument applies if
$\xi = \tau^* \circ \tau$ instead, but the other way round.

If the smallest ribbon is instead 2 direct triangles, for example
\[\input{B1_ii_b.tikz}\]
then consider $T^{g^2}_{\tau_2} \circ T^{g^1}_{\tau_1}\vac$, for any $g^1, g^2 \in G$, $T^{g^2}_{\tau_2} \circ T^{g^1}_{\tau_1}\vac \in \CL(s_0, s_1)$ iff
$g^1 = g^2$, by considering commutation with $A(v_3)$. The same applies for different orientations, and the same argument for dual triangles. For example
\[\input{B1_ii_c.tikz}\]
with $h^1 = h^2$ by considering $B(p_3)$. That $\{F^{h, g}_{\tau_2 \circ \tau_1}\ |\ h, g
\in G \}$ spans $\CL(s_0,s_1)$ is then immediate, where either the first or second variable is surplus respectively.

(iii)
For any $s_0, s_1$ which are distance 3 or further, we follow similar arguments but with an extended set of chosen edges which characterise the state $|\Psi\>$ along
a chosen ribbon $\xi$ between $s_0$ and $s_1$. Outside of this ribbon, the operator $O$ used to characterise $|\Psi\>$ must act trivially. Unlike the previous cases, it must also act trivially at 
at least one site inside the ribbon too, and we use this to calculate the states.

The last triangle in the ribbon $\xi$ must be either direct or dual, so we cover
a similar splitting of cases into direct and dual as in (i). First we consider the direct non-adjacent
case, $\xi = \tau \circ \xi'$, for example
\[
\input{direct_nonadjacent.tikz}
\]
Assume that $\CL(s_0, s_2)$ is spanned by $\{F^{h, g}_{\xi'}\vac\}$ and observe that $\CL(s_0, s_1)$ is a subspace of the space spanned by $\{T^{g'}_{\tau} \circ
F^{h, g}_{\xi'}\vac\ |\ g',g,h \in G\}$ or $\{F^{e, g'}_{\tau} \circ F^{h, g}_{\xi'}\vac\ |\
g',g,h \in G\}$. Specifically, $\CL(s_0, s_1)$ is the
subspace where  $A(v)$ acts as the identity for $v$ at
the site connecting $\xi'$ and $\tau$. Hence, for any $O \vac \in
\CL(s_0, s_1)$,
\[
O \vac=  \frac{1}{|G|} \sum_{h' \in G} h'\la_v O \vac
\]
which we apply to $O = F^{e, g'}_{\tau} \circ F^{h, g}_{\xi'}$,
\begin{align*}
F^{e, g'}_{\tau} \circ & F^{h, g}_{\xi'} \vac=\frac{1}{|G|} \sum_{h' \in G} h'\la_v F^{e, g'}_{\tau} \circ F^{h, g}_{\xi'} \vac = \frac{1}{|G|} \sum_{h' \in G} F^{e, h' g'}_{\tau} \circ F^{h, g h'{}^{-1}}_{\xi'} \vac\\
&=\frac{1}{|G|} \sum_{f \in G} F^{e,f^{-1}g g'}_{\tau} \circ F^{h, f}_{\xi'} \vac=\sum_{f \in G} F^{f^{-1}h f,
f^{-1}g g'}_{\tau} \circ F^{h, f}_{\xi'}\vac= F^{h, gg'}_{\xi}\vac  \end{align*}
after a change of variables to $f=gh'{}^{-1}$ and then using that 
 $F^{a, b}_{\tau} $ is independent of $a$ for a direct triangle operator. This allows us to recognise $F_\xi$ using (\ref{concat}). Denoting $gg'$ as $g$, it follows that $\CL(s_0,s_1)$ is spanned by $\{F^{h, g}_{\xi}\vac\ |\ h, g \in G\}$ as required. 
 
A similar argument applies for the dual distance 3 case, $\xi =
\tau^* \circ \xi'$. Given for example
\[
\input{dual_nonadjacent.tikz}
\]
we have this time $\CL(s_0, s_1)$ is a subspace of the space spanned by $\{L^{h'}_{\tau^*} \circ
F^{h, g}_{\xi'}\vac\ |\ g,h',h \in G\}$ and such that $B(p)=\delta_e\la_{s_2}$ acts as the identity, where $p$ is the face connecting $\xi'$ and $\tau^*$ so that $p \in s_2$. Then 
\begin{align*}
L^{h'}_{\tau^*} \circ F^{h, g}_{\xi'}\vac&=\delta_e \la_{s_2} L^{h'}_{\tau^*} \circ F^{h, g}_{\xi'}\vac = L^{h'}_{\tau^*} \delta_{h'^{-1}}\la \circ
F^{h, g}_{\xi'} \vac \\
&= L^{h'}_{\tau^*} \circ F^{h, g}_{\xi'} \delta_{h'^{-1}g^{-1}hg} \la_{s_2} \vac
\end{align*}
which only holds if $h' = g^{-1}hg$, so for elements of $\CL(s_0,s_1)$ we need only consider
\begin{align*}  L^{g^{-1}hg}_{\tau^*} \circ F^{h, g}_{\xi'} \vac &= F^{g^{-1}hg, e}_{\tau^*} \circ F^{h, g}_{\xi'} \vac=\sum_{f}\delta_{f,g} F^{f^{-1}hf,e}_{\tau^*}\circ F^{h,g}_{\xi'} \vac\\
&=\sum_{f} F^{f^{-1}hf,f^{-1}g}_{\tau^*} \circ F^{h,g}_{\xi'} \vac=F^{h,g}_{\xi}\vac\end{align*}
on noting that $F^{f^{-1}hf,f^{-1}g}_{\tau^*} = \delta_{e,f^{-1}g} F^{f^{-1}hf,e}_{\tau^*}$ and using (\ref{concat}). Thus, $\CL(s_0, s_1)$ is spanned by $\{F^{h,g}_{\xi}\vac\ |\ h,g \in G\}$ as required. 

\section{Universal Quantum Computation with $D(S_3)$}\label{app:universal_comp}
Here, we outline and comment on further aspects of the logical qubit within $D(S_3)$ in \cite{Woot}. First, we describe a $Z$-basis measurement on the logical qubit. It is claimed in \cite{Cirac, Sim}
that there exist `transport' operations $M^{\tau}_{\xi}$ which move $\tau$ quasiparticles along the lattice deterministically. In particular, these should exist such that
\[M^{\tau}_{-\xi'} W^{\tau}_{\xi} \vac = W^{\tau}_{\xi' \circ \xi} \vac\]
for all composeable open ribbons $\xi,\xi'$. It is beyond our scope to construct  $M^{\tau}_{-\xi'}$ here, but assuming it exists, it is a linear combination of chargeon ribbons, and therefore satisfies (\ref{eq:delta_ribbons}). Taking $-\xi$ to be a ribbon that completes $\xi$ to a closed contractible ribbon, we have 
\[M^{\tau}_{-\xi} W^{\tau}_{\xi} \vac =  W^{\tau}_{(-\xi) \circ \xi}\vac = \vac\]

Hence, referring to $\xi,\xi',\xi''$ in Section~\ref{secS3}, we have that applying $M^{\tau}_{-\xi} M^{\tau}_{-\xi'}$ to $|0_L\>$ and measuring the projector $P_{e, 1}$ at any $s_i$ will always yield 
$\vac$. On the other hand, if we begin with $|1_L\>$, we have
\begin{align*}
M^{\tau}_{-\xi} M^{\tau}_{-\xi'} |1_L\> &= M^{\tau}_{-\xi} M^{\tau}_{-\xi'} W^{\sigma}_{\xi''} W^{\tau}_{\xi'} W^{\tau}_{\xi}\vac\\
&= W^{\sigma}_{\xi''}M^{\tau}_{-\xi} M^{\tau}_{-\xi'}W^{\tau}_{\xi'} W^{\tau}_{\xi}\vac\\
&= W^{\sigma}_{\xi''}\vac
\end{align*}
by (\ref{eq:delta_ribbons}), and so applying $P_{e,1}$ at $s_0$, $s_1$ will return $0$. The operation $M^{\tau}_{-\xi} M^{\tau}_{-\xi'}$ followed by measuring $P_{e,1} \la_{s_0}$, say, therefore constitutes a destructive $Z$-basis measurement on the logical qubit: it tells us whether the qubit was in state $|0_L\>$ or $|1_L\>$,
but at the cost of taking us out of the degenerate subspace.

Now consider two distant groups of 4 $\tau$ quasiparticles labelled $a$, $b$:
\[\input{ds3_4.tikz}\]
where group $a$ is as before and group $b$ is a parallel copy with parallel notation. Entanglement between $a$, $b$ is achieved with the gate
\[K_{a,b}:=\frac{1}{2}(\id_a\otimes\id_b+X_a\otimes\id_b+\id_a\otimes X_b - X_a \otimes X_b)\]
where $X_a, X_b$ are the logical operators on the respective qubits.
$K_{a,b}$ has the following representations as a quantum circuit and a ZX-diagram respectively:
\[\input{Kab.tikz}\]
In terms of ribbon operators, this is:
\[K_{a,b} = \frac{1}{2} (\id_a\otimes\id_b+W^{\sigma}_{\xi''_a}\otimes\id_b+\id_a\otimes W^{\sigma}_{\xi''_b} - W^{\sigma}_{\xi''_a} \otimes W^{\sigma}_{\xi''_b})\]
by straighforward substitution. Note that, while $K_{a,b}$ is an entangling operation between the two logical qubits, it only acts along ribbons $\xi''_a, \xi''_b$, and doesn't require ribbons between the two qubits, and
must rely on the large entangled state on the lattice to transmit information. As $K_{a,b}$ requires only the ribbons $\xi''_a, \xi''_b$, it keeps the state within the combined degenerate 
subspace where there are $\tau$ quasiparticles at all sites $s_0, s_1,\cdots,s_7$.

A logical Hadamard can be performed non-deterministically on qubit $a$ using an ancillary qubit. We initialise the ancilla with $|0_b\>$, apply $K_{a,b}$ and then perform a $Z$-basis measurement on qubit $a$.
This teleports the state $|\psi\>$ on qubit $a$ to $H_b|\psi\>$ on qubit $b$, with a possible additional $Z_b$ factor depending on the measurement outcome.
This is obvious from a short calculation with the ZX-calculus \cite{CD}. Consider branch 1, where the measurement results in $\<0_a|$:
\[\input{ds3_5.tikz}\]
and branch 2, where the measurement gives $\<1_a|$:
\[\input{ds3_6.tikz}\]
If we reach branch 2, the process is repeated until the Hadamard alone is implemented (this is quite inefficient).

Equipped with the logical Hadamard and $X$ rotations, we can reach anywhere on the Bloch sphere, and the addition of the entangling gate $K_{a,b}$ allows the implementation of any unitary \cite[Sec~4.5.2]{Niel}.
We note that several other schemes for universal computation using representations of $D(S_3)$ have been described in \cite{Cui2}, although the formulation is categorical rather than
in terms of the quantum double on a lattice. We do not know whether these categorical schemes can be implemented on the lattice.

\end{document}

%% file: basic.tikzstyles
\tikzstyle{boundary vertex}=[inner sep=0mm, minimum size=1mm, shape=circle, draw=black, fill=black]

\tikzstyle{arrow}=[->]
\tikzstyle{red_arrow}=[->, draw=red]
\tikzstyle{cyan_arrow}=[->, draw=cyan]
\tikzstyle{red_dash}=[-, dashed, draw=red]

%% file: quantum.tikzstyles
\tikzstyle{gate}=[shape=rectangle, text height=1.5ex, text depth=0.25ex, yshift=0.5mm, fill=white, draw=black, minimum height=5mm, yshift=-0.5mm, minimum width=5mm, font={\small}, tikzit category=circuit]
\tikzstyle{big gate}=[shape=rectangle, text height=1.5ex, text depth=0.25ex, yshift=0.5mm, fill=white, draw=black, minimum height=10mm, yshift=-0.5mm, minimum width=5mm, font={\small}, tikzit category=circuit]
\tikzstyle{Z dot}=[inner sep=0mm, minimum size=2mm, shape=circle, draw=black, fill={rgb,255: red,221; green,255; blue,221}, tikzit category=zx]
\tikzstyle{Z phase dot}=[minimum size=5mm, font={\footnotesize\boldmath}, shape=rectangle, rounded corners=2mm, inner sep=0.2mm, outer sep=-2mm, scale=0.8, tikzit shape=circle, draw=black, fill={rgb,255: red,221; green,255; blue,221}, tikzit draw=blue, tikzit category=zx]
\tikzstyle{X dot}=[Z dot, shape=circle, draw=black, fill={rgb,255: red,255; green,136; blue,136}, tikzit category=zx]
\tikzstyle{X phase dot}=[Z phase dot, tikzit shape=circle, tikzit draw=blue, fill={rgb,255: red,255; green,136; blue,136}, font={\footnotesize\boldmath}, tikzit category=zx]
\tikzstyle{hadamard}=[fill=yellow, draw=black, shape=rectangle, inner sep=0.6mm, minimum height=1.5mm, minimum width=1.5mm, tikzit category=zx]
\tikzstyle{paulibox}=[fill={rgb,255: red,221; green,221; blue,255}, draw=black, shape=rectangle, inner sep=0.6mm, minimum height=5mm, minimum width=5mm, font={\footnotesize}, text height=1.5ex, text depth=0.25ex, tikzit category=zx]
\tikzstyle{vertex}=[inner sep=0mm, minimum size=1mm, shape=circle, draw=black, fill=black, tikzit category=misc]
\tikzstyle{vertex set}=[inner sep=0mm, minimum size=1mm, shape=circle, draw=black, fill=white, font={\footnotesize\boldmath}, tikzit category=misc]
\tikzstyle{small black dot}=[fill=black, draw=black, shape=circle, inner sep=0pt, minimum width=1.2mm, tikzit category=circuit]
\tikzstyle{cnot ctrl}=[fill=black, draw=black, shape=circle, inner sep=0pt, minimum width=1.2mm, tikzit category=circuit]
\tikzstyle{cnot targ}=[fill=white, draw=white, shape=circle, tikzit category=circuit, label={center:$\oplus$}, inner sep=0pt, minimum width=2.1mm, tikzit fill={rgb,255: red,102; green,204; blue,255}, tikzit draw=black]
\tikzstyle{ket}=[fill=white, draw=black, shape=regular polygon, regular polygon sides=3, regular polygon rotate=-30, scale=0.7, inner sep=1pt, tikzit category=circuit, tikzit shape=rectangle, tikzit fill=green]
\tikzstyle{bra}=[fill=white, draw=black, shape=regular polygon, regular polygon sides=3, regular polygon rotate=30, scale=0.7, inner sep=1pt, tikzit category=circuit, tikzit shape=rectangle, tikzit fill=red]
\tikzstyle{scalar}=[shape=rectangle, text height=1.5ex, text depth=0.25ex, yshift=0.5mm, fill=white, draw=black, minimum height=5mm, yshift=-0.5mm, minimum width=5mm, font={\small}]
\tikzstyle{clabel}=[fill=white, draw=none, shape=rectangle, tikzit fill={rgb,255: red,56; green,255; blue,242}, font={\footnotesize}, inner sep=1pt, tikzit category=labels]
\tikzstyle{empty diagram}=[draw={gray!40!white}, dashed, shape=rectangle, minimum width=1cm, minimum height=1cm, tikzit category=misc]

\tikzstyle{hadamard edge}=[-, dashed, dash pattern=on 2pt off 0.5pt, thick, draw={rgb,255: red,68; green,136; blue,255}]
\tikzstyle{box edge}=[-, dashed, dash pattern=on 2pt off 0.5pt, thick, draw={rgb,255: red,203; green,192; blue,225}]
\tikzstyle{brace edge}=[-, tikzit draw=blue, decorate, decoration={brace,amplitude=1mm,raise=-1mm}]
\tikzstyle{diredge}=[->]
\tikzstyle{double edge}=[-, double, shorten <=-1mm, shorten >=-1mm, double distance=2pt]
\tikzstyle{gray edge}=[-, {gray!60!white}]
\tikzstyle{pointer edge}=[->, very thick, gray]
\tikzstyle{boldedge}=[-, line width=1.6pt, shorten <=-0.17mm, shorten >=-0.17mm]

%% file: toric_exampleA.tikz
\begin{tikzpicture}
	\begin{pgfonlayer}{nodelayer}
		\node [style=boundary vertex] (0) at (0, 0) {};
		\node [style=boundary vertex] (1) at (0, 2) {};
		\node [style=boundary vertex] (2) at (2, 0) {};
		\node [style=boundary vertex] (3) at (2, 2) {};
		\node [style=boundary vertex] (4) at (-2, 2) {};
		\node [style=boundary vertex] (5) at (-2, 0) {};
		\node [style=none] (6) at (0.5, 0.75) {$|s\>$};
		\node [style=none] (7) at (-1, 1) {$p_1$};
		\node [style=none] (8) at (1, 1.5) {$p_2$};
		\node [style=boundary vertex] (9) at (0, 4) {};
		\node [style=none] (10) at (-0.5, 2.5) {$v_2$};
		\node [style=none] (12) at (-0.5, -0.5) {$v_1$};
		\node [style=none] (14) at (1, 2.5) {$|t\>$};
		\node [style=boundary vertex] (15) at (0, 4) {};
		\node [style=none] (16) at (-0.5, 3.25) {$|u\>$};
		\node [style=boundary vertex] (17) at (2, 4) {};
		\node [style=none] (18) at (1, 3.25) {$p_3$};
		\node [style=none] (19) at (-0.5, 4.5) {$v_3$};
	\end{pgfonlayer}
	\begin{pgfonlayer}{edgelayer}
		\draw [style=arrow] (5) to (4);
		\draw [style=arrow] (4) to (1);
		\draw [style=arrow] (1) to (3);
		\draw [style=arrow] (5) to (0);
		\draw [style=arrow] (0) to (1);
		\draw [style=arrow] (0) to (2);
		\draw [style=arrow] (2) to (3);
		\draw [style=arrow] (1) to (9);
		\draw [style=arrow] (15) to (17);
		\draw [style=arrow] (3) to (17);
	\end{pgfonlayer}
\end{tikzpicture}

%% file: toric_exampleB.tikz
\begin{tikzpicture}
	\begin{pgfonlayer}{nodelayer}
		\node [style=boundary vertex] (0) at (0, 0) {};
		\node [style=boundary vertex] (1) at (0, 2) {};
		\node [style=boundary vertex] (2) at (2, 0) {};
		\node [style=boundary vertex] (3) at (2, 2) {};
		\node [style=boundary vertex] (4) at (-2, 2) {};
		\node [style=boundary vertex] (5) at (-2, 0) {};
		\node [style=none] (6) at (0.5, 0.75) {$|s\>$};
		\node [style=none] (7) at (-1, 1) {$p_1$};
		\node [style=none] (8) at (1, 1.5) {$p_2$};
		\node [style=boundary vertex] (9) at (0, 4) {};
		\node [style=none] (10) at (-0.5, 2.5) {$v_2$};
		\node [style=none] (12) at (-0.5, -0.5) {$v_1$};
		\node [style=none] (14) at (1, 2.5) {$|t\>$};
		\node [style=boundary vertex] (15) at (0, 4) {};
		\node [style=none] (16) at (-0.5, 3.25) {$|u\>$};
		\node [style=boundary vertex] (17) at (2, 4) {};
		\node [style=none] (18) at (1, 3.25) {$p_3$};
		\node [style=none] (19) at (-0.5, 4.5) {$v_3$};
		\node [style=none] (20) at (-3, 1) {$q^{-js}$};
	\end{pgfonlayer}
	\begin{pgfonlayer}{edgelayer}
		\draw [style=arrow] (5) to (4);
		\draw [style=arrow] (4) to (1);
		\draw [style=arrow] (1) to (3);
		\draw [style=arrow] (5) to (0);
		\draw [style=arrow] (0) to (1);
		\draw [style=arrow] (0) to (2);
		\draw [style=arrow] (2) to (3);
		\draw [style=arrow] (1) to (9);
		\draw [style=arrow] (15) to (17);
		\draw [style=arrow] (3) to (17);
	\end{pgfonlayer}
\end{tikzpicture}

%% file: toric_exampleC.tikz
\begin{tikzpicture}
	\begin{pgfonlayer}{nodelayer}
		\node [style=boundary vertex] (0) at (0, -2) {};
		\node [style=boundary vertex] (1) at (0, 0) {};
		\node [style=boundary vertex] (2) at (2, -2) {};
		\node [style=boundary vertex] (3) at (2, 0) {};
		\node [style=boundary vertex] (4) at (-2, 0) {};
		\node [style=boundary vertex] (5) at (-2, -2) {};
		\node [style=none] (6) at (1, -1.25) {$|s-i\>$};
		\node [style=none] (7) at (-1, -1) {$p_1$};
		\node [style=none] (8) at (1, -0.5) {$p_2$};
		\node [style=boundary vertex] (9) at (0, 2) {};
		\node [style=none] (10) at (-0.5, 0.5) {$v_2$};
		\node [style=none] (11) at (-0.5, -2.5) {$v_1$};
		\node [style=none] (12) at (1, 0.5) {$|t\>$};
		\node [style=boundary vertex] (13) at (0, 2) {};
		\node [style=none] (14) at (-0.5, 1.25) {$|u\>$};
		\node [style=boundary vertex] (15) at (2, 2) {};
		\node [style=none] (16) at (1, 1.25) {$p_3$};
		\node [style=none] (17) at (-0.5, 2.5) {$v_3$};
		\node [style=none] (18) at (-3, -1) {$q^{-js}$};
	\end{pgfonlayer}
	\begin{pgfonlayer}{edgelayer}
		\draw [style=arrow] (5) to (4);
		\draw [style=arrow] (4) to (1);
		\draw [style=arrow] (1) to (3);
		\draw [style=arrow] (5) to (0);
		\draw [style=arrow] (0) to (1);
		\draw [style=arrow] (0) to (2);
		\draw [style=arrow] (2) to (3);
		\draw [style=arrow] (1) to (9);
		\draw [style=arrow] (13) to (15);
		\draw [style=arrow] (3) to (15);
	\end{pgfonlayer}
\end{tikzpicture}

%% file: toric_exampleD.tikz
\begin{tikzpicture}
	\begin{pgfonlayer}{nodelayer}
		\node [style=boundary vertex] (0) at (0, -2) {};
		\node [style=boundary vertex] (1) at (0, 0) {};
		\node [style=boundary vertex] (2) at (2, -2) {};
		\node [style=boundary vertex] (3) at (2, 0) {};
		\node [style=boundary vertex] (4) at (-2, 0) {};
		\node [style=boundary vertex] (5) at (-2, -2) {};
		\node [style=none] (6) at (1, -1.25) {$|s-i\>$};
		\node [style=none] (7) at (-1, -1) {$p_1$};
		\node [style=none] (8) at (1, -0.5) {$p_2$};
		\node [style=boundary vertex] (9) at (0, 2) {};
		\node [style=none] (10) at (-0.5, 0.5) {$v_2$};
		\node [style=none] (11) at (-0.5, -2.5) {$v_1$};
		\node [style=none] (12) at (1, 0.5) {$|t+i\>$};
		\node [style=boundary vertex] (13) at (0, 2) {};
		\node [style=none] (14) at (-0.5, 1.25) {$|u\>$};
		\node [style=boundary vertex] (15) at (2, 2) {};
		\node [style=none] (16) at (1, 1.25) {$p_3$};
		\node [style=none] (17) at (-0.5, 2.5) {$v_3$};
		\node [style=none] (18) at (-3, -1) {$q^{-js}$};
	\end{pgfonlayer}
	\begin{pgfonlayer}{edgelayer}
		\draw [style=arrow] (5) to (4);
		\draw [style=arrow] (4) to (1);
		\draw [style=arrow] (1) to (3);
		\draw [style=arrow] (5) to (0);
		\draw [style=arrow] (0) to (1);
		\draw [style=arrow] (0) to (2);
		\draw [style=arrow] (2) to (3);
		\draw [style=arrow] (1) to (9);
		\draw [style=arrow] (13) to (15);
		\draw [style=arrow] (3) to (15);
	\end{pgfonlayer}
\end{tikzpicture}

%% file: toric_exampleE.tikz
\begin{tikzpicture}
	\begin{pgfonlayer}{nodelayer}
		\node [style=boundary vertex] (0) at (0, -2) {};
		\node [style=boundary vertex] (1) at (0, 0) {};
		\node [style=boundary vertex] (2) at (2, -2) {};
		\node [style=boundary vertex] (3) at (2, 0) {};
		\node [style=boundary vertex] (4) at (-2, 0) {};
		\node [style=boundary vertex] (5) at (-2, -2) {};
		\node [style=none] (6) at (1, -1.25) {$|s-i\>$};
		\node [style=none] (7) at (-1, -1) {$p_1$};
		\node [style=none] (8) at (1, -0.5) {$p_2$};
		\node [style=boundary vertex] (9) at (0, 2) {};
		\node [style=none] (10) at (-0.5, 0.5) {$v_2$};
		\node [style=none] (11) at (-0.5, -2.5) {$v_1$};
		\node [style=none] (12) at (1, 0.5) {$|t+i\>$};
		\node [style=boundary vertex] (13) at (0, 2) {};
		\node [style=none] (14) at (-0.5, 1.25) {$|u\>$};
		\node [style=boundary vertex] (15) at (2, 2) {};
		\node [style=none] (16) at (1, 1.25) {$p_3$};
		\node [style=none] (17) at (-0.5, 2.5) {$v_3$};
		\node [style=none] (18) at (-3.5, 0) {$q^{-j(s+u)}$};
	\end{pgfonlayer}
	\begin{pgfonlayer}{edgelayer}
		\draw [style=arrow] (5) to (4);
		\draw [style=arrow] (4) to (1);
		\draw [style=arrow] (1) to (3);
		\draw [style=arrow] (5) to (0);
		\draw [style=arrow] (0) to (1);
		\draw [style=arrow] (0) to (2);
		\draw [style=arrow] (2) to (3);
		\draw [style=arrow] (1) to (9);
		\draw [style=arrow] (13) to (15);
		\draw [style=arrow] (3) to (15);
	\end{pgfonlayer}
\end{tikzpicture}

%% file: lattice_innerproduct.tikz
\begin{tikzpicture}
	\begin{pgfonlayer}{nodelayer}
		\node [style=boundary vertex] (0) at (-4, 2) {};
		\node [style=boundary vertex] (1) at (-4, 0) {};
		\node [style=boundary vertex] (2) at (-2, 2) {};
		\node [style=boundary vertex] (3) at (-2, 0) {};
		\node [style=boundary vertex] (4) at (0, 2) {};
		\node [style=boundary vertex] (5) at (0, 0) {};
		\node [style=boundary vertex] (6) at (0, -2) {};
		\node [style=boundary vertex] (7) at (2, 0) {};
		\node [style=none] (8) at (-3, 2.5) {$h'^1$};
		\node [style=none] (9) at (-1, 2.5) {$h'^2$};
		\node [style=none] (10) at (0.5, 1) {$h'^3$};
		\node [style=none] (11) at (1, 0.5) {$h'^4$};
		\node [style=none] (12) at (-4.5, 0.5) {$g'^1$};
		\node [style=none] (13) at (-2.5, 0.5) {$g'^2$};
		\node [style=none] (14) at (-1, -0.5) {$g'^3$};
		\node [style=none] (15) at (-0.5, -1.5) {$g'^4$};
		\node [style=none] (16) at (-6.5, 1) {$|\psi'\>$};
		\node [style=none] (17) at (-5.5, 1) {$:=$};
	\end{pgfonlayer}
	\begin{pgfonlayer}{edgelayer}
		\draw [style=arrow] (1) to (0);
		\draw [style=arrow] (0) to (2);
		\draw [style=arrow] (3) to (2);
		\draw [style=arrow] (2) to (4);
		\draw [style=arrow] (3) to (5);
		\draw [style=arrow] (6) to (5);
		\draw [style=arrow] (5) to (7);
		\draw [style=arrow] (5) to (4);
	\end{pgfonlayer}
\end{tikzpicture}

%% file: toric_maxentangle_v3.tikz
\begin{tikzpicture}
	\begin{pgfonlayer}{nodelayer}
		\node [style=boundary vertex] (1) at (0, 0) {};
		\node [style=boundary vertex] (2) at (2, 0) {};
		\node [style=boundary vertex] (4) at (0, -2) {};
		\node [style=boundary vertex] (5) at (2, 2) {};
		\node [style=none] (7) at (1, -1) {};
		\node [style=none] (8) at (1, 1) {};
		\node [style=none] (9) at (3, 1) {};
		\node [style=boundary vertex] (10) at (0, 2) {};
		\node [style=boundary vertex] (11) at (2, -2) {};
		\node [style=boundary vertex] (12) at (4, 0) {};
		\node [style=boundary vertex] (13) at (4, 2) {};
		\node [style=none] (14) at (0.5, -1) {$s_0$};
		\node [style=none] (15) at (3.25, 1.75) {$s_1$};
		\node [style=none] (18) at (1, 0.5) {$|f\>$};
		\node [style=none] (19) at (-0.5, 1.25) {$|k\>$};
		\node [style=none] (20) at (1, 2.5) {$|l\>$};
		\node [style=none] (21) at (2.5, 0.75) {$|p\>$};
		\node [style=none] (22) at (3, 2.5) {$|m\>$};
		\node [style=none] (24) at (5.75, 0) {$=$};
		\node [style=none] (25) at (-4.75, 0) {$|{\rm Bell};\xi\>=\sum_{i} F_\xi^{i,0}$};
		\node [style=none] (26) at (9.5, 0) {$\delta_{0}(k+l+m)\sum_{i}$};
		\node [style=boundary vertex] (28) at (15, 0) {};
		\node [style=boundary vertex] (29) at (17, 0) {};
		\node [style=boundary vertex] (31) at (15, -2) {};
		\node [style=boundary vertex] (32) at (17, 2) {};
		\node [style=none] (34) at (16, -1) {};
		\node [style=none] (35) at (16, 1) {};
		\node [style=none] (36) at (18, 1) {};
		\node [style=boundary vertex] (37) at (15, 2) {};
		\node [style=boundary vertex] (38) at (17, -2) {};
		\node [style=boundary vertex] (39) at (19, 0) {};
		\node [style=boundary vertex] (40) at (19, 2) {};
		\node [style=none] (41) at (15.5, -1) {$s_0$};
		\node [style=none] (42) at (18.25, 1.75) {$s_1$};
		\node [style=none] (45) at (16, 0.5) {$|f+i\>$};
		\node [style=none] (46) at (14.5, 1.25) {$|k\>$};
		\node [style=none] (48) at (18, 0.75) {$|p-i\>$};
		\node [style=none] (50) at (16, 2.5) {$|l\>$};
		\node [style=none] (51) at (18, 2.5) {$|m\>$};
		\node [style=none] (52) at (-0.5, -1) {$|a\>$};
		\node [style=none] (53) at (1, -2.5) {$|b\>$};
		\node [style=none] (54) at (2.5, -1) {$|c\>$};
		\node [style=none] (55) at (14.5, -1) {$|a\>$};
		\node [style=none] (56) at (16, -2.5) {$|b\>$};
		\node [style=none] (57) at (17.5, -1) {$|c\>$};
		\node [style=boundary vertex] (58) at (-2, 0) {};
		\node [style=boundary vertex] (59) at (13, 0) {};
		\node [style=none] (60) at (-1, 0.5) {$|d\>$};
		\node [style=none] (61) at (14, 0.5) {$|d\>$};
	\end{pgfonlayer}
	\begin{pgfonlayer}{edgelayer}
		\draw [style={red_dash}] (1) to (7.center);
		\draw [style={red_dash}] (1) to (8.center);
		\draw [style={red_dash}] (10) to (8.center);
		\draw [style={red_dash}] (8.center) to (5);
		\draw [style={red_dash}] (5) to (9.center);
		\draw [style=arrow] (1) to (2);
		\draw [style=arrow] (2) to (12);
		\draw [style=arrow] (4) to (1);
		\draw [style=arrow] (4) to (11);
		\draw [style=arrow] (11) to (2);
		\draw [style=arrow] (1) to (10);
		\draw [style=arrow] (10) to (5);
		\draw [style=arrow] (2) to (5);
		\draw [style=arrow] (5) to (13);
		\draw [style=arrow] (12) to (13);
		\draw [style={red_dash}] (9.center) to (13);
		\draw [style={red_dash}] (28) to (34.center);
		\draw [style={red_dash}] (28) to (35.center);
		\draw [style={red_dash}] (37) to (35.center);
		\draw [style={red_dash}] (35.center) to (32);
		\draw [style={red_dash}] (32) to (36.center);
		\draw [style=arrow] (28) to (29);
		\draw [style=arrow] (29) to (39);
		\draw [style=arrow] (31) to (28);
		\draw [style=arrow] (31) to (38);
		\draw [style=arrow] (38) to (29);
		\draw [style=arrow] (28) to (37);
		\draw [style=arrow] (37) to (32);
		\draw [style=arrow] (29) to (32);
		\draw [style=arrow] (32) to (40);
		\draw [style=arrow] (39) to (40);
		\draw [style={red_dash}] (36.center) to (40);
		\draw [style=arrow] (58) to (1);
		\draw [style=arrow] (59) to (28);
	\end{pgfonlayer}
\end{tikzpicture}

%% file: toric_creation.tikz
\begin{tikzpicture}
	\begin{pgfonlayer}{nodelayer}
		\node [style=boundary vertex] (0) at (0, 0) {};
		\node [style=boundary vertex] (1) at (0, 2) {};
		\node [style=boundary vertex] (2) at (2, 0) {};
		\node [style=boundary vertex] (3) at (2, 2) {};
		\node [style=boundary vertex] (4) at (-2, 2) {};
		\node [style=boundary vertex] (5) at (-2, 0) {};
		\node [style=none] (6) at (0.5, 0.75) {$|s\>$};
		\node [style=boundary vertex] (9) at (0, 4) {};
		\node [style=none] (14) at (1.25, 2.5) {$|t\>$};
		\node [style=boundary vertex] (15) at (0, 4) {};
		\node [style=none] (16) at (-0.5, 3.25) {$|u\>$};
		\node [style=boundary vertex] (17) at (2, 4) {};
		\node [style=none] (18) at (-1, 1) {};
		\node [style=none] (19) at (1, 1) {};
		\node [style=none] (20) at (-1.25, 0.5) {$s_0$};
		\node [style=none] (21) at (1.25, 1.5) {$s_1$};
	\end{pgfonlayer}
	\begin{pgfonlayer}{edgelayer}
		\draw [style=arrow] (5) to (4);
		\draw [style=arrow] (4) to (1);
		\draw [style=arrow] (1) to (3);
		\draw [style=arrow] (5) to (0);
		\draw [style=arrow] (0) to (1);
		\draw [style=arrow] (0) to (2);
		\draw [style=arrow] (2) to (3);
		\draw [style=arrow] (1) to (9);
		\draw [style=arrow] (15) to (17);
		\draw [style=arrow] (3) to (17);
		\draw [style={red_dash}] (18.center) to (0);
		\draw [style={red_dash}] (19.center) to (1);
		\draw [style={red_dash}] (18.center) to (1);
	\end{pgfonlayer}
\end{tikzpicture}

%% file: toric_creation2.tikz
\begin{tikzpicture}
	\begin{pgfonlayer}{nodelayer}
		\node [style=boundary vertex] (0) at (0, -2) {};
		\node [style=boundary vertex] (1) at (0, 0) {};
		\node [style=boundary vertex] (2) at (2, -2) {};
		\node [style=boundary vertex] (3) at (2, 0) {};
		\node [style=boundary vertex] (4) at (-2, 0) {};
		\node [style=boundary vertex] (5) at (-2, -2) {};
		\node [style=none] (6) at (0.5, -1.25) {$|s\>$};
		\node [style=boundary vertex] (7) at (0, 2) {};
		\node [style=none] (8) at (1.25, 0.5) {$|t\>$};
		\node [style=boundary vertex] (9) at (0, 2) {};
		\node [style=none] (10) at (-0.5, 1.25) {$|u\>$};
		\node [style=boundary vertex] (11) at (2, 2) {};
		\node [style=none] (12) at (-1, -1) {};
		\node [style=none] (13) at (1, -1) {};
		\node [style=none] (14) at (-1.25, -1.5) {$s_0$};
		\node [style=none] (15) at (1.25, -0.5) {$s_1$};
		\node [style=none] (16) at (-3.5, 0) {$W_{\xi}^{i,j}$};
		\node [style=none] (17) at (3.5, 0) {$=$};
		\node [style=boundary vertex] (18) at (11, -2) {};
		\node [style=boundary vertex] (19) at (11, 0) {};
		\node [style=boundary vertex] (20) at (13, -2) {};
		\node [style=boundary vertex] (21) at (13, 0) {};
		\node [style=boundary vertex] (22) at (9, 0) {};
		\node [style=boundary vertex] (23) at (9, -2) {};
		\node [style=none] (24) at (12, -1.25) {$|s - i\>$};
		\node [style=boundary vertex] (25) at (11, 2) {};
		\node [style=none] (26) at (12.25, 0.5) {$|t\>$};
		\node [style=boundary vertex] (27) at (11, 2) {};
		\node [style=none] (28) at (10.5, 1.25) {$|u\>$};
		\node [style=boundary vertex] (29) at (13, 2) {};
		\node [style=none] (30) at (10, -1) {};
		\node [style=none] (31) at (12, -1) {};
		\node [style=none] (32) at (9.75, -1.5) {$s_0$};
		\node [style=none] (33) at (12.25, -0.5) {$s_1$};
		\node [style=none] (34) at (6.5, 0) {$\sum_k q^{-jk} \delta_k(s)$};
	\end{pgfonlayer}
	\begin{pgfonlayer}{edgelayer}
		\draw [style=arrow] (5) to (4);
		\draw [style=arrow] (4) to (1);
		\draw [style=arrow] (1) to (3);
		\draw [style=arrow] (5) to (0);
		\draw [style=arrow] (0) to (1);
		\draw [style=arrow] (0) to (2);
		\draw [style=arrow] (2) to (3);
		\draw [style=arrow] (1) to (7);
		\draw [style=arrow] (9) to (11);
		\draw [style=arrow] (3) to (11);
		\draw [style={red_dash}] (12.center) to (0);
		\draw [style={red_dash}] (13.center) to (1);
		\draw [style={red_dash}] (12.center) to (1);
		\draw [style=arrow] (23) to (22);
		\draw [style=arrow] (22) to (19);
		\draw [style=arrow] (19) to (21);
		\draw [style=arrow] (23) to (18);
		\draw [style=arrow] (18) to (19);
		\draw [style=arrow] (18) to (20);
		\draw [style=arrow] (20) to (21);
		\draw [style=arrow] (19) to (25);
		\draw [style=arrow] (27) to (29);
		\draw [style=arrow] (21) to (29);
		\draw [style={red_dash}] (30.center) to (18);
		\draw [style={red_dash}] (31.center) to (19);
		\draw [style={red_dash}] (30.center) to (19);
	\end{pgfonlayer}
\end{tikzpicture}

%% file: toric_transport.tikz
\begin{tikzpicture}
	\begin{pgfonlayer}{nodelayer}
		\node [style=boundary vertex] (0) at (0, -2) {};
		\node [style=boundary vertex] (1) at (0, 0) {};
		\node [style=boundary vertex] (2) at (2, -2) {};
		\node [style=boundary vertex] (3) at (2, 0) {};
		\node [style=boundary vertex] (4) at (-2, 0) {};
		\node [style=boundary vertex] (5) at (-2, -2) {};
		\node [style=boundary vertex] (7) at (0, 2) {};
		\node [style=none] (8) at (1.25, 0.5) {$|t\>$};
		\node [style=boundary vertex] (9) at (0, 2) {};
		\node [style=none] (10) at (-0.5, 1.25) {$|u\>$};
		\node [style=boundary vertex] (11) at (2, 2) {};
		\node [style=none] (12) at (-1, -1) {};
		\node [style=none] (13) at (1, -1) {};
		\node [style=none] (14) at (-1.25, -1.5) {$s_0$};
		\node [style=none] (15) at (1.25, -0.5) {$s_1$};
		\node [style=none] (16) at (1, 1) {};
		\node [style=none] (17) at (1.25, 1.5) {$s_2$};
	\end{pgfonlayer}
	\begin{pgfonlayer}{edgelayer}
		\draw [style=arrow] (5) to (4);
		\draw [style=arrow] (4) to (1);
		\draw [style=arrow] (1) to (3);
		\draw [style=arrow] (5) to (0);
		\draw [style=arrow] (0) to (1);
		\draw [style=arrow] (0) to (2);
		\draw [style=arrow] (2) to (3);
		\draw [style=arrow] (1) to (7);
		\draw [style=arrow] (9) to (11);
		\draw [style=arrow] (3) to (11);
		\draw [style={red_dash}] (12.center) to (0);
		\draw [style={red_dash}] (13.center) to (1);
		\draw [style={red_dash}] (9) to (16.center);
	\end{pgfonlayer}
\end{tikzpicture}

%% file: toric_transport2.tikz
\begin{tikzpicture}
	\begin{pgfonlayer}{nodelayer}
		\node [style=boundary vertex] (0) at (0, -2) {};
		\node [style=boundary vertex] (1) at (0, 0) {};
		\node [style=boundary vertex] (2) at (2, -2) {};
		\node [style=boundary vertex] (3) at (2, 0) {};
		\node [style=boundary vertex] (4) at (-2, 0) {};
		\node [style=boundary vertex] (5) at (-2, -2) {};
		\node [style=boundary vertex] (7) at (0, 2) {};
		\node [style=none] (8) at (1.25, 0.5) {$|t\>$};
		\node [style=boundary vertex] (9) at (0, 2) {};
		\node [style=none] (10) at (-0.5, 1.25) {$|u\>$};
		\node [style=boundary vertex] (11) at (2, 2) {};
		\node [style=none] (13) at (1, -1) {};
		\node [style=none] (15) at (1.25, -0.5) {$s_1$};
		\node [style=none] (16) at (1, 1) {};
		\node [style=none] (17) at (1.25, 1.5) {$s_2$};
		\node [style=none] (18) at (-3.75, 0) {$W_{\xi'}^{i,j}$};
		\node [style=none] (19) at (3.5, 0) {$=$};
		\node [style=none] (20) at (5, 0) {$q^{-ju}$};
		\node [style=boundary vertex] (21) at (8.25, -2) {};
		\node [style=boundary vertex] (22) at (8.25, 0) {};
		\node [style=boundary vertex] (23) at (10.25, -2) {};
		\node [style=boundary vertex] (24) at (10.25, 0) {};
		\node [style=boundary vertex] (25) at (6.25, 0) {};
		\node [style=boundary vertex] (26) at (6.25, -2) {};
		\node [style=boundary vertex] (28) at (8.25, 2) {};
		\node [style=none] (29) at (9.25, 0.5) {$|t + i\>$};
		\node [style=boundary vertex] (30) at (8.25, 2) {};
		\node [style=none] (31) at (7.75, 1.25) {$|u\>$};
		\node [style=boundary vertex] (32) at (10.25, 2) {};
		\node [style=none] (33) at (9.25, -1) {};
		\node [style=none] (34) at (9.5, -0.5) {$s_1$};
		\node [style=none] (35) at (9.25, 1) {};
		\node [style=none] (36) at (9.5, 1.5) {$s_2$};
	\end{pgfonlayer}
	\begin{pgfonlayer}{edgelayer}
		\draw [style=arrow] (5) to (4);
		\draw [style=arrow] (4) to (1);
		\draw [style=arrow] (1) to (3);
		\draw [style=arrow] (5) to (0);
		\draw [style=arrow] (0) to (1);
		\draw [style=arrow] (0) to (2);
		\draw [style=arrow] (2) to (3);
		\draw [style=arrow] (1) to (7);
		\draw [style=arrow] (9) to (11);
		\draw [style=arrow] (3) to (11);
		\draw [style={red_dash}] (13.center) to (1);
		\draw [style={red_dash}] (9) to (16.center);
		\draw [style={red_dash}] (1) to (16.center);
		\draw [style=arrow] (26) to (25);
		\draw [style=arrow] (25) to (22);
		\draw [style=arrow] (22) to (24);
		\draw [style=arrow] (26) to (21);
		\draw [style=arrow] (21) to (22);
		\draw [style=arrow] (21) to (23);
		\draw [style=arrow] (23) to (24);
		\draw [style=arrow] (22) to (28);
		\draw [style=arrow] (30) to (32);
		\draw [style=arrow] (24) to (32);
		\draw [style={red_dash}] (33.center) to (22);
		\draw [style={red_dash}] (30) to (35.center);
		\draw [style={red_dash}] (22) to (35.center);
	\end{pgfonlayer}
\end{tikzpicture}

%% file: toric_concatenation_v2.tikz
\begin{tikzpicture}
	\begin{pgfonlayer}{nodelayer}
		\node [style=boundary vertex] (0) at (0, -2) {};
		\node [style=boundary vertex] (1) at (0, 0) {};
		\node [style=boundary vertex] (2) at (2, -2) {};
		\node [style=boundary vertex] (3) at (2, 0) {};
		\node [style=boundary vertex] (4) at (-2, 0) {};
		\node [style=boundary vertex] (5) at (-2, -2) {};
		\node [style=none] (6) at (0.5, -1.25) {$|s\>$};
		\node [style=boundary vertex] (7) at (0, 2) {};
		\node [style=none] (8) at (1.25, 0.5) {$|t\>$};
		\node [style=boundary vertex] (9) at (0, 2) {};
		\node [style=none] (10) at (-0.5, 1.25) {$|u\>$};
		\node [style=boundary vertex] (11) at (2, 2) {};
		\node [style=none] (12) at (-1, -1) {};
		\node [style=none] (13) at (1, -1) {};
		\node [style=none] (14) at (-1.25, -1.5) {$s_0$};
		\node [style=none] (15) at (1.25, -0.5) {$s_1$};
		\node [style=none] (16) at (1, 1) {};
		\node [style=none] (17) at (1.25, 1.5) {$s_2$};
		\node [style=none] (18) at (-4, 0) {$W^{i,j}_{\xi' \circ \xi}$};
		\node [style=none] (19) at (3, 0) {$=$};
		\node [style=boundary vertex] (20) at (8.5, -2) {};
		\node [style=boundary vertex] (21) at (8.5, 0) {};
		\node [style=boundary vertex] (22) at (10.5, -2) {};
		\node [style=boundary vertex] (23) at (10.5, 0) {};
		\node [style=boundary vertex] (24) at (6.5, 0) {};
		\node [style=boundary vertex] (25) at (6.5, -2) {};
		\node [style=boundary vertex] (27) at (8.5, 2) {};
		\node [style=boundary vertex] (29) at (8.5, 2) {};
		\node [style=none] (30) at (8, 1.25) {$|u\>$};
		\node [style=boundary vertex] (31) at (10.5, 2) {};
		\node [style=none] (32) at (7.5, -1) {};
		\node [style=none] (33) at (9.5, -1) {};
		\node [style=none] (34) at (7.25, -1.5) {$s_0$};
		\node [style=none] (35) at (9.75, -0.5) {$s_1$};
		\node [style=none] (36) at (9.5, 1) {};
		\node [style=none] (37) at (9.75, 1.5) {$s_2$};
		\node [style=none] (38) at (9.5, 0.5) {$|t + i\>$};
		\node [style=none] (39) at (9.5, -1.25) {$|s - i\>$};
		\node [style=none] (40) at (4.75, 0) {$q^{-j(u+s)}$};
	\end{pgfonlayer}
	\begin{pgfonlayer}{edgelayer}
		\draw [style=arrow] (5) to (4);
		\draw [style=arrow] (4) to (1);
		\draw [style=arrow] (1) to (3);
		\draw [style=arrow] (5) to (0);
		\draw [style=arrow] (0) to (1);
		\draw [style=arrow] (0) to (2);
		\draw [style=arrow] (2) to (3);
		\draw [style=arrow] (1) to (7);
		\draw [style=arrow] (9) to (11);
		\draw [style=arrow] (3) to (11);
		\draw [style={red_dash}] (12.center) to (0);
		\draw [style={red_dash}] (13.center) to (1);
		\draw [style={red_dash}] (9) to (16.center);
		\draw [style={red_dash}] (12.center) to (1);
		\draw [style={red_dash}] (1) to (16.center);
		\draw [style=arrow] (25) to (24);
		\draw [style=arrow] (24) to (21);
		\draw [style=arrow] (21) to (23);
		\draw [style=arrow] (25) to (20);
		\draw [style=arrow] (20) to (21);
		\draw [style=arrow] (20) to (22);
		\draw [style=arrow] (22) to (23);
		\draw [style=arrow] (21) to (27);
		\draw [style=arrow] (29) to (31);
		\draw [style=arrow] (23) to (31);
		\draw [style={red_dash}] (32.center) to (20);
		\draw [style={red_dash}] (33.center) to (21);
		\draw [style={red_dash}] (29) to (36.center);
		\draw [style={red_dash}] (32.center) to (21);
		\draw [style={red_dash}] (21) to (36.center);
	\end{pgfonlayer}
\end{tikzpicture}

%% file: braiding1.tikz
\begin{tikzpicture}
	\begin{pgfonlayer}{nodelayer}
		\node [style=boundary vertex] (0) at (0, -2) {};
		\node [style=boundary vertex] (1) at (0, 0) {};
		\node [style=boundary vertex] (2) at (2, -2) {};
		\node [style=boundary vertex] (3) at (2, 0) {};
		\node [style=boundary vertex] (4) at (-2, 0) {};
		\node [style=boundary vertex] (5) at (-2, -2) {};
		\node [style=none] (6) at (3, 0.5) {$|n\>$};
		\node [style=none] (8) at (1, 0.5) {$|m\>$};
		\node [style=none] (10) at (-3, 0.5) {$|k\>$};
		\node [style=none] (12) at (-3, -1) {};
		\node [style=none] (13) at (1, -1) {};
		\node [style=none] (14) at (-3.5, -1) {$s_0$};
		\node [style=none] (15) at (1.5, -1) {$s_1$};
		\node [style=boundary vertex] (16) at (-4, 0) {};
		\node [style=boundary vertex] (17) at (-4, -2) {};
		\node [style=none] (18) at (-1, 0.5) {$|l\>$};
		\node [style=boundary vertex] (20) at (4, 0) {};
		\node [style=boundary vertex] (21) at (4, -2) {};
		\node [style=boundary vertex] (22) at (6, -2) {};
		\node [style=boundary vertex] (23) at (6, 0) {};
		\node [style=none] (25) at (-1, -1) {};
		\node [style=none] (26) at (2.5, -1) {$|o\>$};
		\node [style=none] (27) at (3, -2.5) {$|p\>$};
		\node [style=none] (28) at (4.5, -1.25) {$|q\>$};
	\end{pgfonlayer}
	\begin{pgfonlayer}{edgelayer}
		\draw [style=arrow] (5) to (4);
		\draw [style=arrow] (4) to (1);
		\draw [style=arrow] (1) to (3);
		\draw [style=arrow] (5) to (0);
		\draw [style=arrow] (0) to (1);
		\draw [style=arrow] (0) to (2);
		\draw [style=arrow] (2) to (3);
		\draw [style=arrow] (17) to (5);
		\draw [style=arrow] (17) to (16);
		\draw [style=arrow] (16) to (4);
		\draw [style={red_dash}] (13.center) to (3);
		\draw [style=arrow] (3) to (20);
		\draw [style=arrow] (2) to (21);
		\draw [style=arrow] (21) to (20);
		\draw [style=arrow] (21) to (22);
		\draw [style=arrow] (20) to (23);
		\draw [style=arrow] (22) to (23);
		\draw [style={red_dash}] (4) to (25.center);
		\draw [style={red_dash}] (25.center) to (1);
		\draw [style={red_dash}] (1) to (13.center);
		\draw [style={red_dash}] (16) to (12.center);
		\draw [style={red_dash}] (12.center) to (4);
	\end{pgfonlayer}
\end{tikzpicture}

%% file: braiding2.tikz
\begin{tikzpicture}
	\begin{pgfonlayer}{nodelayer}
		\node [style=boundary vertex] (0) at (-0.5, -2) {};
		\node [style=boundary vertex] (1) at (-0.5, 0) {};
		\node [style=boundary vertex] (2) at (1.5, -2) {};
		\node [style=boundary vertex] (3) at (1.5, 0) {};
		\node [style=boundary vertex] (4) at (-2.5, 0) {};
		\node [style=boundary vertex] (5) at (-2.5, -2) {};
		\node [style=none] (6) at (2.5, 0.5) {$|n\>$};
		\node [style=none] (8) at (0.5, 0.5) {$Z^j|m\>$};
		\node [style=none] (10) at (-3.5, 0.5) {$Z^j|k\>$};
		\node [style=boundary vertex] (16) at (-4.5, 0) {};
		\node [style=boundary vertex] (17) at (-4.5, -2) {};
		\node [style=none] (18) at (-1.5, 0.5) {$Z^j|l\>$};
		\node [style=boundary vertex] (20) at (3.5, 0) {};
		\node [style=boundary vertex] (21) at (3.5, -2) {};
		\node [style=boundary vertex] (22) at (5.5, -2) {};
		\node [style=boundary vertex] (23) at (5.5, 0) {};
		\node [style=none] (26) at (2, -1) {$|o\>$};
		\node [style=none] (27) at (2.5, -2.5) {$|p\>$};
		\node [style=none] (28) at (4, -1.25) {$|q\>$};
	\end{pgfonlayer}
	\begin{pgfonlayer}{edgelayer}
		\draw [style=arrow] (5) to (4);
		\draw [style=arrow] (4) to (1);
		\draw [style=arrow] (1) to (3);
		\draw [style=arrow] (5) to (0);
		\draw [style=arrow] (0) to (1);
		\draw [style=arrow] (0) to (2);
		\draw [style=arrow] (2) to (3);
		\draw [style=arrow] (17) to (5);
		\draw [style=arrow] (17) to (16);
		\draw [style=arrow] (16) to (4);
		\draw [style=arrow] (3) to (20);
		\draw [style=arrow] (2) to (21);
		\draw [style=arrow] (21) to (20);
		\draw [style=arrow] (21) to (22);
		\draw [style=arrow] (20) to (23);
		\draw [style=arrow] (22) to (23);
	\end{pgfonlayer}
\end{tikzpicture}

%% file: braiding3.tikz
\begin{tikzpicture}
	\begin{pgfonlayer}{nodelayer}
		\node [style=boundary vertex] (0) at (-2, -2) {};
		\node [style=boundary vertex] (1) at (-2, 0) {};
		\node [style=boundary vertex] (2) at (0, -2) {};
		\node [style=boundary vertex] (3) at (0, 0) {};
		\node [style=boundary vertex] (4) at (-4, 0) {};
		\node [style=boundary vertex] (5) at (-4, -2) {};
		\node [style=none] (8) at (-1, 0.5) {$Z^j|m\>$};
		\node [style=none] (10) at (-5, 0.5) {$Z^j|k\>$};
		\node [style=boundary vertex] (12) at (-6, 0) {};
		\node [style=boundary vertex] (13) at (-6, -2) {};
		\node [style=none] (14) at (-3, 0.5) {$Z^j|l\>$};
		\node [style=boundary vertex] (16) at (2, 0) {};
		\node [style=boundary vertex] (17) at (2, -2) {};
		\node [style=boundary vertex] (18) at (4, -2) {};
		\node [style=boundary vertex] (19) at (4, 0) {};
		\node [style=none] (23) at (2.5, -1.25) {$|q\>$};
		\node [style=none] (25) at (3.5, -1) {$s_2$};
		\node [style=none] (26) at (1.5, -1) {$s_3$};
		\node [style=none] (27) at (1, -1) {};
		\node [style=none] (28) at (3, -1) {};
		\node [style=none] (29) at (1, 0.5) {$|n\>$};
		\node [style=none] (30) at (0.5, -1) {$|o\>$};
		\node [style=none] (31) at (1, -2.5) {$|p\>$};
	\end{pgfonlayer}
	\begin{pgfonlayer}{edgelayer}
		\draw [style=arrow] (5) to (4);
		\draw [style=arrow] (4) to (1);
		\draw [style=arrow] (1) to (3);
		\draw [style=arrow] (5) to (0);
		\draw [style=arrow] (0) to (1);
		\draw [style=arrow] (0) to (2);
		\draw [style=arrow] (2) to (3);
		\draw [style=arrow] (13) to (5);
		\draw [style=arrow] (13) to (12);
		\draw [style=arrow] (12) to (4);
		\draw [style=arrow] (3) to (16);
		\draw [style=arrow] (2) to (17);
		\draw [style=arrow] (17) to (16);
		\draw [style=arrow] (17) to (18);
		\draw [style=arrow] (16) to (19);
		\draw [style=arrow] (18) to (19);
		\draw [style={red_dash}] (27.center) to (16);
		\draw [style={red_dash}] (28.center) to (19);
		\draw [style={red_dash}] (28.center) to (16);
	\end{pgfonlayer}
\end{tikzpicture}

%% file: braiding4.tikz
\begin{tikzpicture}
	\begin{pgfonlayer}{nodelayer}
		\node [style=boundary vertex] (0) at (-1, -2) {};
		\node [style=boundary vertex] (1) at (-1, 0) {};
		\node [style=boundary vertex] (2) at (1, -2) {};
		\node [style=boundary vertex] (3) at (1, 0) {};
		\node [style=boundary vertex] (4) at (-3, 0) {};
		\node [style=boundary vertex] (5) at (-3, -2) {};
		\node [style=none] (8) at (0, 0.5) {$Z^j|m\>$};
		\node [style=none] (10) at (-4, 0.5) {$Z^j|k\>$};
		\node [style=boundary vertex] (12) at (-5, 0) {};
		\node [style=boundary vertex] (13) at (-5, -2) {};
		\node [style=none] (14) at (-2, 0.5) {$Z^j|l\>$};
		\node [style=boundary vertex] (16) at (3, 0) {};
		\node [style=boundary vertex] (17) at (3, -2) {};
		\node [style=boundary vertex] (18) at (5, -2) {};
		\node [style=boundary vertex] (19) at (5, 0) {};
		\node [style=none] (23) at (4, -1.25) {$X^{i}|q\>$};
		\node [style=none] (24) at (2, 0.5) {$|n\>$};
		\node [style=none] (25) at (1.5, -1) {$|o\>$};
		\node [style=none] (26) at (2, -2.5) {$|p\>$};
	\end{pgfonlayer}
	\begin{pgfonlayer}{edgelayer}
		\draw [style=arrow] (5) to (4);
		\draw [style=arrow] (4) to (1);
		\draw [style=arrow] (1) to (3);
		\draw [style=arrow] (5) to (0);
		\draw [style=arrow] (0) to (1);
		\draw [style=arrow] (0) to (2);
		\draw [style=arrow] (2) to (3);
		\draw [style=arrow] (13) to (5);
		\draw [style=arrow] (13) to (12);
		\draw [style=arrow] (12) to (4);
		\draw [style=arrow] (3) to (16);
		\draw [style=arrow] (2) to (17);
		\draw [style=arrow] (17) to (16);
		\draw [style=arrow] (17) to (18);
		\draw [style=arrow] (16) to (19);
		\draw [style=arrow] (18) to (19);
	\end{pgfonlayer}
\end{tikzpicture}

%% file: braiding5.tikz
\begin{tikzpicture}
	\begin{pgfonlayer}{nodelayer}
		\node [style=boundary vertex] (0) at (-2, -2) {};
		\node [style=boundary vertex] (1) at (-2, 0) {};
		\node [style=boundary vertex] (2) at (0, -2) {};
		\node [style=boundary vertex] (3) at (0, 0) {};
		\node [style=boundary vertex] (4) at (-4, 0) {};
		\node [style=boundary vertex] (5) at (-4, -2) {};
		\node [style=none] (6) at (1, 0.5) {$|n\>$};
		\node [style=none] (8) at (-1, 0.5) {$Z^j|m\>$};
		\node [style=none] (10) at (-5, 0.5) {$Z^j|k\>$};
		\node [style=boundary vertex] (12) at (-6, 0) {};
		\node [style=boundary vertex] (13) at (-6, -2) {};
		\node [style=none] (14) at (-3, 0.5) {$Z^j|l\>$};
		\node [style=boundary vertex] (16) at (2, 0) {};
		\node [style=boundary vertex] (17) at (2, -2) {};
		\node [style=boundary vertex] (18) at (4, -2) {};
		\node [style=boundary vertex] (19) at (4, 0) {};
		\node [style=none] (21) at (0.5, -1) {$|o\>$};
		\node [style=none] (22) at (1, -2.5) {$|p\>$};
		\node [style=none] (23) at (3, -1.25) {$X^{i}|q\>$};
		\node [style=none] (24) at (-1, -1) {};
		\node [style=none] (25) at (-1, 1) {};
		\node [style=none] (26) at (1, 1) {};
		\node [style=none] (27) at (3, 1) {};
		\node [style=none] (28) at (3, -1) {};
		\node [style=none] (29) at (3, -3) {};
		\node [style=none] (30) at (1, -3) {};
		\node [style=none] (31) at (-1, -3) {};
		\node [style=none] (32) at (-1, -0.5) {$s_1$};
	\end{pgfonlayer}
	\begin{pgfonlayer}{edgelayer}
		\draw [style=arrow] (5) to (4);
		\draw [style=arrow] (4) to (1);
		\draw [style=arrow] (1) to (3);
		\draw [style=arrow] (5) to (0);
		\draw [style=arrow] (0) to (1);
		\draw [style=arrow] (0) to (2);
		\draw [style=arrow] (2) to (3);
		\draw [style=arrow] (13) to (5);
		\draw [style=arrow] (13) to (12);
		\draw [style=arrow] (12) to (4);
		\draw [style=arrow] (3) to (16);
		\draw [style=arrow] (2) to (17);
		\draw [style=arrow] (17) to (16);
		\draw [style=arrow] (17) to (18);
		\draw [style=arrow] (16) to (19);
		\draw [style=arrow] (18) to (19);
		\draw [style={red_dash}] (31.center) to (2);
		\draw [style={red_dash}] (2) to (24.center);
		\draw [style={red_dash}] (24.center) to (3);
		\draw [style={red_dash}] (3) to (25.center);
		\draw [style={red_dash}] (3) to (26.center);
		\draw [style={red_dash}] (26.center) to (16);
		\draw [style={red_dash}] (16) to (27.center);
		\draw [style={red_dash}] (16) to (28.center);
		\draw [style={red_dash}] (28.center) to (17);
		\draw [style={red_dash}] (17) to (29.center);
		\draw [style={red_dash}] (17) to (30.center);
		\draw [style={red_dash}] (30.center) to (2);
	\end{pgfonlayer}
\end{tikzpicture}

%% file: braiding6.tikz
\begin{tikzpicture}
	\begin{pgfonlayer}{nodelayer}
		\node [style=boundary vertex] (0) at (-1, -1) {};
		\node [style=boundary vertex] (1) at (-1, 1) {};
		\node [style=boundary vertex] (2) at (1, -1) {};
		\node [style=boundary vertex] (3) at (1, 1) {};
		\node [style=boundary vertex] (4) at (-3, 1) {};
		\node [style=boundary vertex] (5) at (-3, -1) {};
		\node [style=none] (6) at (0, 1.5) {$Z^j|m\>$};
		\node [style=none] (7) at (-4, 1.5) {$Z^j|k\>$};
		\node [style=boundary vertex] (8) at (-5, 1) {};
		\node [style=boundary vertex] (9) at (-5, -1) {};
		\node [style=none] (10) at (-2, 1.5) {$Z^j|l\>$};
		\node [style=boundary vertex] (11) at (3, 1) {};
		\node [style=boundary vertex] (12) at (3, -1) {};
		\node [style=boundary vertex] (13) at (5, -1) {};
		\node [style=boundary vertex] (14) at (5, 1) {};
		\node [style=none] (15) at (4.5, -0.25) {$Z^{j}X^{i}|q\>$};
		\node [style=none] (16) at (2, 1.5) {$Z^{-j}|n\>$};
		\node [style=none] (17) at (0, -0.25) {$Z^{-j}|o\>$};
		\node [style=none] (18) at (2, -1.5) {$Z^j|p\>$};
	\end{pgfonlayer}
	\begin{pgfonlayer}{edgelayer}
		\draw [style=arrow] (5) to (4);
		\draw [style=arrow] (4) to (1);
		\draw [style=arrow] (1) to (3);
		\draw [style=arrow] (5) to (0);
		\draw [style=arrow] (0) to (1);
		\draw [style=arrow] (0) to (2);
		\draw [style=arrow] (2) to (3);
		\draw [style=arrow] (9) to (5);
		\draw [style=arrow] (9) to (8);
		\draw [style=arrow] (8) to (4);
		\draw [style=arrow] (3) to (11);
		\draw [style=arrow] (2) to (12);
		\draw [style=arrow] (12) to (11);
		\draw [style=arrow] (12) to (13);
		\draw [style=arrow] (11) to (14);
		\draw [style=arrow] (13) to (14);
	\end{pgfonlayer}
\end{tikzpicture}

%% file: ds3_1.tikz
\begin{tikzpicture}
	\begin{pgfonlayer}{nodelayer}
		\node [style=boundary vertex] (0) at (-1, -1) {};
		\node [style=boundary vertex] (1) at (-1, 1) {};
		\node [style=boundary vertex] (2) at (1, -1) {};
		\node [style=boundary vertex] (3) at (1, 1) {};
		\node [style=boundary vertex] (4) at (-3, 1) {};
		\node [style=boundary vertex] (5) at (-3, -1) {};
		\node [style=boundary vertex] (7) at (-1, 3) {};
		\node [style=boundary vertex] (9) at (-1, 3) {};
		\node [style=boundary vertex] (11) at (1, 3) {};
		\node [style=boundary vertex] (15) at (3, 3) {};
		\node [style=boundary vertex] (16) at (3, 1) {};
		\node [style=boundary vertex] (17) at (3, -1) {};
		\node [style=boundary vertex] (18) at (3, -3) {};
		\node [style=boundary vertex] (19) at (1, -3) {};
		\node [style=boundary vertex] (20) at (-1, -3) {};
		\node [style=boundary vertex] (21) at (-3, -3) {};
		\node [style=boundary vertex] (22) at (-3, 3) {};
		\node [style=none] (23) at (-2, -2) {};
		\node [style=none] (24) at (2, -2) {};
		\node [style=none] (25) at (0, -2) {};
		\node [style=none] (26) at (-2, -1.5) {$s_0$};
		\node [style=none] (27) at (2, -1.5) {$s_1$};
	\end{pgfonlayer}
	\begin{pgfonlayer}{edgelayer}
		\draw [style=arrow] (5) to (4);
		\draw [style=arrow] (4) to (1);
		\draw [style=arrow] (1) to (3);
		\draw [style=arrow] (5) to (0);
		\draw [style=arrow] (0) to (1);
		\draw [style=arrow] (0) to (2);
		\draw [style=arrow] (2) to (3);
		\draw [style=arrow] (1) to (7);
		\draw [style=arrow] (9) to (11);
		\draw [style=arrow] (3) to (11);
		\draw [style=arrow] (4) to (22);
		\draw [style=arrow] (22) to (9);
		\draw [style=arrow] (11) to (15);
		\draw [style=arrow] (16) to (15);
		\draw [style=arrow] (3) to (16);
		\draw [style=arrow] (21) to (5);
		\draw [style=arrow] (21) to (20);
		\draw [style=arrow] (20) to (0);
		\draw [style=arrow] (20) to (19);
		\draw [style=arrow] (19) to (2);
		\draw [style=arrow] (19) to (18);
		\draw [style=arrow] (18) to (17);
		\draw [style=arrow] (17) to (16);
		\draw [style=arrow] (2) to (17);
		\draw [style={red_dash}] (23.center) to (0);
		\draw [style={red_dash}] (24.center) to (2);
		\draw [style={red_dash}] (0) to (25.center);
		\draw [style={red_dash}] (25.center) to (2);
	\end{pgfonlayer}
\end{tikzpicture}

%% file: ds3_2.tikz
\begin{tikzpicture}
	\begin{pgfonlayer}{nodelayer}
		\node [style=boundary vertex] (0) at (-1, -1) {};
		\node [style=boundary vertex] (1) at (-1, 1) {};
		\node [style=boundary vertex] (2) at (1, -1) {};
		\node [style=boundary vertex] (3) at (1, 1) {};
		\node [style=boundary vertex] (4) at (-3, 1) {};
		\node [style=boundary vertex] (5) at (-3, -1) {};
		\node [style=boundary vertex] (7) at (-1, 3) {};
		\node [style=boundary vertex] (9) at (-1, 3) {};
		\node [style=boundary vertex] (11) at (1, 3) {};
		\node [style=boundary vertex] (15) at (3, 3) {};
		\node [style=boundary vertex] (16) at (3, 1) {};
		\node [style=boundary vertex] (17) at (3, -1) {};
		\node [style=boundary vertex] (18) at (3, -3) {};
		\node [style=boundary vertex] (19) at (1, -3) {};
		\node [style=boundary vertex] (20) at (-1, -3) {};
		\node [style=boundary vertex] (21) at (-3, -3) {};
		\node [style=boundary vertex] (22) at (-3, 3) {};
		\node [style=none] (28) at (-2, 2) {};
		\node [style=none] (29) at (0, 2) {};
		\node [style=none] (30) at (2, 2) {};
		\node [style=none] (31) at (-2, 1.5) {$s_2$};
		\node [style=none] (32) at (2, 1.5) {$s_3$};
	\end{pgfonlayer}
	\begin{pgfonlayer}{edgelayer}
		\draw [style=arrow] (5) to (4);
		\draw [style=arrow] (4) to (1);
		\draw [style=arrow] (1) to (3);
		\draw [style=arrow] (5) to (0);
		\draw [style=arrow] (0) to (1);
		\draw [style=arrow] (0) to (2);
		\draw [style=arrow] (2) to (3);
		\draw [style=arrow] (1) to (7);
		\draw [style=arrow] (9) to (11);
		\draw [style=arrow] (3) to (11);
		\draw [style=arrow] (4) to (22);
		\draw [style=arrow] (22) to (9);
		\draw [style=arrow] (11) to (15);
		\draw [style=arrow] (16) to (15);
		\draw [style=arrow] (3) to (16);
		\draw [style=arrow] (21) to (5);
		\draw [style=arrow] (21) to (20);
		\draw [style=arrow] (20) to (0);
		\draw [style=arrow] (20) to (19);
		\draw [style=arrow] (19) to (2);
		\draw [style=arrow] (19) to (18);
		\draw [style=arrow] (18) to (17);
		\draw [style=arrow] (17) to (16);
		\draw [style=arrow] (2) to (17);
		\draw [style={red_dash}] (28.center) to (1);
		\draw [style={red_dash}] (1) to (29.center);
		\draw [style={red_dash}] (29.center) to (3);
		\draw [style={red_dash}] (3) to (30.center);
	\end{pgfonlayer}
\end{tikzpicture}

%% file: ds3_3.tikz
\begin{tikzpicture}
	\begin{pgfonlayer}{nodelayer}
		\node [style=boundary vertex] (0) at (-1, -1) {};
		\node [style=boundary vertex] (1) at (-1, 1) {};
		\node [style=boundary vertex] (2) at (1, -1) {};
		\node [style=boundary vertex] (3) at (1, 1) {};
		\node [style=boundary vertex] (4) at (-3, 1) {};
		\node [style=boundary vertex] (5) at (-3, -1) {};
		\node [style=boundary vertex] (6) at (-1, 3) {};
		\node [style=boundary vertex] (7) at (-1, 3) {};
		\node [style=boundary vertex] (8) at (1, 3) {};
		\node [style=boundary vertex] (9) at (3, 3) {};
		\node [style=boundary vertex] (10) at (3, 1) {};
		\node [style=boundary vertex] (11) at (3, -1) {};
		\node [style=boundary vertex] (12) at (3, -3) {};
		\node [style=boundary vertex] (13) at (1, -3) {};
		\node [style=boundary vertex] (14) at (-1, -3) {};
		\node [style=boundary vertex] (15) at (-3, -3) {};
		\node [style=boundary vertex] (16) at (-3, 3) {};
		\node [style=none] (17) at (-2, 2) {};
		\node [style=none] (20) at (-2, 1.5) {$s_2$};
		\node [style=none] (21) at (-2, -2) {};
		\node [style=none] (22) at (-2, 0) {};
		\node [style=none] (23) at (-2, -1.5) {$s_0$};
	\end{pgfonlayer}
	\begin{pgfonlayer}{edgelayer}
		\draw [style=arrow] (5) to (4);
		\draw [style=arrow] (4) to (1);
		\draw [style=arrow] (1) to (3);
		\draw [style=arrow] (5) to (0);
		\draw [style=arrow] (0) to (1);
		\draw [style=arrow] (0) to (2);
		\draw [style=arrow] (2) to (3);
		\draw [style=arrow] (1) to (6);
		\draw [style=arrow] (7) to (8);
		\draw [style=arrow] (3) to (8);
		\draw [style=arrow] (4) to (16);
		\draw [style=arrow] (16) to (7);
		\draw [style=arrow] (8) to (9);
		\draw [style=arrow] (10) to (9);
		\draw [style=arrow] (3) to (10);
		\draw [style=arrow] (15) to (5);
		\draw [style=arrow] (15) to (14);
		\draw [style=arrow] (14) to (0);
		\draw [style=arrow] (14) to (13);
		\draw [style=arrow] (13) to (2);
		\draw [style=arrow] (13) to (12);
		\draw [style=arrow] (12) to (11);
		\draw [style=arrow] (11) to (10);
		\draw [style=arrow] (2) to (11);
		\draw [style={red_dash}] (17.center) to (1);
		\draw [style={red_dash}] (22.center) to (1);
		\draw [style={red_dash}] (22.center) to (0);
		\draw [style={red_dash}] (0) to (21.center);
	\end{pgfonlayer}
\end{tikzpicture}

%% file: stronglyopen.tikz
\begin{tikzpicture}
	\begin{pgfonlayer}{nodelayer}
		\node [style=boundary vertex] (0) at (0, 0) {};
		\node [style=boundary vertex] (1) at (-2, 0) {};
		\node [style=boundary vertex] (2) at (-2, -2) {};
		\node [style=boundary vertex] (3) at (0, -2) {};
		\node [style=boundary vertex] (4) at (2, -2) {};
		\node [style=boundary vertex] (5) at (2, 0) {};
		\node [style=boundary vertex] (6) at (4, 0) {};
		\node [style=boundary vertex] (7) at (4, 2) {};
		\node [style=boundary vertex] (8) at (2, 2) {};
		\node [style=none] (10) at (-1, -1) {};
		\node [style=none] (11) at (1, -1) {};
		\node [style=none] (12) at (3, -1) {};
		\node [style=none] (13) at (3, 1) {};
		\node [style=none] (14) at (-1.5, -1) {$s_0$};
		\node [style=none] (15) at (4, 1.25) {$s_n$};
	\end{pgfonlayer}
	\begin{pgfonlayer}{edgelayer}
		\draw [style={red_dash}] (1) to (10.center);
		\draw [style={red_dash}] (10.center) to (0);
		\draw [style={red_dash}] (0) to (11.center);
		\draw [style={red_dash}] (11.center) to (5);
		\draw [style={red_dash}] (5) to (12.center);
		\draw [style={red_dash}] (5) to (13.center);
		\draw [style={red_dash}] (13.center) to (8);
		\draw [style=arrow] (2) to (1);
		\draw [style=arrow] (1) to (0);
		\draw [style=arrow] (0) to (5);
		\draw [style=arrow] (5) to (8);
		\draw [style=arrow] (8) to (7);
		\draw [style=arrow] (5) to (6);
		\draw [style=arrow] (2) to (3);
		\draw [style=arrow] (3) to (0);
		\draw [style=arrow] (3) to (4);
		\draw [style=arrow] (4) to (5);
		\draw [style={red_dash}] (13.center) to (7);
	\end{pgfonlayer}
\end{tikzpicture}

%% file: NOTstronglyopen1.tikz
\begin{tikzpicture}
	\begin{pgfonlayer}{nodelayer}
		\node [style=boundary vertex] (0) at (0, 0) {};
		\node [style=boundary vertex] (1) at (-2, 0) {};
		\node [style=boundary vertex] (2) at (-2, -2) {};
		\node [style=boundary vertex] (3) at (0, -2) {};
		\node [style=boundary vertex] (4) at (2, -2) {};
		\node [style=boundary vertex] (5) at (2, 0) {};
		\node [style=boundary vertex] (6) at (4, 0) {};
		\node [style=none] (9) at (-1, -1) {};
		\node [style=none] (10) at (1, -1) {};
		\node [style=none] (11) at (3, -1) {};
		\node [style=boundary vertex] (12) at (2, 2) {};
		\node [style=boundary vertex] (13) at (0, 2) {};
		\node [style=boundary vertex] (14) at (0, -4) {};
		\node [style=none] (15) at (3, 1) {};
		\node [style=none] (16) at (1, 1) {};
		\node [style=none] (17) at (1, -3) {};
		\node [style=boundary vertex] (18) at (4, 2) {};
		\node [style=none] (19) at (-1.5, -1) {$s_0$};
		\node [style=none] (20) at (1, -3.75) {$s_n$};
		\node [style=none] (21) at (0.25, -0.75) {$s_2$};
		\node [style=none] (22) at (1, -0.25) {$s_3$};
	\end{pgfonlayer}
	\begin{pgfonlayer}{edgelayer}
		\draw [style={red_dash}] (1) to (9.center);
		\draw [style={red_dash}] (9.center) to (0);
		\draw [style={red_dash}] (0) to (10.center);
		\draw [style={red_dash}] (10.center) to (5);
		\draw [style={red_dash}] (5) to (11.center);
		\draw [style=arrow] (2) to (1);
		\draw [style=arrow] (1) to (0);
		\draw [style=arrow] (0) to (5);
		\draw [style=arrow] (5) to (6);
		\draw [style=arrow] (2) to (3);
		\draw [style=arrow] (3) to (0);
		\draw [style=arrow] (3) to (4);
		\draw [style=arrow] (4) to (5);
		\draw [style={red_dash}] (11.center) to (6);
		\draw [style={red_dash}] (6) to (15.center);
		\draw [style={red_dash}] (15.center) to (12);
		\draw [style={red_dash}] (12) to (16.center);
		\draw [style={red_dash}] (16.center) to (13);
		\draw [style={red_dash}] (0) to (16.center);
		\draw [style={red_dash}] (3) to (10.center);
		\draw [style={red_dash}] (3) to (17.center);
		\draw [style={red_dash}] (14) to (17.center);
		\draw [style={red_dash}] (15.center) to (18);
		\draw [style=arrow] (14) to (3);
		\draw [style=arrow] (0) to (13);
		\draw [style=arrow] (13) to (12);
		\draw [style=arrow] (5) to (12);
		\draw [style=arrow] (12) to (18);
		\draw [style=arrow] (6) to (18);
	\end{pgfonlayer}
\end{tikzpicture}

%% file: NOTstronglyopen2.tikz
\begin{tikzpicture}
	\begin{pgfonlayer}{nodelayer}
		\node [style=boundary vertex] (0) at (0, 0) {};
		\node [style=boundary vertex] (1) at (-2, 0) {};
		\node [style=boundary vertex] (2) at (-2, -2) {};
		\node [style=boundary vertex] (3) at (0, -2) {};
		\node [style=boundary vertex] (4) at (2, -2) {};
		\node [style=boundary vertex] (5) at (2, 0) {};
		\node [style=boundary vertex] (6) at (4, 0) {};
		\node [style=none] (9) at (-1, -1) {};
		\node [style=none] (10) at (1, -1) {};
		\node [style=none] (11) at (3, -1) {};
		\node [style=boundary vertex] (12) at (2, 2) {};
		\node [style=boundary vertex] (18) at (4, 2) {};
		\node [style=none] (19) at (-1.5, -1) {$s_0$};
		\node [style=none] (21) at (2.25, 0.75) {$s_2$};
		\node [style=none] (22) at (3.75, 1) {$s_3$};
		\node [style=none] (23) at (3, 1) {};
		\node [style=none] (24) at (5, 1) {};
		\node [style=none] (25) at (3, 3) {};
		\node [style=none] (26) at (3, 5) {};
		\node [style=boundary vertex] (27) at (4, 4) {};
		\node [style=none] (28) at (3.75, 5) {$s_n$};
		\node [style=boundary vertex] (29) at (2, 4) {};
	\end{pgfonlayer}
	\begin{pgfonlayer}{edgelayer}
		\draw [style={red_dash}] (1) to (9.center);
		\draw [style={red_dash}] (9.center) to (0);
		\draw [style={red_dash}] (0) to (10.center);
		\draw [style={red_dash}] (10.center) to (5);
		\draw [style={red_dash}] (5) to (11.center);
		\draw [style=arrow] (2) to (1);
		\draw [style=arrow] (1) to (0);
		\draw [style=arrow] (0) to (5);
		\draw [style=arrow] (5) to (6);
		\draw [style=arrow] (2) to (3);
		\draw [style=arrow] (3) to (0);
		\draw [style=arrow] (3) to (4);
		\draw [style=arrow] (4) to (5);
		\draw [style=arrow] (5) to (12);
		\draw [style=arrow] (12) to (18);
		\draw [style=arrow] (6) to (18);
		\draw [style={red_dash}] (5) to (23.center);
		\draw [style={red_dash}] (23.center) to (6);
		\draw [style={red_dash}] (6) to (24.center);
		\draw [style={red_dash}] (24.center) to (18);
		\draw [style={red_dash}] (18) to (23.center);
		\draw [style={red_dash}] (25.center) to (18);
		\draw [style={red_dash}] (25.center) to (27);
		\draw [style={red_dash}] (27) to (26.center);
		\draw [style=arrow] (12) to (29);
		\draw [style=arrow] (18) to (27);
		\draw [style=arrow] (29) to (27);
	\end{pgfonlayer}
\end{tikzpicture}

%% file: border_example2.tikz
\begin{tikzpicture}
	\begin{pgfonlayer}{nodelayer}
		\node [style=boundary vertex] (0) at (-2, 2) {};
		\node [style=boundary vertex] (1) at (-2, 0) {};
		\node [style=boundary vertex] (2) at (0, 0) {};
		\node [style=boundary vertex] (3) at (0, -2) {};
		\node [style=boundary vertex] (4) at (2, -2) {};
		\node [style=boundary vertex] (5) at (2, 0) {};
		\node [style=boundary vertex] (6) at (-2, -2) {};
		\node [style=boundary vertex] (7) at (0, 2) {};
		\node [style=boundary vertex] (8) at (2, 2) {};
		\node [style=none] (30) at (-2.5, 1.25) {$g^3$};
		\node [style=none] (31) at (-1, 2.5) {$g^1$};
		\node [style=none] (32) at (1, 2.5) {$g^2$};
		\node [style=none] (33) at (0.5, 1.25) {$g^4$};
		\node [style=none] (34) at (2.5, 1.25) {$g^5$};
		\node [style=none] (35) at (-1, -2.5) {$g^{11}$};
		\node [style=none] (36) at (1, -2.5) {$g^{12}$};
		\node [style=none] (37) at (-1, 0.5) {$g^6$};
		\node [style=none] (38) at (1, 0.5) {$g^7$};
		\node [style=none] (39) at (-2.5, -1) {$g^8$};
		\node [style=none] (40) at (0.5, -1) {$g^9$};
		\node [style=none] (41) at (2.5, -1) {$g^{10}$};
	\end{pgfonlayer}
	\begin{pgfonlayer}{edgelayer}
		\draw [style=arrow] (2) to (5);
		\draw [style=arrow] (4) to (5);
		\draw [style=arrow] (6) to (1);
		\draw [style=arrow] (6) to (3);
		\draw [style=arrow] (2) to (7);
		\draw [style=arrow] (5) to (8);
		\draw [style=arrow] (7) to (8);
		\draw [style=arrow] (3) to (4);
		\draw [style=arrow] (3) to (2);
		\draw [style=arrow] (1) to (2);
		\draw [style=arrow] (1) to (0);
		\draw [style=arrow] (0) to (7);
		\draw [style={red_arrow}, bend left=345] (4) to (3);
		\draw [style={red_arrow}, bend left=15] (3) to (2);
		\draw [style={red_arrow}, bend left=15] (2) to (1);
		\draw [style={red_arrow}, bend right=15] (1) to (0);
		\draw [style={red_arrow}, bend right=15] (0) to (7);
	\end{pgfonlayer}
\end{tikzpicture}

%% file: border_composition.tikz
\begin{tikzpicture}
	\begin{pgfonlayer}{nodelayer}
		\node [style=boundary vertex] (0) at (-5, 1) {};
		\node [style=boundary vertex] (1) at (-5, -1) {};
		\node [style=boundary vertex] (2) at (-3, -1) {};
		\node [style=boundary vertex] (5) at (-1, -1) {};
		\node [style=boundary vertex] (7) at (-3, 1) {};
		\node [style=boundary vertex] (8) at (-1, 1) {};
		\node [style=none] (9) at (0, 0) {$=$};
		\node [style=boundary vertex] (10) at (1, 1) {};
		\node [style=boundary vertex] (11) at (1, -1) {};
		\node [style=boundary vertex] (12) at (3, -1) {};
		\node [style=boundary vertex] (13) at (5, -1) {};
		\node [style=boundary vertex] (14) at (3, 1) {};
		\node [style=boundary vertex] (15) at (5, 1) {};
		\node [style=none] (16) at (-4, 0) {$p_1$};
		\node [style=none] (17) at (-2, 0) {$p_2$};
		\node [style=none] (18) at (2, 0) {$p_1$};
		\node [style=none] (19) at (4, 0) {$p_2$};
	\end{pgfonlayer}
	\begin{pgfonlayer}{edgelayer}
		\draw [style={red_arrow}, bend left=345] (2) to (1);
		\draw [style={red_arrow}, bend right=15] (1) to (0);
		\draw [style={red_arrow}, bend right=15] (0) to (7);
		\draw [style={red_arrow}, bend right] (7) to (2);
		\draw [style={cyan_arrow}, bend right] (2) to (7);
		\draw [style={cyan_arrow}, bend right=15] (7) to (8);
		\draw [style={cyan_arrow}, bend right=15] (8) to (5);
		\draw [style={cyan_arrow}, bend left=345] (5) to (2);
		\draw [style={red_arrow}, bend right=15] (14) to (15);
		\draw [style={red_arrow}, bend right=15] (15) to (13);
		\draw [style={red_arrow}, bend left=345] (13) to (12);
		\draw [style={red_arrow}, bend left=345] (12) to (11);
		\draw [style={red_arrow}, bend right=15] (11) to (10);
		\draw [style={red_arrow}, bend right=15] (10) to (14);
		\draw [style=arrow] (2) to (7);
		\draw [style=arrow] (12) to (14);
		\draw [style=arrow] (1) to (2);
		\draw [style=arrow] (2) to (5);
		\draw [style=arrow] (5) to (8);
		\draw [style=arrow] (7) to (8);
		\draw [style=arrow] (1) to (0);
		\draw [style=arrow] (0) to (7);
		\draw [style=arrow] (11) to (10);
		\draw [style=arrow] (10) to (14);
		\draw [style=arrow] (14) to (15);
		\draw [style=arrow] (13) to (15);
		\draw [style=arrow] (12) to (13);
		\draw [style=arrow] (11) to (12);
	\end{pgfonlayer}
\end{tikzpicture}

%% file: direct_basecase.tikz
\begin{tikzpicture}
	\begin{pgfonlayer}{nodelayer}
		\node [style=boundary vertex] (0) at (-1, 0) {};
		\node [style=boundary vertex] (1) at (1, 0) {};
		\node [style=none] (2) at (0, -1) {};
		\node [style=none] (3) at (-0.75, -0.75) {$s_0$};
		\node [style=none] (4) at (0.75, -0.75) {$s_1$};
		\node [style=none] (5) at (0, 0.5) {$g^1$};
		\node [style=boundary vertex] (6) at (1, 2) {};
		\node [style=boundary vertex] (7) at (3, 0) {};
		\node [style=boundary vertex] (8) at (1, -2) {};
		\node [style=boundary vertex] (9) at (-1, 2) {};
		\node [style=boundary vertex] (10) at (-3, 0) {};
		\node [style=boundary vertex] (11) at (-1, -2) {};
		\node [style=none] (12) at (0, 1.25) {$p_2$};
	\end{pgfonlayer}
	\begin{pgfonlayer}{edgelayer}
		\draw [style={red_dash}] (0) to (2.center);
		\draw [style={red_dash}] (2.center) to (1);
		\draw [style=arrow] (0) to (1);
		\draw [style=arrow] (1) to (6);
		\draw [style=arrow] (1) to (7);
		\draw [style=arrow] (9) to (6);
		\draw [style=arrow] (0) to (9);
		\draw [style=arrow] (10) to (0);
		\draw [style=arrow] (11) to (0);
		\draw [style=arrow] (8) to (1);
		\draw [style=arrow] (11) to (8);
	\end{pgfonlayer}
\end{tikzpicture}

%% file: dual_basecase.tikz
\begin{tikzpicture}
	\begin{pgfonlayer}{nodelayer}
		\node [style=boundary vertex] (0) at (0, 0) {};
		\node [style=boundary vertex] (1) at (0, 2) {};
		\node [style=boundary vertex] (2) at (-2, 0) {};
		\node [style=boundary vertex] (3) at (0, -2) {};
		\node [style=boundary vertex] (4) at (-2, -2) {};
		\node [style=boundary vertex] (5) at (2, -2) {};
		\node [style=boundary vertex] (6) at (2, 0) {};
		\node [style=none] (7) at (-1, -1) {};
		\node [style=none] (8) at (1, -1) {};
		\node [style=none] (9) at (-1, -0.5) {$s_0$};
		\node [style=none] (10) at (1, -0.5) {$s_1$};
		\node [style=none] (11) at (0.25, -1.25) {$g^2$};
		\node [style=boundary vertex] (12) at (0, -4) {};
		\node [style=none] (13) at (-0.5, -2.5) {$v_2$};
	\end{pgfonlayer}
	\begin{pgfonlayer}{edgelayer}
		\draw [style=arrow] (0) to (1);
		\draw [style=arrow] (2) to (0);
		\draw [style=arrow] (3) to (0);
		\draw [style=arrow] (4) to (2);
		\draw [style=arrow] (4) to (3);
		\draw [style=arrow] (3) to (5);
		\draw [style=arrow] (5) to (6);
		\draw [style=arrow] (0) to (6);
		\draw [style={red_dash}] (7.center) to (0);
		\draw [style={red_dash}] (0) to (8.center);
		\draw [style=arrow] (12) to (3);
	\end{pgfonlayer}
\end{tikzpicture}

%% file: B1_ii_a.tikz
\begin{tikzpicture}
	\begin{pgfonlayer}{nodelayer}
		\node [style=boundary vertex] (0) at (0, -2) {};
		\node [style=boundary vertex] (1) at (0, 0) {};
		\node [style=boundary vertex] (2) at (2, -2) {};
		\node [style=boundary vertex] (3) at (2, 0) {};
		\node [style=boundary vertex] (4) at (-2, 0) {};
		\node [style=boundary vertex] (5) at (-2, -2) {};
		\node [style=none] (12) at (-1, -1) {};
		\node [style=none] (13) at (1, -1) {};
		\node [style=none] (14) at (-1, -0.5) {$s_0$};
		\node [style=none] (15) at (1.75, -0.75) {$s_1$};
		\node [style=none] (16) at (-0.5, -1.25) {$g^1$};
		\node [style=none] (17) at (1, 0.5) {$g^2$};
	\end{pgfonlayer}
	\begin{pgfonlayer}{edgelayer}
		\draw [style=arrow] (5) to (4);
		\draw [style=arrow] (4) to (1);
		\draw [style=arrow] (1) to (3);
		\draw [style=arrow] (5) to (0);
		\draw [style=arrow] (0) to (1);
		\draw [style=arrow] (0) to (2);
		\draw [style=arrow] (2) to (3);
		\draw [style={red_dash}] (12.center) to (1);
		\draw [style={red_dash}] (13.center) to (3);
		\draw [style={red_dash}] (1) to (13.center);
	\end{pgfonlayer}
\end{tikzpicture}

%% file: B1_ii_b.tikz
\begin{tikzpicture}
	\begin{pgfonlayer}{nodelayer}
		\node [style=boundary vertex] (0) at (0, -2) {};
		\node [style=boundary vertex] (1) at (0, 0) {};
		\node [style=boundary vertex] (2) at (2, -2) {};
		\node [style=boundary vertex] (3) at (2, 0) {};
		\node [style=none] (13) at (1, -1) {};
		\node [style=none] (15) at (0.25, -0.75) {$s_0$};
		\node [style=none] (16) at (1.25, -1.75) {$s_1$};
		\node [style=none] (17) at (2.75, 0.25) {$v_3$};
	\end{pgfonlayer}
	\begin{pgfonlayer}{edgelayer}
		\draw [style=arrow] (1) to (3);
		\draw [style=arrow] (0) to (1);
		\draw [style=arrow] (0) to (2);
		\draw [style=arrow] (2) to (3);
		\draw [style={red_dash}] (13.center) to (3);
		\draw [style={red_dash}] (1) to (13.center);
		\draw [style={red_dash}] (13.center) to (2);
	\end{pgfonlayer}
\end{tikzpicture}

%% file: B1_ii_c.tikz
\begin{tikzpicture}
	\begin{pgfonlayer}{nodelayer}
		\node [style=boundary vertex] (1) at (-1, 1) {};
		\node [style=boundary vertex] (3) at (1, 1) {};
		\node [style=none] (4) at (0, 0) {};
		\node [style=none] (5) at (-2.5, 0.25) {$s_0$};
		\node [style=none] (7) at (0.25, -0.5) {$p_3$};
		\node [style=boundary vertex] (9) at (-1, -1) {};
		\node [style=none] (11) at (-2, 0) {};
		\node [style=none] (12) at (0, 2) {};
		\node [style=none] (13) at (-0.5, 2.25) {$s_1$};
		\node [style=boundary vertex] (14) at (-3, 1) {};
		\node [style=boundary vertex] (15) at (-1, 3) {};
	\end{pgfonlayer}
	\begin{pgfonlayer}{edgelayer}
		\draw [style=arrow] (1) to (3);
		\draw [style={red_dash}] (1) to (4.center);
		\draw [style=arrow] (9) to (1);
		\draw [style={red_dash}] (11.center) to (1);
		\draw [style={red_dash}] (1) to (12.center);
		\draw [style=arrow] (14) to (1);
		\draw [style=arrow] (1) to (15);
	\end{pgfonlayer}
\end{tikzpicture}

%% file: direct_nonadjacent.tikz
\begin{tikzpicture}
	\begin{pgfonlayer}{nodelayer}
		\node [style=boundary vertex] (0) at (-1, 0) {};
		\node [style=boundary vertex] (1) at (1, 0) {};
		\node [style=none] (2) at (0, -1) {};
		\node [style=boundary vertex] (8) at (1, -2) {};
		\node [style=boundary vertex] (10) at (-3, 0) {};
		\node [style=boundary vertex] (11) at (-1, -2) {};
		\node [style=none] (12) at (-2, -1) {};
		\node [style=none] (13) at (-4, -1) {};
		\node [style=none] (14) at (-4, -3) {};
		\node [style=none] (15) at (-6, -3) {};
		\node [style=boundary vertex] (16) at (-3, -2) {};
		\node [style=boundary vertex] (17) at (-5, -2) {};
		\node [style=boundary vertex] (18) at (-5, 0) {};
		\node [style=boundary vertex] (19) at (-3, -4) {};
		\node [style=boundary vertex] (20) at (-5, -4) {};
		\node [style=boundary vertex] (21) at (-7, -2) {};
		\node [style=boundary vertex] (22) at (-7, -4) {};
		\node [style=none] (23) at (-6, -2.25) {$s_0$};
		\node [style=none] (24) at (-0.75, -0.75) {$s_2$};
		\node [style=none] (25) at (0.75, -0.75) {$s_1$};
		\node [style=none] (26) at (-1, 0.5) {$v$};
		\node [style=none] (27) at (0, -0.5) {$\tau$};
		\node [style=none] (28) at (-4.5, -2.75) {$\xi'$};
	\end{pgfonlayer}
	\begin{pgfonlayer}{edgelayer}
		\draw [style={red_dash}] (0) to (2.center);
		\draw [style={red_dash}] (2.center) to (1);
		\draw [style=arrow] (0) to (1);
		\draw [style=arrow] (10) to (0);
		\draw [style=arrow] (11) to (0);
		\draw [style=arrow] (8) to (1);
		\draw [style=arrow] (11) to (8);
		\draw [style={red_dash}] (14.center) to (19);
		\draw [style={red_dash}] (14.center) to (16);
		\draw [style={red_dash}] (16) to (13.center);
		\draw [style={red_dash}] (13.center) to (10);
		\draw [style={red_dash}] (10) to (12.center);
		\draw [style={red_dash}] (12.center) to (0);
		\draw [style={red_dash}] (20) to (14.center);
		\draw [style={red_dash}] (15.center) to (20);
		\draw [style={red_dash}] (15.center) to (22);
		\draw [style={red_dash}] (15.center) to (21);
		\draw [style=arrow] (21) to (17);
		\draw [style=arrow] (17) to (18);
		\draw [style=arrow] (18) to (10);
		\draw [style=arrow] (17) to (16);
		\draw [style=arrow] (16) to (10);
		\draw [style=arrow] (16) to (11);
		\draw [style=arrow] (19) to (16);
		\draw [style=arrow] (20) to (19);
		\draw [style=arrow] (22) to (21);
		\draw [style=arrow] (22) to (20);
		\draw [style=arrow] (20) to (17);
	\end{pgfonlayer}
\end{tikzpicture}

%% file: dual_nonadjacent.tikz
\begin{tikzpicture}
	\begin{pgfonlayer}{nodelayer}
		\node [style=boundary vertex] (0) at (0, 1) {};
		\node [style=boundary vertex] (1) at (0, -1) {};
		\node [style=boundary vertex] (2) at (2, -1) {};
		\node [style=boundary vertex] (3) at (2, 1) {};
		\node [style=boundary vertex] (4) at (4, 1) {};
		\node [style=boundary vertex] (5) at (4, -1) {};
		\node [style=boundary vertex] (6) at (2, -3) {};
		\node [style=boundary vertex] (7) at (4, -3) {};
		\node [style=boundary vertex] (8) at (6, -1) {};
		\node [style=boundary vertex] (9) at (6, -3) {};
		\node [style=boundary vertex] (10) at (6, -5) {};
		\node [style=boundary vertex] (11) at (8, -5) {};
		\node [style=boundary vertex] (12) at (8, -3) {};
		\node [style=boundary vertex] (13) at (8, -1) {};
		\node [style=none] (14) at (7, -4) {};
		\node [style=none] (15) at (7, -2) {};
		\node [style=none] (16) at (5, -2) {};
		\node [style=none] (17) at (3, -2) {};
		\node [style=none] (18) at (3, 0) {};
		\node [style=none] (19) at (1, 0) {};
		\node [style=none] (20) at (0.25, 0.25) {$s_0$};
		\node [style=none] (21) at (7.75, -3.75) {$s_1$};
		\node [style=none] (22) at (7.75, -2.25) {$s_2$};
		\node [style=none] (23) at (6.5, -2.5) {$p$};
		\node [style=none] (24) at (3, -0.5) {$\xi'$};
		\node [style=none] (25) at (7.25, -3.25) {$\tau^*$};
	\end{pgfonlayer}
	\begin{pgfonlayer}{edgelayer}
		\draw [style={red_dash}] (8) to (15.center);
		\draw [style={red_dash}] (16.center) to (8);
		\draw [style={red_dash}] (5) to (16.center);
		\draw [style={red_dash}] (17.center) to (5);
		\draw [style={red_dash}] (17.center) to (2);
		\draw [style={red_dash}] (2) to (18.center);
		\draw [style={red_dash}] (18.center) to (3);
		\draw [style={red_dash}] (19.center) to (3);
		\draw [style={red_dash}] (19.center) to (0);
		\draw [style=arrow] (1) to (0);
		\draw [style=arrow] (0) to (3);
		\draw [style=arrow] (3) to (4);
		\draw [style=arrow] (5) to (4);
		\draw [style=arrow] (1) to (2);
		\draw [style=arrow] (2) to (3);
		\draw [style=arrow] (2) to (5);
		\draw [style=arrow] (5) to (8);
		\draw [style=arrow] (8) to (13);
		\draw [style=arrow] (6) to (2);
		\draw [style=arrow] (6) to (7);
		\draw [style=arrow] (7) to (5);
		\draw [style=arrow] (7) to (9);
		\draw [style=arrow] (9) to (8);
		\draw [style=arrow] (9) to (12);
		\draw [style=arrow] (12) to (13);
		\draw [style=arrow] (10) to (9);
		\draw [style=arrow] (10) to (11);
		\draw [style=arrow] (11) to (12);
		\draw [style={red_dash}] (15.center) to (13);
		\draw [style={red_dash}] (15.center) to (12);
		\draw [style={red_dash}] (14.center) to (12);
	\end{pgfonlayer}
\end{tikzpicture}

%% file: ds3_4.tikz
\begin{tikzpicture}
	\begin{pgfonlayer}{nodelayer}
		\node [style=boundary vertex] (0) at (-2, -1) {};
		\node [style=boundary vertex] (1) at (-2, 1) {};
		\node [style=boundary vertex] (2) at (0, -1) {};
		\node [style=boundary vertex] (3) at (0, 1) {};
		\node [style=boundary vertex] (4) at (-4, 1) {};
		\node [style=boundary vertex] (5) at (-4, -1) {};
		\node [style=boundary vertex] (7) at (-2, 3) {};
		\node [style=boundary vertex] (9) at (-2, 3) {};
		\node [style=boundary vertex] (11) at (0, 3) {};
		\node [style=boundary vertex] (15) at (2, 3) {};
		\node [style=boundary vertex] (16) at (2, 1) {};
		\node [style=boundary vertex] (17) at (2, -1) {};
		\node [style=boundary vertex] (18) at (2, -3) {};
		\node [style=boundary vertex] (19) at (0, -3) {};
		\node [style=boundary vertex] (20) at (-2, -3) {};
		\node [style=boundary vertex] (21) at (-4, -3) {};
		\node [style=boundary vertex] (22) at (-4, 3) {};
		\node [style=none] (28) at (-3, 2) {};
		\node [style=none] (30) at (1, 2) {};
		\node [style=none] (31) at (-3, 1.5) {$s_2$};
		\node [style=none] (32) at (1, 1.5) {$s_3$};
		\node [style=boundary vertex] (33) at (10, -1) {};
		\node [style=boundary vertex] (34) at (10, 1) {};
		\node [style=boundary vertex] (35) at (12, -1) {};
		\node [style=boundary vertex] (36) at (12, 1) {};
		\node [style=boundary vertex] (37) at (8, 1) {};
		\node [style=boundary vertex] (38) at (8, -1) {};
		\node [style=boundary vertex] (39) at (10, 3) {};
		\node [style=boundary vertex] (40) at (10, 3) {};
		\node [style=boundary vertex] (41) at (12, 3) {};
		\node [style=boundary vertex] (42) at (14, 3) {};
		\node [style=boundary vertex] (43) at (14, 1) {};
		\node [style=boundary vertex] (44) at (14, -1) {};
		\node [style=boundary vertex] (45) at (14, -3) {};
		\node [style=boundary vertex] (46) at (12, -3) {};
		\node [style=boundary vertex] (47) at (10, -3) {};
		\node [style=boundary vertex] (48) at (8, -3) {};
		\node [style=boundary vertex] (49) at (8, 3) {};
		\node [style=none] (50) at (9, 2) {};
		\node [style=none] (52) at (13, 2) {};
		\node [style=none] (53) at (9, 1.5) {$s_6$};
		\node [style=none] (54) at (13, 1.5) {$s_7$};
		\node [style=none] (59) at (-3, -2) {};
		\node [style=none] (60) at (1, -2) {};
		\node [style=none] (61) at (-3, -2.5) {$s_0$};
		\node [style=none] (62) at (1, -2.5) {$s_1$};
		\node [style=none] (63) at (9, -2) {};
		\node [style=none] (64) at (13, -2) {};
		\node [style=none] (65) at (9, -2.5) {$s_4$};
		\node [style=none] (66) at (13, -2.5) {$s_5$};
		\node [style=none] (67) at (5, 0) {$\cdots$};
		\node [style=none] (68) at (4, -3) {};
		\node [style=none] (69) at (4, -1) {};
		\node [style=none] (70) at (4, 1) {};
		\node [style=none] (71) at (4, 3) {};
		\node [style=none] (72) at (6, 3) {};
		\node [style=none] (73) at (6, 1) {};
		\node [style=none] (74) at (6, -1) {};
		\node [style=none] (75) at (6, -3) {};
		\node [style=none] (76) at (-1, 4) {$a$};
		\node [style=none] (77) at (11, 4) {$b$};
	\end{pgfonlayer}
	\begin{pgfonlayer}{edgelayer}
		\draw [style=arrow] (5) to (4);
		\draw [style=arrow] (4) to (1);
		\draw [style=arrow] (1) to (3);
		\draw [style=arrow] (5) to (0);
		\draw [style=arrow] (0) to (1);
		\draw [style=arrow] (0) to (2);
		\draw [style=arrow] (2) to (3);
		\draw [style=arrow] (1) to (7);
		\draw [style=arrow] (9) to (11);
		\draw [style=arrow] (3) to (11);
		\draw [style=arrow] (4) to (22);
		\draw [style=arrow] (22) to (9);
		\draw [style=arrow] (11) to (15);
		\draw [style=arrow] (16) to (15);
		\draw [style=arrow] (3) to (16);
		\draw [style=arrow] (21) to (5);
		\draw [style=arrow] (21) to (20);
		\draw [style=arrow] (20) to (0);
		\draw [style=arrow] (20) to (19);
		\draw [style=arrow] (19) to (2);
		\draw [style=arrow] (19) to (18);
		\draw [style=arrow] (18) to (17);
		\draw [style=arrow] (17) to (16);
		\draw [style=arrow] (2) to (17);
		\draw [style=arrow] (38) to (37);
		\draw [style=arrow] (37) to (34);
		\draw [style=arrow] (34) to (36);
		\draw [style=arrow] (38) to (33);
		\draw [style=arrow] (33) to (34);
		\draw [style=arrow] (33) to (35);
		\draw [style=arrow] (35) to (36);
		\draw [style=arrow] (34) to (39);
		\draw [style=arrow] (40) to (41);
		\draw [style=arrow] (36) to (41);
		\draw [style=arrow] (37) to (49);
		\draw [style=arrow] (49) to (40);
		\draw [style=arrow] (41) to (42);
		\draw [style=arrow] (43) to (42);
		\draw [style=arrow] (36) to (43);
		\draw [style=arrow] (48) to (38);
		\draw [style=arrow] (48) to (47);
		\draw [style=arrow] (47) to (33);
		\draw [style=arrow] (47) to (46);
		\draw [style=arrow] (46) to (35);
		\draw [style=arrow] (46) to (45);
		\draw [style=arrow] (45) to (44);
		\draw [style=arrow] (44) to (43);
		\draw [style=arrow] (35) to (44);
		\draw [style=arrow] (18) to (68.center);
		\draw [style=arrow] (17) to (69.center);
		\draw [style=arrow] (16) to (70.center);
		\draw [style=arrow] (15) to (71.center);
		\draw [style=arrow] (75.center) to (48);
		\draw [style=arrow] (74.center) to (38);
		\draw [style=arrow] (73.center) to (37);
		\draw [style=arrow] (72.center) to (49);
		\draw [style={red_dash}] (28.center) to (1);
		\draw [style={red_dash}] (3) to (30.center);
		\draw [style={red_dash}] (59.center) to (0);
		\draw [style={red_dash}] (2) to (60.center);
		\draw [style={red_dash}] (63.center) to (33);
		\draw [style={red_dash}] (35) to (64.center);
		\draw [style={red_dash}] (36) to (52.center);
		\draw [style={red_dash}] (50.center) to (34);
	\end{pgfonlayer}
\end{tikzpicture}

%% file: Kab.tikz
\begin{tikzpicture}
	\begin{pgfonlayer}{nodelayer}
		\node [style=none] (0) at (-4, 2) {};
		\node [style=none] (1) at (-4, 0) {};
		\node [style=none] (2) at (0, 2) {};
		\node [style=none] (3) at (0, 0) {};
		\node [style=cnot ctrl] (8) at (-2, 2) {};
		\node [style=cnot ctrl] (9) at (-2, 0) {};
		\node [style=gate] (10) at (-3, 2) {$H$};
		\node [style=gate] (11) at (-3, 0) {$H$};
		\node [style=gate] (12) at (-1, 0) {$H$};
		\node [style=gate] (13) at (-1, 2) {$H$};
		\node [style=none] (14) at (-3.75, -0.75) {$b$};
		\node [style=none] (15) at (-3.75, 2.75) {$a$};
		\node [style=none] (16) at (4, 2) {};
		\node [style=none] (17) at (4, 0) {};
		\node [style=none] (18) at (6, 2) {};
		\node [style=none] (19) at (6, 0) {};
		\node [style=none] (26) at (4.5, -0.75) {$b$};
		\node [style=none] (27) at (4.5, 2.75) {$a$};
		\node [style=X dot] (28) at (5, 2) {};
		\node [style=X dot] (29) at (5, 0) {};
		\node [style=hadamard] (30) at (5, 1) {};
	\end{pgfonlayer}
	\begin{pgfonlayer}{edgelayer}
		\draw (9) to (8);
		\draw (0.center) to (10);
		\draw (10) to (8);
		\draw (8) to (13);
		\draw (13) to (2.center);
		\draw (1.center) to (11);
		\draw (11) to (9);
		\draw (9) to (12);
		\draw (12) to (3.center);
		\draw (16.center) to (28);
		\draw (28) to (18.center);
		\draw (28) to (30);
		\draw (30) to (29);
		\draw (29) to (17.center);
		\draw (29) to (19.center);
	\end{pgfonlayer}
\end{tikzpicture}

%% file: ds3_5.tikz
\begin{tikzpicture}
	\begin{pgfonlayer}{nodelayer}
		\node [style=none] (0) at (-1, 2) {};
		\node [style=none] (3) at (1, 0) {};
		\node [style=none] (4) at (-0.5, -0.75) {$b$};
		\node [style=none] (5) at (-0.5, 2.75) {$a$};
		\node [style=X dot] (6) at (0, 2) {};
		\node [style=X dot] (7) at (0, 0) {};
		\node [style=hadamard] (8) at (0, 1) {};
		\node [style=X dot] (9) at (-1, 0) {};
		\node [style=X dot] (10) at (1, 2) {};
		\node [style=none] (11) at (2, 1) {$=$};
		\node [style=none] (12) at (3, 2) {};
		\node [style=none] (13) at (5, 0) {};
		\node [style=hadamard] (14) at (4, 1) {};
		\node [style=none] (15) at (3.25, 2.75) {$a$};
		\node [style=none] (16) at (4.75, -0.75) {$b$};
	\end{pgfonlayer}
	\begin{pgfonlayer}{edgelayer}
		\draw (0.center) to (6);
		\draw (6) to (8);
		\draw (8) to (7);
		\draw (7) to (3.center);
		\draw (9) to (7);
		\draw (6) to (10);
		\draw [in=-180, out=0, looseness=1.75] (12.center) to (13.center);
	\end{pgfonlayer}
\end{tikzpicture}

%% file: ds3_6.tikz
\begin{tikzpicture}
	\begin{pgfonlayer}{nodelayer}
		\node [style=none] (0) at (-3, 2) {};
		\node [style=none] (1) at (-1, 0) {};
		\node [style=none] (2) at (-2.5, -0.75) {$b$};
		\node [style=none] (3) at (-2.5, 2.75) {$a$};
		\node [style=X dot] (4) at (-2, 2) {};
		\node [style=X dot] (5) at (-2, 0) {};
		\node [style=hadamard] (6) at (-2, 1) {};
		\node [style=X dot] (7) at (-3, 0) {};
		\node [style=none] (9) at (0, 1) {$=$};
		\node [style=none] (10) at (1, 2) {};
		\node [style=none] (11) at (3, 0) {};
		\node [style=hadamard] (12) at (2, 1) {};
		\node [style=none] (13) at (1.25, 2.75) {$a$};
		\node [style=none] (14) at (2.75, -0.75) {$b$};
		\node [style=X phase dot] (15) at (-1, 2) {$\pi$};
		\node [style=Z phase dot] (16) at (3, 0) {$\pi$};
		\node [style=none] (17) at (4, 0) {};
	\end{pgfonlayer}
	\begin{pgfonlayer}{edgelayer}
		\draw (0.center) to (4);
		\draw (4) to (6);
		\draw (6) to (5);
		\draw (5) to (1.center);
		\draw (7) to (5);
		\draw [in=-180, out=0, looseness=1.75] (10.center) to (11.center);
		\draw (4) to (15);
		\draw (16) to (17.center);
	\end{pgfonlayer}
\end{tikzpicture}